\documentclass[12pt]{iopart}
\usepackage[british]{babel}
\usepackage{mathtools}
\usepackage{iopams} 
\usepackage{amsmath,bm}
\usepackage{graphicx}
\usepackage{epstopdf}
\usepackage{esint}
\usepackage{hyperref,cite}
\usepackage{enumerate}
\usepackage{amsthm}
\usepackage{color}
\usepackage{xspace}

\begin{document}
\title[]{Reconfiguration, swelling and tagged monomer dynamics of a single polymer chain in Gaussian and non-Gaussian active baths}
\author{Koushik Goswami$^1$, Subhasish Chaki$^2$ and Rajarshi Chakrabarti*}
\address{$^1$ Institute of Physics \& Astronomy, University of Potsdam, Karl-Liebknecht-Stra$\beta$e 24/25,  14476 Potsdam, Germany
\\
$^2$ Department of Materials Science and Engineering, University of Illinois, 1304 West Green Street, Urbana, Illinois 61801, USA
\\
*Department of Chemistry, Indian Institute of Technology Bombay, Mumbai, Powai 400076, India. *E-mail: rajarshi@chem.iitb.ac.in}

\begin{abstract}
\noindent 
In this topical review, we give an overview of the structure and dynamics of a single polymer chain in active baths, Gaussian or non-Gaussian. The review begins with the discussion of single flexible or semiflexible linear polymer chains subjected to two noises, thermal and active. The active noise has either Gaussian or non-Gaussian distribution but has a memory, accounting for the persistent motion of the active bath particles. This finite persistence makes the reconfiguration dynamics of the chain slow as compared to the purely thermal case and the chain swells. The active noise also results superdiffusive or ballistic motion of the tagged monomer. We present all the calculations in details but mainly focus on the analytically exact or almost exact results on the topic, as obtained from our group in recent years. In addition, we briefly mention important works of other groups and include some of our new results. The review concludes with pointing out the implications of polymer chains in active bath in biologically relevant context and its future directions.
\end{abstract}

\maketitle
\section{Introduction \label{intro}}
\noindent Self-propulsion is a major achievement of biological evolution and is 
essential for the survival of living species such as bacteria, cells, algae and other micro-organisms \cite{wu2000particle, bechinger2016active,de2017single, krishnamurthy2016micrometre}. Locomotion in search for food, orientation toward light, and spreading of the own species are only possible due to self-propulsion \cite{de2017single}. On the mesoscopic level, it has been believed that the self-propulsion is achieved due to the additional input of energy from the environment, or the internal conversion of chemical energy and thus, the equilibrium properties such as detailed balance, zero net energy flux and fluctuation-dissipation theorem fail to hold. There is a plethora of examples of self-propelled systems in nature, usually termed as ``active matter'', extending from individual entity such as motile cell, spermatozoa, swimming bacteria to collective existence such as swarming of bees, schooling of fishes, flocking of birds, to name a few \cite{czirok2000collective}. In the search of nutrients  bacteria swims around  using  its flagella; a kinesin motor  protein carrying cargo  walks on the microtubule track and its movement is powered by ATP hydrolysis; an  enzyme which is catalytically active shows enhanced diffusion and usually moves  towards the lesser concentration gradient of substrate (antichemotaxis). However, over the past decade, artificial prototypes of active matter  have also been developed and have enriched the field. For example, a Pt-silica Janus particle in $H_2O_2$ solution self-propels and exhibits nonequilibrium behavior due to diffusiophoresis, just like a bacteria \cite{PhysRevE.88.032304}. The physics of microswimmers is fairly different from that we experience in our macroscopic world \cite{elgeti2015physics}. For bees, fishes and birds, inertia is dominating over the viscous forces and for microswimmers like bacteria, it is the other way around. For microswimmers, the motion predominantly occurs at low Reynolds number and is greatly influenced by the athermal fluctuations arising from some active processes fueled by ATP hydrolysis \cite{bechinger2016active}. They propel in a persistent manner and show enhanced motion in the presence of activity as the active fluctuations usually has long-lived correlations compared to the thermal one, and the propulsion and activity are often coupled \cite{jepson2013enhanced,valeriani2011colloids,PhysRevE.91.062124,ramaswamy2017active,fodor2018statistical}.

\noindent Here our main focus is to discuss a tiny portion of the active matter panorama: single polymer chain under active fluctuations \cite{doi:10.1063/1.4891095,chelakkot2014flagellar,kaiser2015does,babel2016dynamics, chaki2022polymer, Samanta_2016, singhjphysa,osmanovic2017dynamics, ghosh2014dynamics, winkler2020physics, peterson,ghosh2022}. As mentioned earlier, many cellular processes are controlled by the activity of its constituents such as nucleic acids (RNA and DNA), proteins and enzymes which are naturally occurring polymeric systems.  In contrast to the macroscopic systems, the size of the above-said polymers are of mesoscopic scale, and so their motion  predominantly occurs at low Reynolds number and is greatly influenced by the fluctuations arising from their surroundings and hydrodynamic interactions.  For the connected particles as in polymeric chains, understanding the collective or conformational behavior is essential to derive physical variables, and their mechanical properties such as shape, size and deformability often determine the behavior. Unlike the colloidal particles, the coupling between the activity and the conformational modulations of polymers gives rise to novel structural features of individual polymers. For example, Spermatozoa typically move by a snake-like wiggling of their flexible tail or myosin motors walking along actin filaments generate power strokes that affect the conformations of the cytoskeletal network. In the context of chromosomes, it has been experimentally shown that the activity can give rise to the coherent motions of chromatin loci on micron length scale \cite{weber2012nonthermal,zidovska2013micron,javer2013short}. However, in the recent past years, the studies of active polymeric systems have seen a great upsurge; some studies have considered the motion of a flexible  chain comprised of passive beads in an active environment \cite{vandebroek2015dynamics,doi:10.1063/1.4891095,Shin_2015,Samanta_2016,doi:10.1063/1.5086152,mousavi2021active, biswas2017, eisenstecken2016conformational}; a similar prototypical realization is a chain of connected active particles in a thermal bath \cite{osmanovic2017dynamics,osmanovic2018properties}.  One of its upgradation is to introduce rigidity in the chain backbone which results its net movement in a preferred direction and so it is often referred to as polar active system \cite{martin2019active,winkler2020physics}.   Other studies include the flow-mediated relaxation, linear viscoelastic response, emerging beat patterns, dynamics of cyclic active polymers, activity-induced loop extrusion, stress-induced propulsion of filaments etc \cite{mousavi2019active,PhysRevLett.109.158302,PhysRevLett.126.097801}. 
\\
\\
Understanding the timescales in which a biopolymer such as peptide or protein undergoes conformational rearrangement is of great relevance in the scientific community. In dense liquid, conformational changes in polymer are driven by thermal fluctuations and thus follow diffusive dynamics of the chain. A theoretical framework for describing these conformational transitions are provided by the Kramers' model of barrier crossing rate where the rate of folding  depends exponentially on the height of the folding free energy barrier, with a prefactor representing the ``attempt frequency" of crossing the barrier. This prefactor can be related to the rate of reconfiguration (or ``reconfiguration time") of unfolded and nonnative conformations at which the protein can diffusively explore its conformational space \cite{soranno2012quantifying}. The fluorescence correlation spectroscopy (FCS) experiments have been used to study the reconfiguration time \cite{makarov2010spatiotemporal}.  The basic principle relies on the  F\"{o}rster resonance energy transfer between a pair of dyes attached to distant monomers along a polymer chain \cite{widengren1998fluorescence,PhysRevE.73.041919,makarov2010spatiotemporal}. By monitoring the intensity autocorrelation function in the fluorescence intensities of the donor and acceptor, which is intimately related to the fluctuations in the distance between the donor and acceptor, one can obtain the reconfiguration time. Hence, investigating the reconfiguration time of active polymer is highly important as it is an essential measure of the characteristic time scale of conformational changes of a single chain driven by active fluctuations.  \\ \\


\noindent The review is arranged as follows. Section \ref{intro} is devoted to the introduction, while section \ref{sec2} deals with the models of single polymer chains, namely flexible and semiflexible. Section \ref{single particle} discusses the dynamics of a harmonically confined particle in active baths, Gaussian as well as non-Gaussian. In section \ref{sec4}, equation of motions of single polymer chains in active baths are analytically treated and discussed in detail. Section \ref{sec-5} and \ref{sec-6} respectively deal with the conformational properties and the dynamics of the chain. In this review, we focus primarily on the reconfiguration dynamics, which is treated in section \ref{sec-7} of the chain, the tagged monomer motion (section \ref{sec-6}) and conformational properties of the chain (section \ref{sec-5}). In particular, we restrict ourselves to analytically exact results. In section \ref{conclusion}, we give an overview of the current status of the topic and point out future directions. Throughout the paper, we have adopted a pedagogical style of writing, keeping in mind that the readers outside the community of active matter and polymer physics, with interests in biological physics can follow and appreciate the review.

\section{ Models of a single polymer  \label{model} \label{sec2}}
\noindent When it comes to a many-body problem, such as  polymers, where  neighbouring interactions are important, the single-particle description (given in Sec. \ref{single particle}) cannot be employed directly. In the following section we briefly outline  the models used to capture the conformational and dynamical properties of polymers. 

\subsection{Flexible polymer}
\noindent The simplest description for a polymer is Rouse chain \cite{doi1988theory}  which is composed of $N$ number of beads (or monomers) connected by simple harmonic springs with spring constant $\mu$, $\mu=\frac{3 k_B T }{b^2},$ $b$ is the Kuhn length and $k_B T$ is the thermal energy. In this model, the interaction among monomers is ``local" in the sense that it persists only up to the adjacent monomers which are linked to each other by the elastic force \cite{doi1988theory}.  So the total energy is given by $\mathcal{H}(\bm{r}_1,\bm{r}_2,\cdots,\bm{r}_N)=\frac{\mu}{2}\sum_{i=2}^{N}\left(\bm{r}_i-\bm{r}_{i-1}\right)^2,$ which, in the continuum limit, can be expressed as   
\begin{align}
    \mathcal{H}[\bm{r}]=\frac{\mu}{2}\int_{0}^{N}dn\,\left(\frac{\partial \bm{r}(n,t)}{\partial n}\right)^2,
\end{align}
 where $\bm{r}(n,t)$ is the position of the $n^{th}$ bead at time $t.$ As the chain is suspended in a thermal bath, each monomer is subjected to random thermal force. Therefore, in the free draining limit (here we ignore the hydrodynamics interaction), the equation of motion (EOM) of the  $n^{th}$ monomer can be expressed as \cite{doi1988theory}
 \begin{align}
\gamma \frac{\partial}{\partial t}\bm{r}(n,t)=-\frac{\delta \mathcal{H}[\bm{r}]}{\delta \bm{r}}+\bm{\eta}(n,t)=\mu\frac{\partial^2 \bm{r}(n,t)}{\partial n^2}+\bm{\eta}(n,t),\label{rouse_dynamics}
 \end{align}
 with the boundary conditions for the end beads:
 $$\frac{\partial \bm{r}(n,t)}{\partial n}\Big|_{n=0,N}=0.$$
 Here, $\bm{\eta}(n, t)$ is the thermal noise acting on the $n^{th}$ monomer and $\gamma $  is the friction coefficient per unit bead. Usually, $\bm{\eta}(n, t)$ is  considered as the white Gaussian noise with the following properties: $\langle \bm{\eta}_i(n, t)\rangle=0,$ and $\langle \bm{\eta}_i(m, t) \bm{\eta}_j(n, t')\rangle=2\gamma k_B T \delta_{ij}\delta(m-n)\delta(t-t'),$ where $\{i,j\}\in \{x,y,z\}.$
 \\
 \\
 In the presence of an active noise $\sigma(t)$ \cite{szamel2014self, Samanta_2016, nandi2017nonequilibrium,chaki2018,goswami2019diffusion, chaki2019effects,Goswami2019}, one requires to  include the force term into the dynamics, and therefore, Eq. (\ref{rouse_dynamics}) modifies to 
\begin{align}
     \gamma \frac{\partial}{\partial t}\bm{r}(n,t)=\mu\frac{\partial^2 \bm{r}(n,t)}{\partial n^2}+\bm{\eta}(n,t)+\bm{\sigma}(n,t)\label{rouse_dynamics}.
\end{align}
Here, the active force on the $n^{th}$ monomer is denoted by  $\bm{\sigma}(n,t),$  which  keeps the system out of equilibrium as it violates the fluctuation-dissipation theorem. The noise $\bm{\sigma}(n,t)$ is assumed as a zero-mean, stationary and non-Markovian process, and its contributions on different monomer positions are uncorrelated, $i.e.,$ $$\langle \bm{\sigma}_i(m, t)\rangle=0,\,\;\langle \bm{\sigma}_i(m, t) \bm{\sigma}_j(n, t')\rangle=\Gamma \delta_{ij}\delta(m-n)\mathcal{C}(|t-t'|),$$  where $\mathcal{C}(t)$ is the correlation function in time with noise amplitude $\Gamma.$ Note that there are several realizations of the active noise and its respective statistical properties are discussed in Sec. \ref{single particle}.
\subsection{Semiflexible polymer}
\noindent Unlike a flexible chain,   a semiflexible chain has a finite bending rigidity which put some restrictions in bond angles, and so it involves an energetic cost to bend the backbone of the chain.  In the continuum limit, such chain is usually described by a space curve $\bm{r}(s,t),$ where $s$ denotes the coordinate along the contour, and for a polymer of length $L,$ $s\in (0, L).$ Note that $\bm{r}(s,t)$ is well defined, $i.e.,$ it is continuous and differentiable at each $s.$  For this system, the total energy is given by \cite{saito1967statistical,PhysRevLett.77.5389,ranjith2002dynamics}
\begin{align}
    \mathcal{H}[\bm{r}]=\frac{1}{2}\int_{0}^{L}ds\,\left[\kappa\left(\frac{\partial^2 \bm{r}(s,t)}{\partial s^2}\right)^2+\mu(s)\left(\frac{\partial \bm{r}(s,t)}{\partial s}\right)^2\right],
\end{align}
 where the first term inside the bracket corresponds to the bending energy with the bending constant $\kappa,$ \cite{doi1988theory} and the second one, as mentioned in the case of flexible polymer, captures the stretching contributions determined by the stretching coefficient $\mu(s).$  
 In an active bath, the dynamics of $\bm{r}(s,t)$ is governed by the  overdamped EOM \cite{aragon1985dynamics,ghosh2014dynamics}
 \begin{align}
\gamma \frac{\partial}{\partial t}\bm{r}(s,t)=-\kappa\frac{\partial^4 \bm{r}(s,t)}{\partial s^4}+\frac{\partial }{\partial s}\left(\mu(s)\frac{\partial \bm{r}(s,t)}{\partial s}\right)+\bm{\eta}(s,t)+\bm{\sigma}(s,t),\label{semiflexible_dynamics}
 \end{align}
 with the following boundary conditions:
 \begin{align}
 &\frac{\partial }{\partial s}[\mu(s=0)\bm{r}(s=0,t)]=\kappa\frac{\partial^3 }{\partial s^3}\bm{r}(s= 0,t)=0,\nonumber\\
 & \frac{\partial }{\partial s}[\mu(s=L)\bm{r}(s=L,t)]=\kappa\frac{\partial^3 }{\partial s^3}\bm{r}(s= L,t)=0.\label{bcs_semiflexible}
 \end{align}
In Eq. (\ref{semiflexible_dynamics}), the two  noise terms are   $\bm{\eta}(s,t)$  and $\bm{\sigma}(s,t),$ and they hold the same meaning as the previous case. In this model, there exists differential tensions along the contour of the chain subjected to some  nonequilibrium forcing. Therefore $\mu(s)$ can be interpreted as the tension along the contour which preserves  the local constraint: 
$
\Big\langle \left(\frac{\partial \bm{r}(s,t)}{\partial s}\right)^2\Big\rangle=1,$ where $\frac{\partial \bm{r}(s,t)}{\partial s}=\hat{u}$ denotes the unit tangent vector at contour-location $s.$  The two-point  correlation  between  tangent vectors decays over a characteristic lengthscale $l_p$, $viz.$, $\langle \hat{u}(s)\hat{u}(s')\rangle=e^{-\frac{|s-s'|}{l_p}}$, $l_p$ is called the persistence length.  Note that polymers with higher persistence lengths are more rigid, that is to say, $l_p \propto \kappa.$   Now as a simple approximation, the mean tension can be assumed throughout the contour, so one can   replace $\mu(s)$ by its average value $\mu.$ For such a case, the stiffness constant $\kappa$ is related to the persistence length $l_p$ via $\kappa=\frac32 l_p k_B T,$  and $\mu=\frac{3\,k_B T}{2 l_p}$  \cite{winkler1994models}. \\ 

\begin{figure}
\centering
\includegraphics[width=0.65\linewidth]{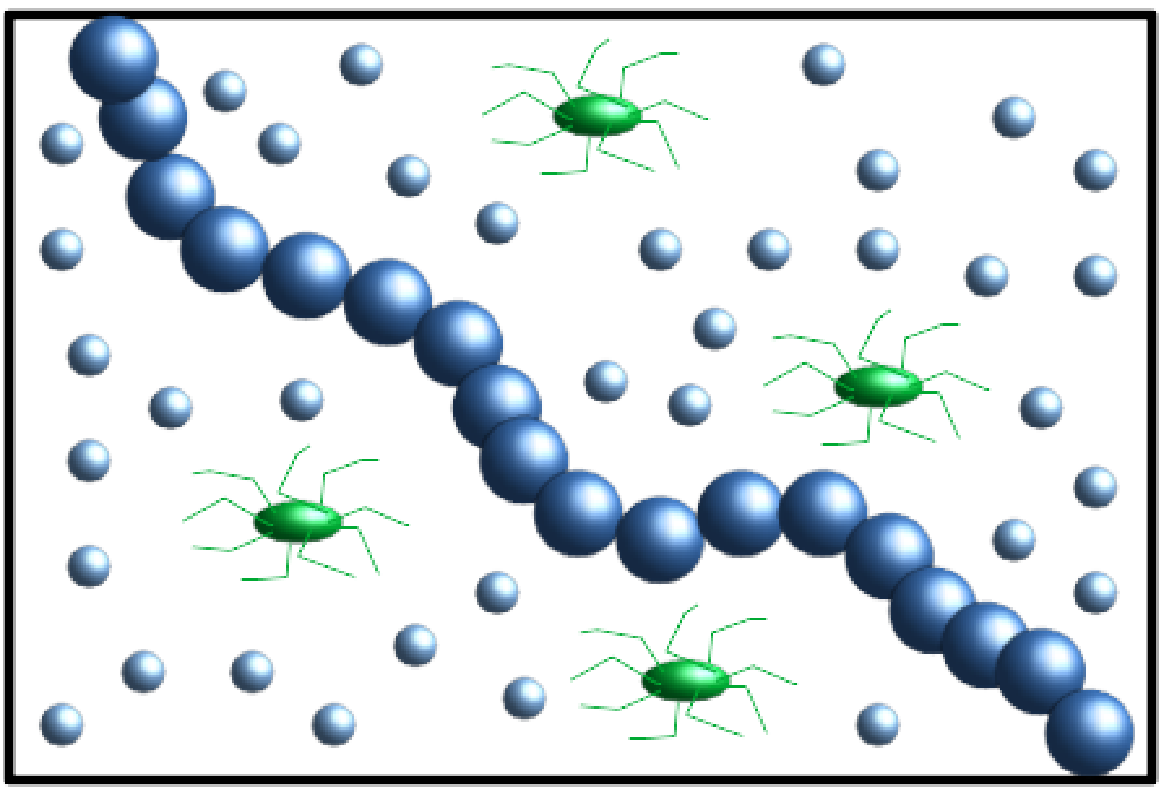}
\caption{A cartoon showing a polymer (blue curve) in a bath containing active particles such as flagellated bacteria (green). The blue small circles represent water molecules.\label{figure1}  }
\includegraphics[width=0.65\linewidth]{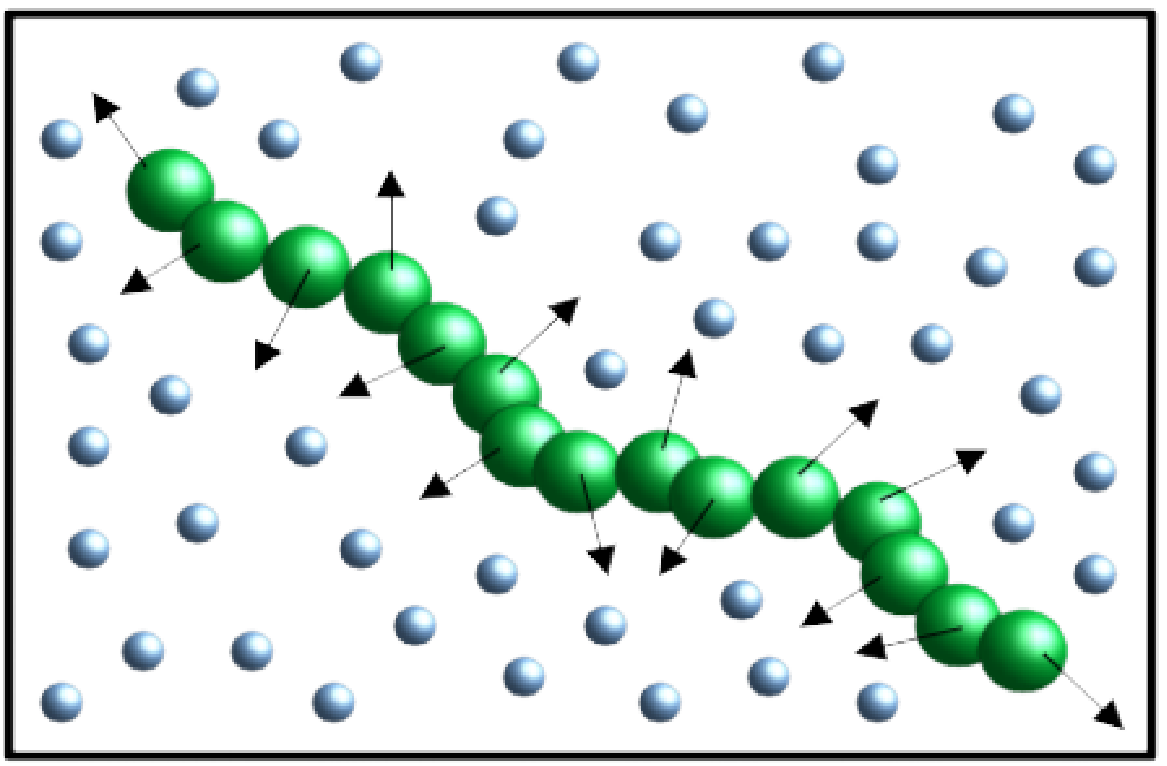}
\caption{A representation of an active polymer (green curve) in a thermal bath. The arrows show the direction of the self-propulsion. \label{figure2}  }
\label{polymer-cartoon}
 \end{figure}

\noindent It is worth noting here that a slightly different description of the semiflexible polymer was proposed by Winkler's group \cite{winkler1994models,harnau1995dynamic}. In their model, the local constraint of the tangent vector was dropped and it was replaced by a more relaxed (and global) condition: $\int_{0}^{L}ds\,\Big\langle\left(\frac{\partial \bm{r}(s,t)}{\partial s}\right)^2\Big\rangle=L.$ This introduces some degrees of  internal elasticity in the system, and thus some distinct conformational and dynamical properties have been observed \cite{eisenstecken2017internal,winkler2020physics}. Nevertheless,  the present model [described by Eqs. (\ref{semiflexible_dynamics}) and (\ref{bcs_semiflexible})] captures the key essence of the semiflexibility, that is, the interplay between rigidity and elasticity. Also, it is analytically more tractable as the formalism follows straightforwardly similar to the case of flexible one, and thus it is more handy from a theorist's perspective. For example, see Ref. \cite{ghosh2014dynamics} where this model is used to describe active fluctuations of a semiflexible polymer (figures \ref{figure1} and \ref{figure2}).

\section{ Diffusion of a harmonically confined  particle  in active bath  \label{single particle}}
 \noindent In this section, we shall digress from the main topic, but it may be useful for our latter discussion. Here we consider the the dynamics of a single passive particle in a harmonic potential of the form:  $U_p(x)=\frac{\lambda}{2}x_p^2,$ where the subscript $p$  is used to label a specific (say, $p-$th) particle (it may be more clear in the context of polymer dynamics). Apart from thermal fluctuations described by $\eta_p(t)$, the particle is driven by the active noise  $\sigma_p$ which arises due to the interaction with the active particles present in the surroundings \cite{szamel2014self, Samanta_2016,goswami2019diffusion}. In the overdamped limit, the dynamics can be described by the following stochastic equation \cite{chaki2018,chaki2019effects,goswami2019heat,goswami2021work}: 
 \begin{align}
 \frac{\partial}{\partial t}x_p(t)=-\frac{1}{\tau_p}x_p(t)+\frac{1}{\gamma_p}\eta_p(t)+\frac{1}{\gamma_p}\sigma_p(t)\label{langevin} ,
\end{align}
where the relaxation time is $\tau_p=\frac{\gamma_p}{\lambda},$ and $\gamma_p$ is the friction coefficient which is related to the thermal diffusivity $D_T$ via the Einstein relation: $D_T\,\gamma_p=k_B T.$ However, the noise $\sigma_p(t)$ does not follow the fluctuation-dissipation theorem (FDT) and so the particle is driven away from equilibrium. $\sigma_p(t)$ can be realized in different ways as discussed in the next subsections, but one of the key features that distinguish it from the thermal noise is its characteristic timescale $\tau_A$ which is finite unlike  zero  in the case of the thermal fluctuations. This generates persistent  motion,  which means that the particle moves on an average in a particular direction over a time period $\tau_A.$ 
\\
\\
Once the properties of the noises are given, the solution of Eq. (\ref{langevin}) can be found. Here we consider the noises to be random, acting on the particle without any directional bias, which implies that $\langle \eta_p\rangle=\langle \sigma_p\rangle=0.$ Also, the dynamics of two noises are decoupled, $viz.$, $\langle \eta_p( t) \sigma_p( t')\rangle=0.$ So the average position can be calculated as 
\begin{align}
  \langle x_p(t)\rangle= \frac{1}{\gamma_p}\int_{-\infty}^{t}dt'\,e^{-\frac{(t-t')}{\tau_p}}\left(\langle \eta_p(t')\rangle+\langle \sigma_p(t')\rangle \right)=0,
\end{align}
and the auto-correlation function is 
\begin{align}
\langle x_p(t) x_p(t')\rangle=\frac{1}{\gamma_p^2}\int_{-\infty}^{t}dt_1 \int_{-\infty}^{t'}dt_2\,e^{-\frac{(t-t_1)}{\tau_p}-\frac{(t'-t_2)}{\tau_p}}\left(\langle \eta_p(t_1)\eta_p(t_2)\rangle+\langle \sigma_p(t_1)\sigma_p(t_2)\rangle\right).\label{correlation}
\end{align}
For the free-particle case, Eq. (\ref{langevin}) reduces to
 \begin{align}
 \frac{\partial}{\partial t}x_0(t)=\frac{1}{\gamma_0}\eta_0(t)+\frac{1}{\gamma_0}\sigma_0(t)\label{langevin0}.
\end{align}
 To distinguish it from the confined case, we take $p=0$ to represent the free-particle dynamics.
From the above, we can write  $x_0(t)=x_0(0)+ \frac{1}{\gamma_0}\int_{0}^{t}dt'\,\left(\eta_0(t')+\sigma_0(t')\right).$ Consequently, one can obtain $\left<x_0(t)\right>=\left<x_0(0)\right>+ \frac{1}{\gamma_0}\int_{0}^{t}dt'\,\left(\left<\eta_0(t')\right>+\left<\sigma_0(t')\right>\right)=0,$ and the mean square displacement (MSD) is given by 
\begin{align}
    \langle x_0^2(t)\rangle=\frac{1}{\gamma_0^2}\int_{0}^{t}dt_1 \int_{0}^{t}dt_2\,\left(\langle \eta_0(t_1)\eta_0(t_2)\rangle+\langle \sigma_0(t_1)\sigma_0(t_2)\rangle\right).\label{msd_free0}
\end{align}
In the following subsections, we discuss about the models for active noise, and then find out relevant properties of the particle in such baths.

\subsection{Gaussian active bath \label{sec-GCN}}
Here the active noise is modeled as the Gaussian colored noise, which can be realized as \cite{PhysRevLett.117.038103,Goswami2019,chaki2019effects,goswami2022inertial,goswami2022motion}
\begin{align}
    \dot{\sigma_p}(t)=-\frac{1}{\tau_A}\sigma_p(t)+\frac{1}{\tau_A}\eta_w(t),
\end{align}
with $\tau_A$ being the persistence time and $\eta_w$ being the white Gaussian noise, $i.e.$, $\langle \eta_w(t)\eta_w(t')\rangle=2C_{\sigma}\delta(t-t')$ and $\langle \eta_w(t)\rangle=0.$ This is known as the  Ornstein-Uhlenbeck process (OUP), and so it is referred as the OUP model. For such a case, one can easily find the auto-correlation function to be 
$$\langle \sigma_p( t) \sigma_p( t')\rangle=\frac{C_{\sigma_1}}{\tau_A}\,\text{exp}\left[-\frac{|t-t'|}{\tau_A}\right].$$ 
\\
\\
Another way of realizing the active noise is through the modified Ornstein-Uhlenbeck process (MOUP) \cite{chaki2018, chaki2019effects, szamel2014self, Samanta_2016, nandi2017nonequilibrium,goswami2022inertial}for which  the governing equation of $\sigma_p( t)$  can be expressed as
$$\dot{\sigma_p}(t)=-\frac{1}{\tau_A}\sigma_p(t)+\sqrt{\frac{1}{\tau_A}}\eta_w(t),$$ and so it is also exponentially correlated with the auto-correlation function $$\langle \sigma_p( t) \sigma_p( t')\rangle=C_{\sigma_2 }\,\text{exp}\left[-\frac{|t-t'|}{\tau_A}\right].$$
Notice that at $\tau_A\rightarrow 0,$ the correlation vanishes for the above case, but in the OUP model, it becomes delta-correlated. The coefficient of  autocorrelation in the MOUP-noise model is independent of the correlation time $\tau_A,$ and it can be mapped to that of OUP model if one replaces $C_{\sigma_2}$ by $C_{\sigma_1}/\tau_A.$ As mentioned earlier, the thermal noise  has the delta correlation, 
$$\langle \eta_p( t) \eta_p( t')\rangle=C_{\eta}\delta(t-t').$$
Using the above results, one can compute the position-position correlation function for a confined particle from Eq. (\ref{correlation}). In the case of OUP model, it reads \cite{Samanta_2016}
\begin{align}
\langle x_p(t) x_p(t')\rangle&=\frac{1}{\gamma_p^2}\int_{-\infty}^{t}dt_1 \int_{-\infty}^{t'}dt_2\,e^{-\frac{(t-t_1)}{\tau_p}-\frac{(t'-t_2)}{\tau_p}}\left(\langle \eta_p(t_1)\eta_p(t_2)\rangle+\langle \sigma_p(t_1)\sigma_p(t_2)\rangle\right)\nonumber\\
&=\frac{C_{\eta}\tau_p}{2\gamma_p^2}e^{-\frac{|t-t'|}{\tau_p}}+\frac{C_{\sigma_1}}{\gamma_p^2}\frac{ \tau _p^2 \left(\tau _A e^{-\frac{|t-t'|}{\tau _A}}-\tau _p e^{-\frac{|t-t'|}{\tau _p}}\right)}{\tau _A^2-\tau _p^2}.\label{correlationx_Gaussian}
\end{align}
See Appendix. \ref{appen1} for the derivation. 
Without active noise, the correlation is described by a single exponential, whereas the multi-exponential behavior is found in the case of active diffusion.   
Taking $t=t',$ we can compute the mean square displacement (MSD) which is given by 
\begin{align}
 \langle x_p^2(t) \rangle=  \langle x_p^2(0) \rangle=\frac{C_{\eta}\tau_p}{2\gamma_p^2}+\frac{C_{\sigma_1} \tau _p^2}{\gamma_p^2(\tau_A+\tau_p)}\label{msd_gauss_p}.
\end{align}
So the variance takes a stationary value which is inversely proportional to $\gamma_p^2.$ This is obvious as the particle travels a lesser distance in a more viscous medium. Notice that at $\tau_A\rightarrow 0,$ the second term in the RHS of Eq. (\ref{msd_gauss_p}) takes a form which is similar to the thermal part (first term on the RHS) which reconfirms the fact that the active noise behaves like the thermal one in this limit. In the MOUP model,  the MSD can be calculated in a straightforward way from Eqs. (\ref{correlationx_Gaussian}) upon replacing $C_{\sigma_1}$ by $C_{\sigma_2}\tau_A.$ Notice that, the second term vanishes at $\tau_A\rightarrow 0,$ as the active noise is switched off in this model.
\\
\\
 Now we can compute the MSD of a free particle  in the OUP bath. This can be found easily using Eq. (\ref{msd_free0}), and it reads \cite{Samanta_2016}
\begin{align}
 \langle(x_0(t)-x_0(0))^2\rangle &= \frac{1}{\gamma_0^2}\int_{0}^{t}dt_1\int_{0}^{t}dt_2\,\left(\langle\eta_0(t_1)\eta_0(t_2)\rangle+\langle\sigma_0(t_1)\sigma_0(t_2)\rangle\right) \nonumber\\
 &=\frac{C_{\eta}}{ \gamma_0^2}\int_{0}^{t}dt_1\int_{0}^{t}dt_2\,\delta(t_1-t_2)+\frac{2 C_{\sigma_1}}{ \tau_A\gamma_0^2}\int_{0}^{t}dt_1\int_{0}^{t_1}dt_2\,e^{-\frac{|t-t'|}{\tau_A}}\nonumber\\
 &=\frac{C_{\eta}}{ \gamma_0^2}\int_{0}^{t}dt_1+\frac{2 C_{\sigma_1}}{ \gamma_0^2}\int_{0}^{t}dt_1\left(1-e^{-\frac{t_1}{\tau_A}}\right)\nonumber\\
 &=\frac{C_{\eta}}{ \gamma_0^2}t+\frac{2 C_{\sigma_1} \tau_A}{ \gamma_0^2}\left(\frac{t}{\tau_A}-1+e^{-\frac{t}{\tau_A}}\right)\label{msdg_x0}.
\end{align}
In the limit $\tau_A\gg t,$ the MSD approximates to 
$\langle(x_0(t)-x_0(0))^2\rangle \approx \frac{C_{\eta}}{ \gamma_0^2}t+\frac{C_{\sigma_1} }{ \gamma_0^2 \tau_A} t^2,$ which suggests a ballistic behavior. On the other hand, for $\tau_A\ll t,$  the dynamics becomes diffusive with an effective diffusivity $\left(\frac{C_{\eta}}{ \gamma_0^2}+\frac{2 C_{\sigma_1} }{ \gamma_0^2}\right).$

\subsection{Non-Gaussian active bath}
\begin{figure}
\centering
\includegraphics[width=0.65\linewidth]{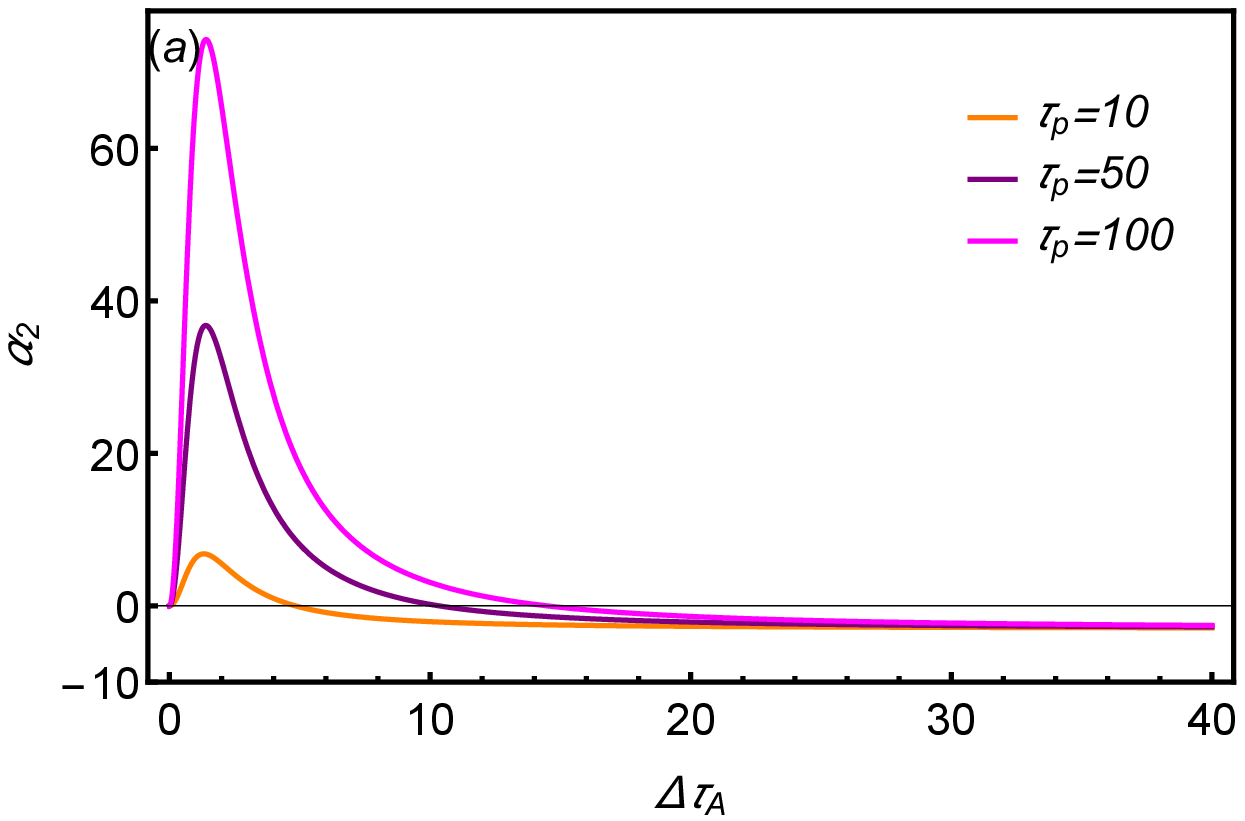}
\includegraphics[width=0.65\linewidth]{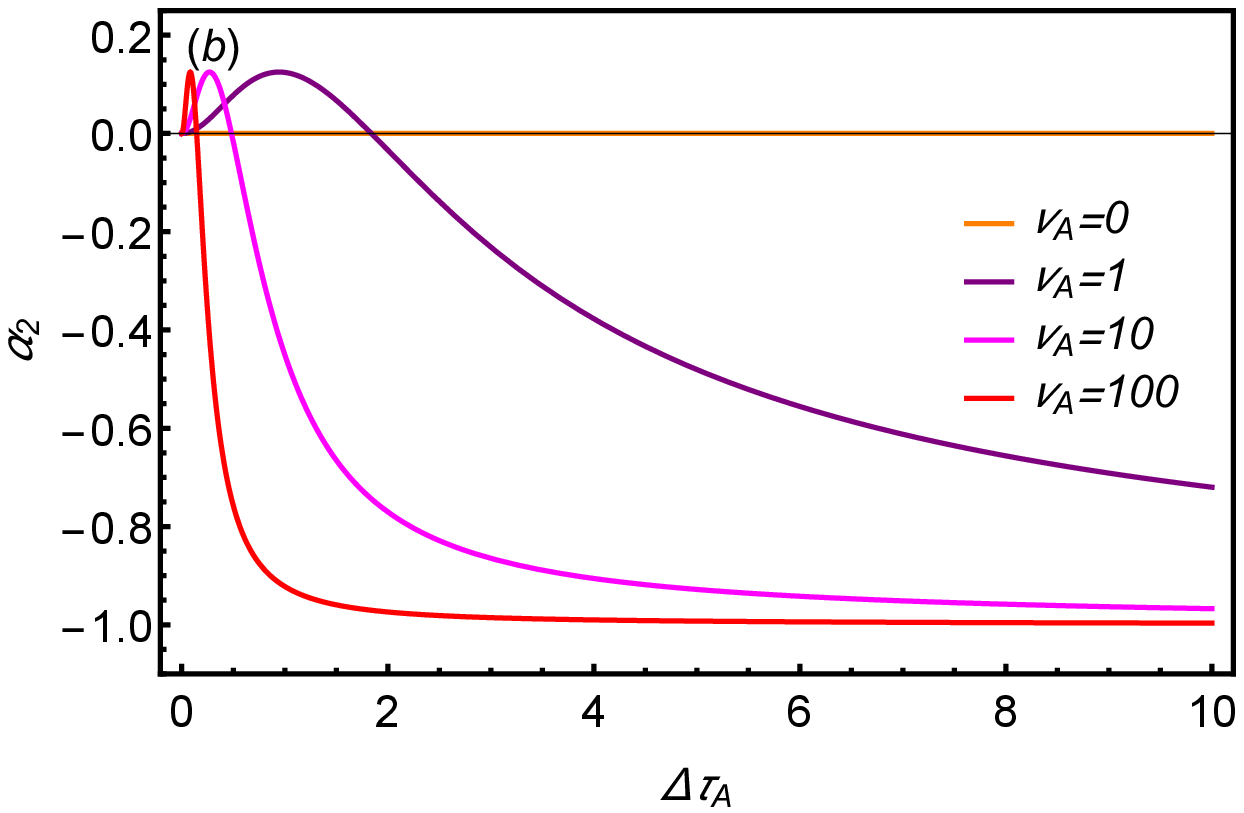}
\caption{Plot of the non-Gaussian parameter as a function of $\Delta \tau_A$ (a) for different values of $\tau_p$ taking $\nu_a=1,$ and (b) for different values of $\nu_A$ taking $\tau_p=1.$ Other parameters are taken as unity. }
     \label{ngp-single}
 \end{figure}

 \noindent The non-Gaussian noise $\sigma_p(t)$ can be realized as a sequence of pulses with width $\Delta \tau_A$ occurring at random times $t_i$ which is exponentially distributed with a constant rate $\nu_A.$ So $\sigma_p(t)$ is referred to as the shot noise which  can be expressed as \cite{chaki2019effects}
\begin{align}
\sigma_p(t)=\sum_i\,h_i\,g(t-t_i)\label{sigmap},
\end{align}
where the amplitude takes only two values, $i.e.$, $h_i=\pm 1$ and the pulse has the form of \cite{doi:10.1063/1.5086152}
\begin{align}
    g(t)=\sigma_A [\Theta(t)-\Theta(t-\Delta \tau_A)],\label{g(t)}
\end{align}
 where $\Theta(t)$ is the usual Heaviside step function.
\\
\\
For reader's convenience, the two-point correlation given by Eq. (\ref{correlation_shot}) is rewritten here,
\begin{align}
 \langle\sigma_p(t)\sigma_p(t')\rangle=\sigma_A^2\nu_A \left(\Delta \tau_A-|t-t'|\right)\,\Theta\left(\Delta \tau_A-|t-t'|\right). \label{correlation_nong_shot} 
\end{align}
For small $\Delta \tau_A$, one can use the Taylor series expansion of $\Theta(t-\Delta \tau_A)$ and then, it is trivial to show $g(t)=\sigma_A \Delta \tau_A \delta(t)$. Using the Eq. \ref{correlation_nong_shot}, one can compute the MSD for a free particle and the second-order correlation for the confined particle as shown in Appendices  \ref{appen3} and  \ref{appen4}. The results are given below. 
The MSD for the free particle is 
\begin{align}
 \langle(x_0(t)-x_0(0))^2\rangle &= \frac{1}{\gamma_0^2}\int_{0}^{t}dt_1\int_{0}^{t}dt_2\,\left(\langle\eta_0(t_1)\eta_0(t_2)\rangle+\langle\sigma_0(t_1)\sigma_0(t_2)\rangle\right) \nonumber\\
 &=\frac{C_{\eta}}{ \gamma_0^2}t+\frac{\sigma_A^2\nu_A}{3\gamma_0^2} \left(\left(\Delta \tau _A-t\right)^3 \Theta \left(\Delta \tau _A-t\right)+  \Delta \tau _A^2(3t-\Delta \tau _A)\right).\label{msd_x0_nong}
 \end{align}
The position-position autocorrelation function  for the harmonically confined particle is given by 
\begin{align}
\langle x_p(t) x_p(t')\rangle &=\frac{1}{\gamma_p^2}\int_{-\infty}^{t}dt_1 \int_{-\infty}^{t'}dt_2\,e^{-\frac{(t-t_1)}{\tau_p}-\frac{(t'-t_2)}{\tau_p}}\left(\langle \eta_p(t_1)\eta_p(t_2)\rangle+\langle \sigma_p(t_1)\sigma_p(t_2)\rangle\right)\nonumber\\
&=\frac{C_{\eta}\tau_p}{2\gamma_p^2}e^{-\frac{(t-t')}{\tau_p}}+\frac{\sigma_A^2}{\gamma_p^2}\tau_p^3\nu_A\,e^{-\frac{|t-t'|}{\tau_p}}\left(\text{cosh}\left(\frac{\Delta \tau_A}{\tau_p}\right)-1\right)\nonumber\\
&+\frac{\sigma_A^2}{\gamma_p^2}\tau_p^2\nu_A\,\Theta\left(\Delta \tau_A-|t-t'|\right)\left(\Delta \tau_A-|t-t'|\right)\left(1-\frac{\text{sinh}\left(\frac{1}{\tau_p}\left(\Delta \tau_A-|t-t'|\right)\right)}{\frac{1}{\tau_p}\left(\Delta \tau_A-|t-t'|\right)}\right).\label{correlation_nong}
\end{align}
The MSD at the stationary state for the confined particle can be calculated taking $t=t'$ in the above equation, and it reads
\begin{align}
\langle x_p^2(t)\rangle &=\frac{C_{\eta}\tau_p}{2\gamma_p^2}+\frac{\sigma_A^2}{\gamma_p^2}\tau_p^3\nu_A\,\left(\text{cosh}\left(\frac{\Delta \tau_A}{\tau_p}\right)-1+\frac{\Delta \tau_A}{\tau_p}-\text{sinh}\left(\frac{\Delta\tau_A}{\tau_p}\right)\right)\nonumber\\
&=\frac{C_{\eta}\tau_p}{2\gamma_p^2}+\frac{\sigma_A^2}{\gamma_p^2}\tau_p^3\nu_A\,\left(\text{exp}\left(-\frac{\Delta \tau_A}{\tau_p}\right)+\frac{\Delta \tau_A}{\tau_p}-1\right).\label{msd_nong}
\end{align}
\\
\\
In a nonequilibrium bath, the distribution may deviate from the Gaussianity which can be estimated from a non-dimensional parameter known as the non-Gaussian parameter (NGP). It denoted as $\alpha_2(t)$ may be defined in one dimension as 
\begin{align}
\alpha_2(t)=\frac13\frac{\langle x^4(t)\rangle}{\langle x^2(t)\rangle^2}-1.\label{ngp1d}
\end{align}
For a Gaussian distribution, $\alpha_2(t)=0;$ any values other than zero indicates the non-Gaussian nature. Using Eqs. (\ref{msd_nong}) and (\ref{x^4_nongauss}), one can obtain the NGP at the stationary state:
\begin{align}
 &\text{The non-Gaussian parameter}=\alpha_2 \nonumber\\
 &=\frac{3\langle x_p^2(t) \rangle^2+3\nu_A\tau_p^5\,\frac{\sigma_A^4}{\gamma_p^4}\left(\text{exp}\left(-\frac{\Delta \tau_A}{\tau_p}\right)+\frac{\Delta \tau_A}{\tau_p}-1\right)\left[1+\nu_A(\tau_p-\Delta \tau_A)-\nu_A\tau_p\text{exp}\left(-\frac{\Delta \tau_A}{\tau_p}\right)\right]}{3\langle x_p^2(t) \rangle^2}-1\nonumber\\
 &=\frac{3\nu_A\tau_p^5\,\frac{\sigma_A^4}{\gamma_p^4}\left(\text{exp}\left(-\frac{\Delta \tau_A}{\tau_p}\right)+\frac{\Delta \tau_A}{\tau_p}-1\right)\left[1+\nu_A(\tau_p-\Delta \tau_A)-\nu_A\tau_p\text{exp}\left(-\frac{\Delta \tau_A}{\tau_p}\right)\right]}{3\left[\frac{C_{\eta}\tau_p}{2\gamma_p^2}+\frac{\sigma_A^2}{\gamma_p^2}\tau_p^3\nu_A\,\left(\text{exp}\left(-\frac{\Delta \tau_A}{\tau_p}\right)+\frac{\Delta \tau_A}{\tau_p}-1\right)\right]^2}
\end{align}
 For $\nu_A=0,$ the active noise is absent and so the distribution is Gaussian. The negative or positive NGP imply that the underlying distribution has less- or more pronounced tails than the Gaussian distribution, respectively. For a harmonically trapped particle, the NGP becomes negative on increasing the activity as shown in Fig. \ref{ngp-single}.  However, for small persistence time, we observe a positive non-Gaussian parameter with increase in the thermal relaxation time ($\tau_p$).

\section{Analytical treatment to EOMs\label{sec4}}
\noindent The standard analytical approach to solve the EOMs  given by Eq. (\ref{rouse_dynamics}) and  Eq. (\ref{semiflexible_dynamics}) is to employ the normal-mode analysis.  In the following, we illustrate the method for two models of polymer. 
 
\subsection{Flexible polymer}
Here $\bm{r}(n,t)$ may be decomposed into time-dependent normal modes $\bm{\psi}(t)$ as \cite{doi1988theory}
\begin{align}
\bm{r}(n,t)=\bm{\psi}_0(t)+2\sum_{p=0}^{\infty} \mathcal{U}_p(n)\bm{\psi}_p(t),\label{eigen_expansion0}
\end{align}
where $\mathcal{U}_p(n)$ is the eigenfunction of the operator, $\mathcal{L}=-\mu\frac{\partial^2 \bm{r}(n,t)}{\partial n^2},$ with eigenvalue $\epsilon_p=\mu\left(\frac{p \pi}{N}\right)^2.$ For the Rouse model,  $\mathcal{U}_p(n)=\text{cos}\left(\frac{p \pi n}{N}\right),$ and form a orthogonal set, $viz.,$ $\int_{0}^{N} dn\,\text{cos}\left(\frac{p \pi n}{N}\right) \text{cos}\left(\frac{q \pi n}{N}\right)=\frac{N}{2}\delta_{p q}.$   Using Eq. (\ref{eigen_expansion0}) in Eq. (\ref{rouse_dynamics}), one may arrive at 
\begin{align}
\gamma_p \frac{\partial}{\partial t}\bm{\psi}_p(t)=-\mu_p\bm{\psi}(t)+\bm{\eta}_p(t)+\bm{\sigma}_p(t),\label{normalrouse_dynamics}
\end{align}
which describes the dynamics of $p^{th}$ Rouse mode. In this mode, $\mu_p=\epsilon_p\frac{\gamma_p}{\gamma},$ and the  friction coefficient is  $\gamma_p,$ and it is given by $\gamma_p=2N \gamma$ for $p\neq 0,$ and $\gamma_0=N \gamma.$ So the relaxation time for the $p^{th}$ mode can be defined as $\tau_p=\frac{\gamma_p}{\mu_p}=\frac{\gamma\,\gamma_p}{\epsilon_p\,\gamma_p},$ which is given by $\tau_p=\frac{\gamma N^2 b^2}{3 k_B T \pi^2 p^2}=\frac{\tau_1}{p^2},$ where the longest relaxation time is $\tau_1=\frac{\gamma N^2 b^2}{3 \pi^2 k_B T  },$ and $\mu_p=\frac{2 N\gamma}{\tau_1}p^2=\left(\frac{6 \pi^2 k_B T}{N b^2}\right)p^2=\mu_1 p^2.$ In Eq. (\ref{normalrouse_dynamics}), the final two  terms represent the mode amplitudes for the two noises. For thermal noise, the second-order correlation of its mode transforms to 
$$\langle \bm{\eta}_{pi}( t) \bm{\eta}_{qj}( t')\rangle=\frac{\gamma_p^2}{N \gamma}(1+\delta_{p0}) k_B T \delta_{ij}\delta_{pq}\delta(t-t').$$
Similarly, the properties of normal modes for the active noise slightly modify, which we mention  on  considering a specific model. Now the solution of Eq. (\ref{normalrouse_dynamics}) can be expressed as \begin{align}
   \bm{\psi}_p(t)= \frac{1}{\gamma_p}\int_{-\infty}^{t}dt'\,e^{-\frac{(t-t')}{\tau_p}}\left(\bm{\eta}_p(t')+\bm{\sigma}_p(t')\right),
\end{align}
for $p>0,$ which corresponds to the case of a harmonically bound particle as discussed in the previous section (see Eq. (\ref{langevin})).  For $p=0$, $
 \bm{\psi}_0(t)= \frac{1}{\gamma_p}\int_{-\infty}^{t}dt'\,\left(\bm{\eta}_p(t')+\bm{\sigma}_p(t')\right),$ which describes a pure  translational motion of the center of mass (COM). This is equivalent to the dynamics of a free particle.

\subsection{Semiflexible polymer}
To solve Eq. (\ref{semiflexible_dynamics}) analytically, $\bm{r}(s,t)$ may  be expressed in terms of orthogonal eigenfunctions $\mathcal{U}_p(s)$ like the previous case, as follows \cite{harnau1995dynamic}:
\begin{align}
\bm{r}(s,t)=\bm{\psi}_0(t)+2\sum_{p=0}^{\infty} \mathcal{U}_p(s)\bm{\psi}_p(t),\label{eigen_expansion}
\end{align}
where $\mathcal{U}_p(s)$ is the solution of the eigenvalue equation
\begin{align}
\left[-\kappa\frac{\partial^4 }{\partial s^4}+\mu\frac{\partial^2 }{\partial s^2}\right]\mathcal{U}_p(s)=-\epsilon_p \mathcal{U}_p(s),\label{eigen_semiflexible}
\end{align}
 which  satisfies the boundary conditions [Eq. (\ref{bcs_semiflexible})]. So the eigenfunction can be easily found to
 $\mathcal{U}_p(s)=\text{cos}\left(\frac{p \pi s}{L}\right),$ with eigenvalue $\epsilon_p=\kappa\left(\frac{p \pi}{L}\right)^4+\mu\left(\frac{p \pi}{L}\right)^2.$ Using Eq. (\ref{eigen_expansion}) in Eq. (\ref{semiflexible_dynamics}), and by virtue of Eq. (\ref{eigen_semiflexible}), it can be shown that the normal mode obeys the equation:  
\begin{align}
    \gamma_p \frac{\partial}{\partial t}\bm{\psi}_p(t)=-\omega_p\bm{\psi}_p(t)+\bm{\eta}_p(t)+\bm{\sigma}_p(t),\label{normalsemi_dynamics}
\end{align}
 where $\omega_p=\epsilon_p(\frac{\gamma_p}{\gamma})$, $\gamma_p=2 L \gamma$ for $p>0$ and $\gamma_p= L \gamma$ for $p=0.$ Notice that, here $\gamma$ is the friction coefficient per unit contour length, which is different from the one used in the Rouse chain. For Rouse chain, $\gamma$ is used to denote its value per unit bead of length $b,$ but for the sake of notational brevity, we keep the same notation for both the cases.  So the relaxation time can be defined as $\tau_p=\frac{\gamma_p}{\omega_p}=\frac{\gamma}{\frac32 l_p k_B T\left(\frac{p^4\pi^4}{L^4}+\frac{p^2\pi^2}{l_p^2L^2}\right)}.$ Here we should mention that in the length-scale  $L\gg l_p,$ the semiflexible polymer model can be well reduced to a Rouse chain as $\kappa \rightarrow 0,$ and the Kuhn length can be defined as $b=2l_p.$ In the Rouse chain, total beads are related to the total length of the polymer as $L=Nb,$ and the viscosity coefficient per unit bead ($\gamma$) may be obtained by multiplying its value per unit contour with the Kuhn length of the polymer. So the relaxation time modifies to the well-known Rouse time, $\tau_p=\frac{\gamma N^2 b^2}{3\pi^2 k_B T p^2}.$
 \\
 \\
The equation of motion (EOM) of a normal mode for any  polymer model described above has a generic form given by Eqs. (\ref{normalrouse_dynamics}) and (\ref{normalsemi_dynamics}. Since, in the present case, there involves no directional bias into the problem, it can be reduced to a one-dimensional EOM which resembles the overdamped Langevin equation for a harmonically confined particle in an active bath as given in Eq. (\ref{langevin}). However, for $p=0,$  it is most accurately described as a free particle [see Eq. (\ref{langevin0})].

\section{Conformational  properties\label{sec-5}}
The radius of gyration ($\langle R_g^2\rangle$) and the mean square end-to-end distance ($\langle R_e^2\rangle$) are good measures for the conformation of a polymer chain.  
One can first define the center-of-mass (COM) position as 
\begin{align}
\bm{r}_0(t)=\frac{1}{N}\int_{0}^{N}dn\,\bm{r}(n,t) =\bm{\psi}_0(t).\label{RCOM}
\end{align} 
The radius of gyration (ROG) can be expressed as 
\begin{align}
  \langle R_g^2\rangle=  \frac{1}{N}\int_{0}^{N}dn\,\langle\left(\bm{r}(n,t)-\bm{r}_0(t)\right)^2\rangle.
\end{align}
By virtue of Eq. (\ref{eigen_expansion0}), it can be rewritten as 
\begin{align}
 \langle R_g^2\rangle =6\sum_{p=1}^{\infty}\,\langle\bm{\psi}_{p}^2(t)\rangle.\label{Rg}   
\end{align}
The end-to-end vector  of a polymer is   $\bm{R}_e(t)=\bm{r}(N,t)-\bm{r}(0,t).$ With the help of Eq. (\ref{eigen_expansion0}), it can be expressed as \begin{align}
\bm{R}_e(t)=2 \sum_{p=1}^{\infty}\text{cos}\left(\frac{p \pi N}{N}\right)\bm{\psi}_p(t)-2 \sum_{p=1}^{\infty}\bm{\psi}_p(t)=-4\sum_{p: \text{odd integers}}^{}\bm{\psi}_p(t).\label{r_end}
\end{align}
So the temporal auto-correlation function of the end-to-end vector can be written as 
\begin{align}
 \phi(t-t')=\langle \bm{R}_e(t)  \bm{R}_e(t') \rangle &= 16 \sum_{p,q: \text{odd integers}}^{}\langle\bm{\psi}_p(t)\bm{\psi}_q(t')\rangle\nonumber\\
 &=16\sum_{p,q: \text{odd integers}}^{}\sum_{\{i,j\}\in\{x,y,z\}}^{}\,\delta_{ij}\delta_{pq}\langle\bm{\psi}_{pi}(t)\bm{\psi}_{qj}(t')\rangle\nonumber\\
 &=48\sum_{p=1}^{\infty}\,\langle\bm{\psi}_{2p-1}(t)\bm{\psi}_{2p-1}(t')\rangle.\label{correlation_flex_end}
\end{align}
So the mean square end-to-end distance  is
\begin{align}
    \phi(0)=\langle \bm{R}_e^2(t)  \rangle=\langle \bm{R}_e^2(0)  \rangle=48\sum_{p=1}^{\infty}\,\langle\bm{\psi}_{2p-1}^2(0)\rangle.\label{msd_flex_xp0}
\end{align}
Note that, for the semiflexible polymer model, the total number of beads $N$ should be replaced with its total contour length $L.$ Otherwise, all expressions shown in this section hold for both models of the polymer. 


\subsection{Gaussian active bath}
\begin{figure}
    \centering
\includegraphics[width=0.65\linewidth]{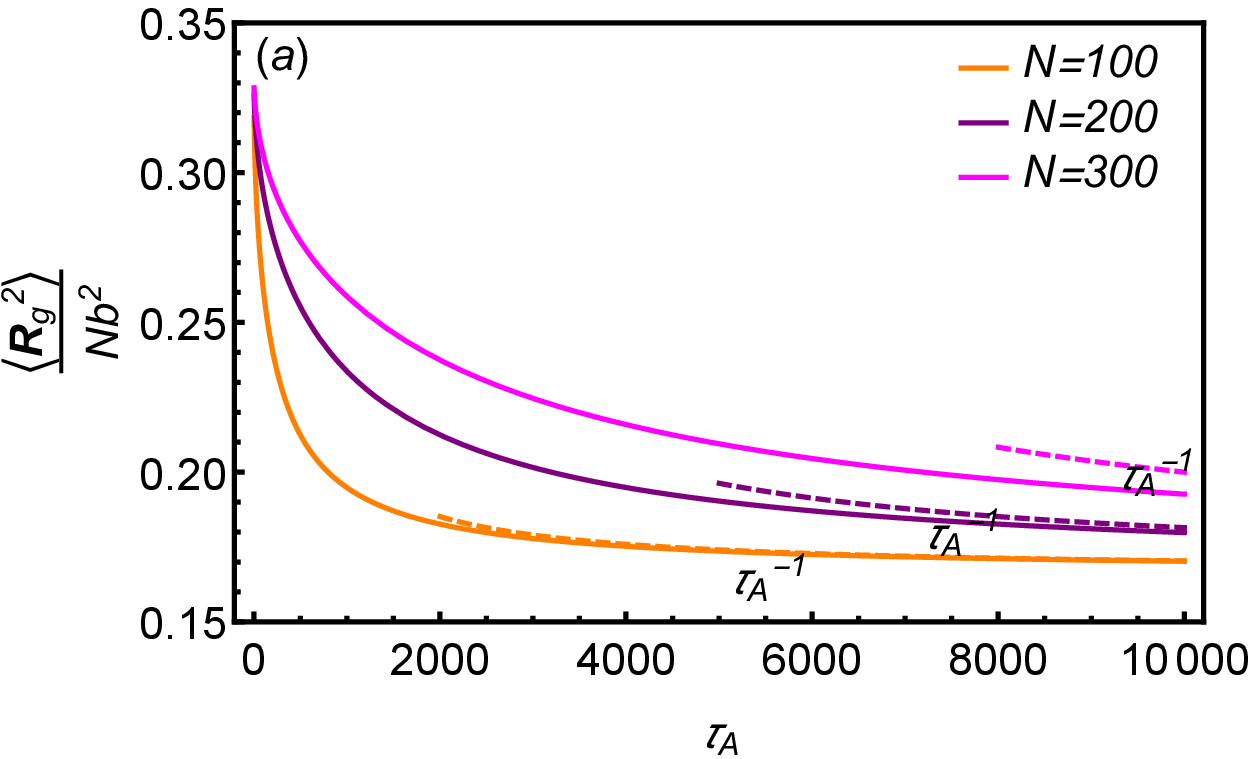}
 \centering
\includegraphics[width=0.65\linewidth]{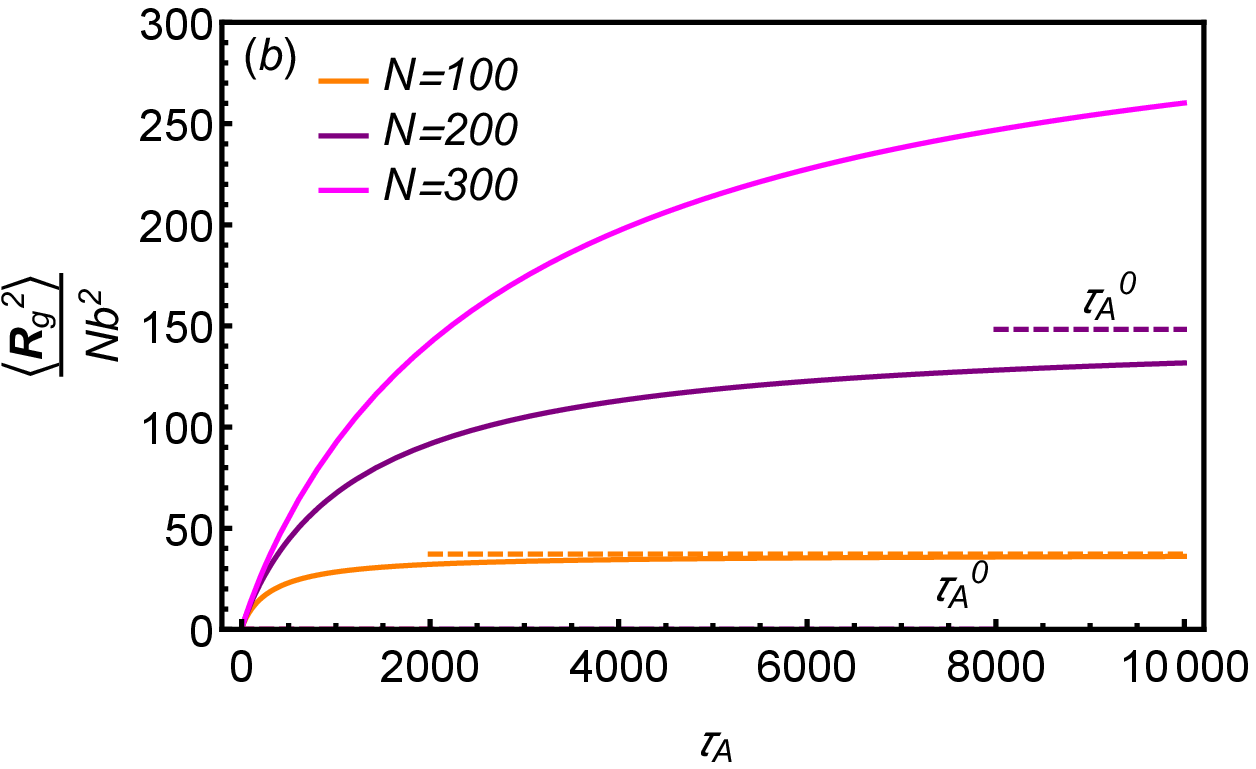}
 \centering
\includegraphics[width=0.65\linewidth]{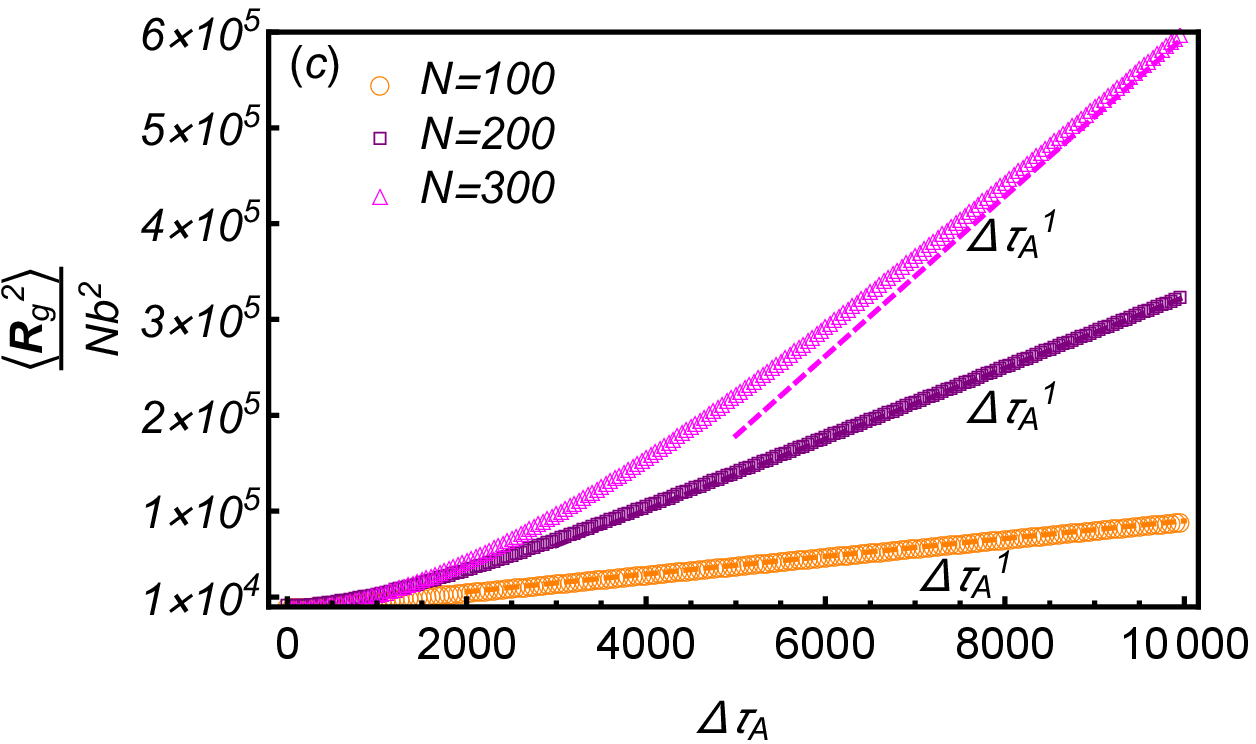}
    \caption{Radius of gyration versus correlation time plot for different lengths (N) of the flexible polymer in the (a) OUP, (b) MOUP and (c) non-Gaussian baths. In this paper, other parameters ($b,\,\gamma,\,C_A,\,\tau_A$) are taken as unity if not specified by other values.  }
    \label{ROG_flex}
\end{figure}

\begin{figure}
    \centering
\includegraphics[width=0.65\linewidth]{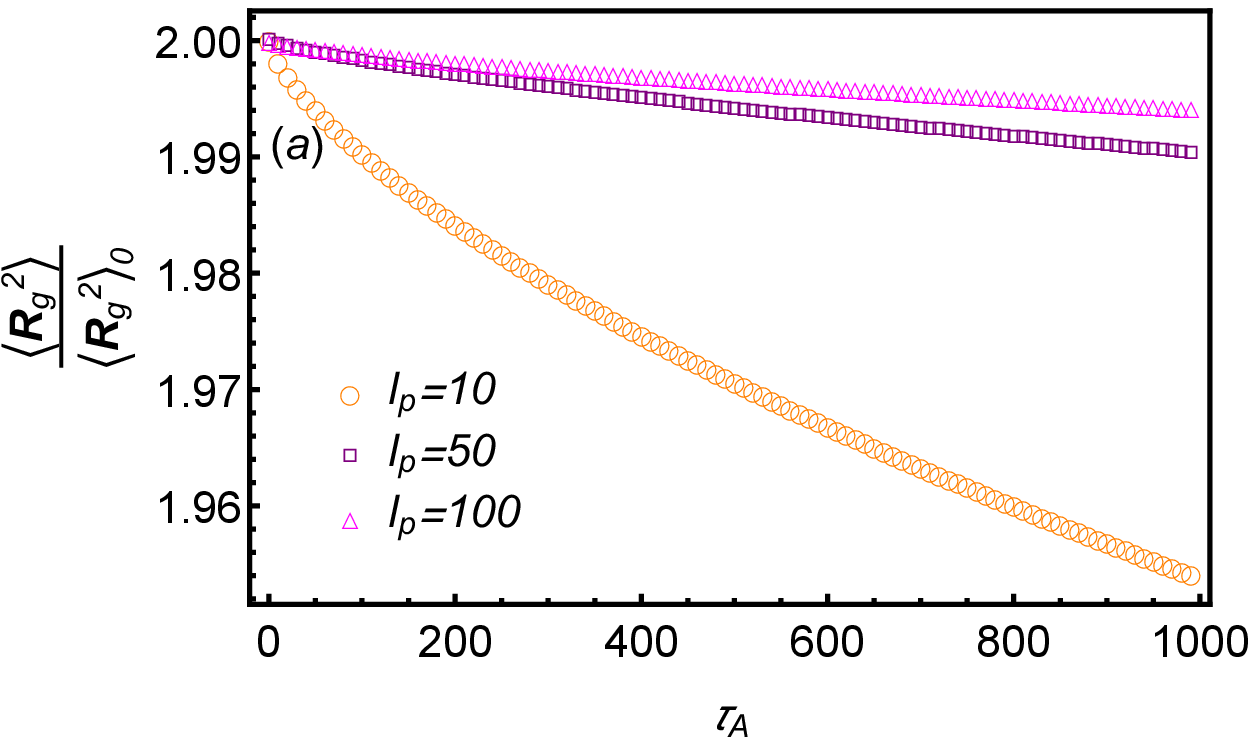}
 \centering
\includegraphics[width=0.65\linewidth]{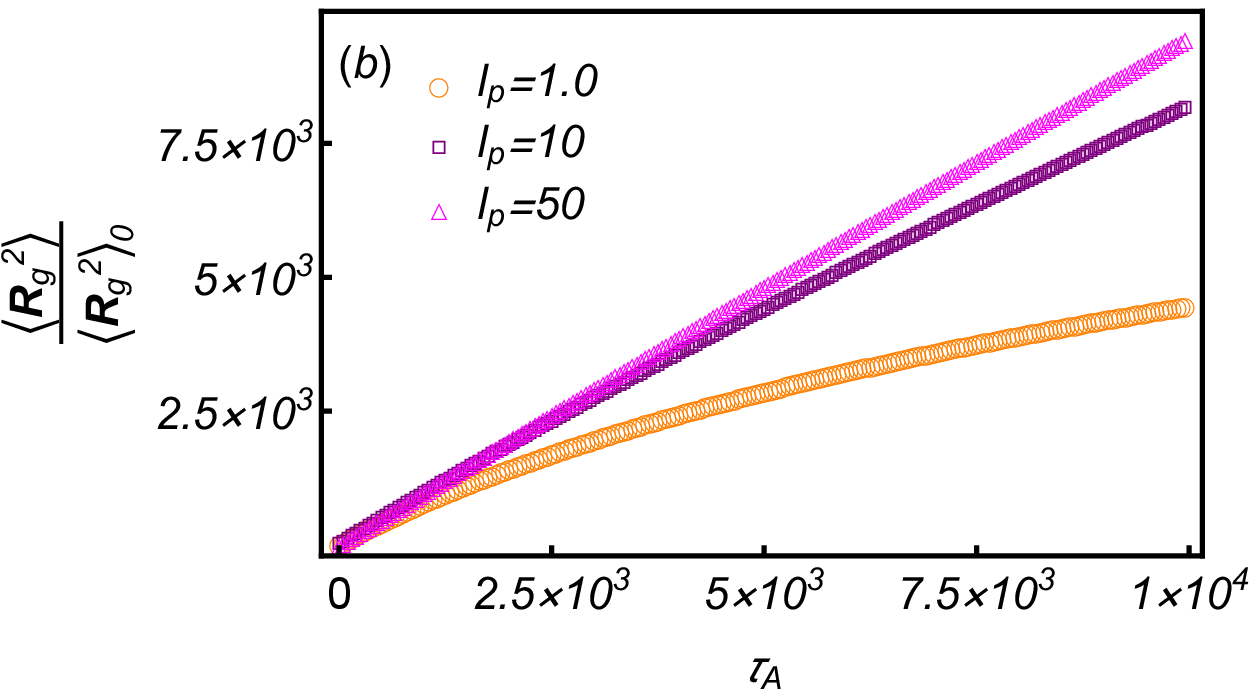}
 \centering
\includegraphics[width=0.65\linewidth]{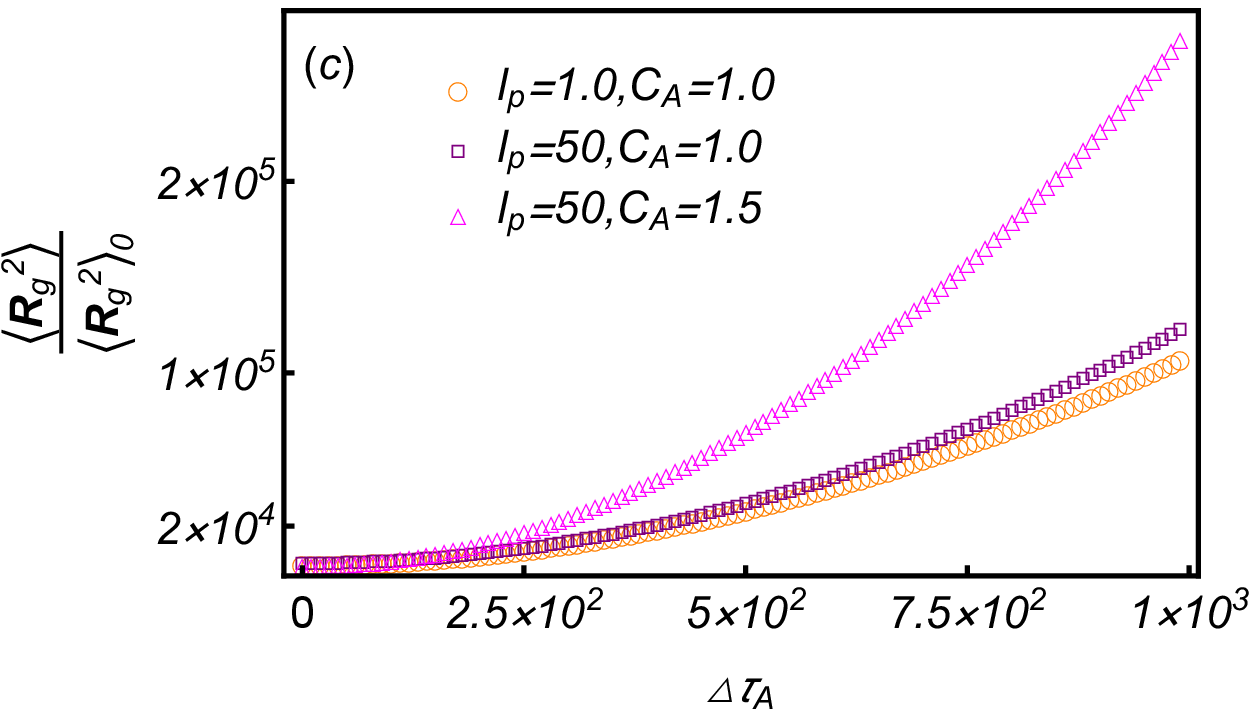}
    \caption{Radius of gyration versus correlation time plot for different persistence lengths of the semiflexible polymer in the (a) OUP, (b) MOUP and (c) non-Gaussian baths. Here, $L=500.$  }
    \label{ROG_semiflex}
\end{figure}

\begin{figure}
    \centering
\includegraphics[width=0.65\linewidth]{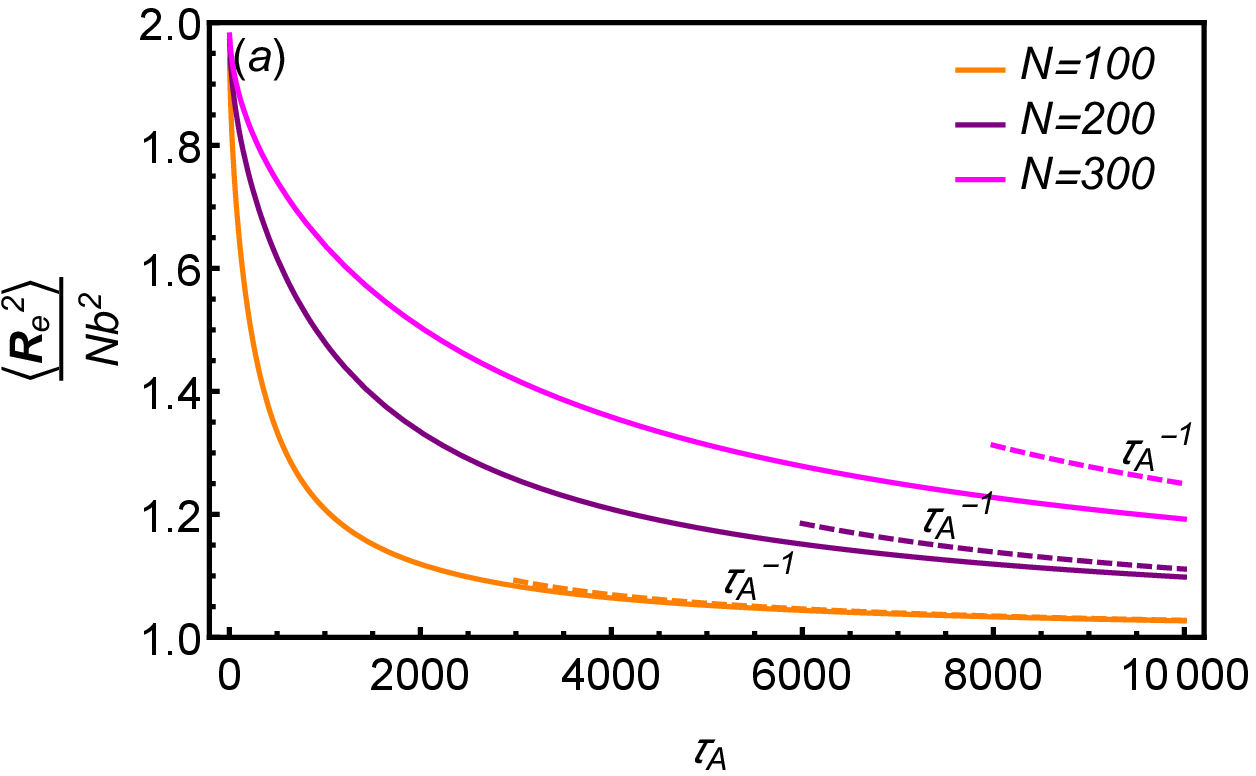}
 \centering
\includegraphics[width=0.65\linewidth]{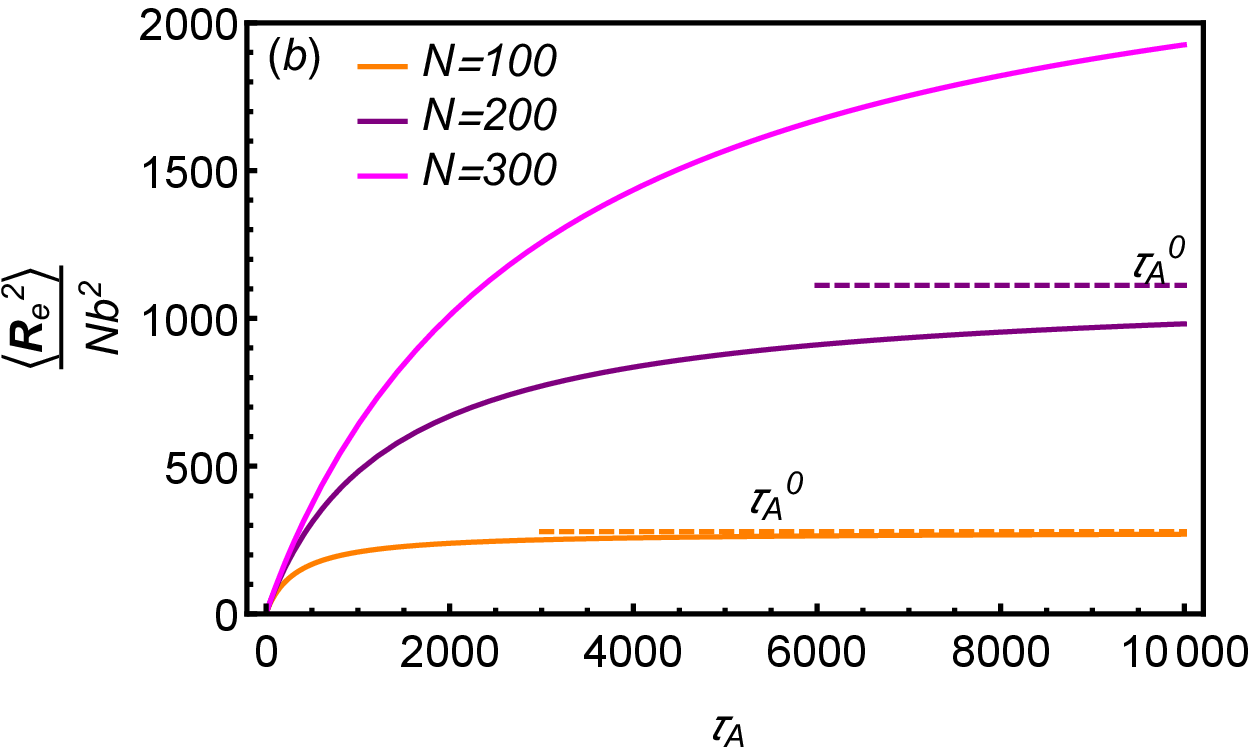}
 \centering
\includegraphics[width=0.65\linewidth]{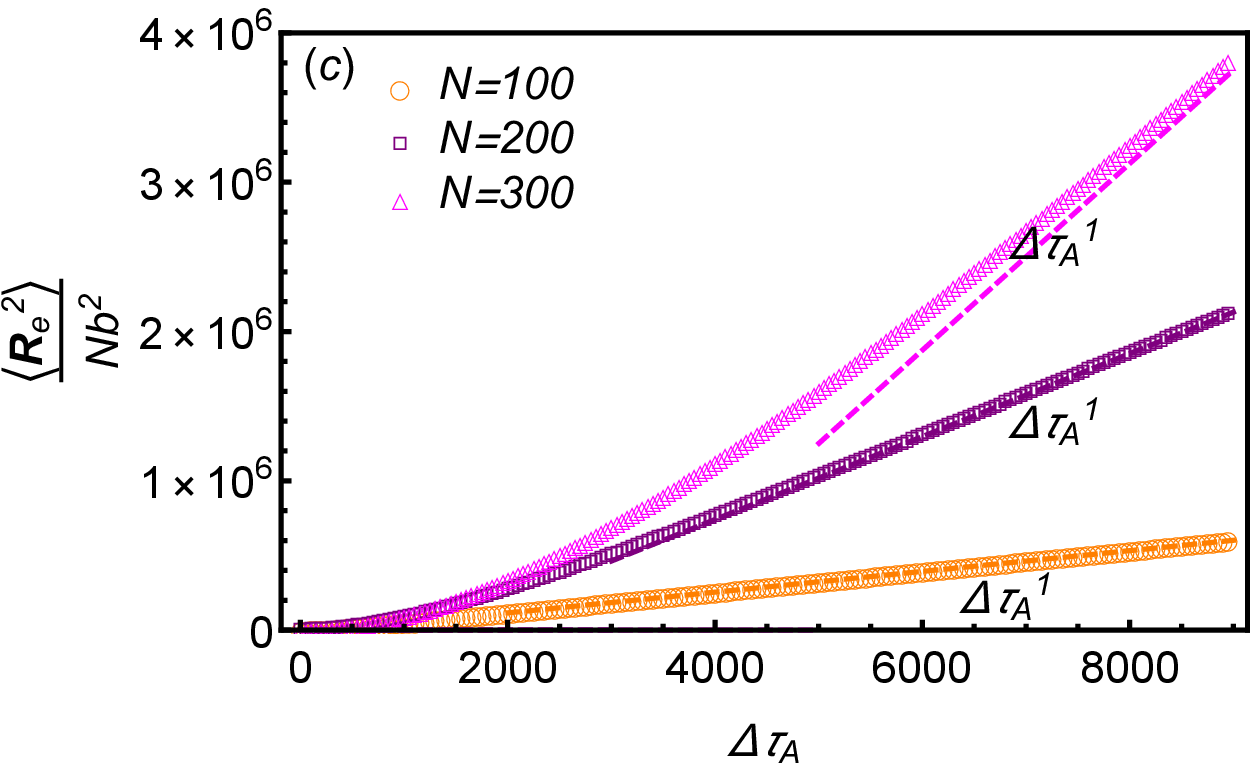}
    \caption{Mean square end-to-end distance versus correlation time plot for different lengths of the flexible polymer in the (a) OUP, (b) MOUP and (c) non-Gaussian baths.  }
    \label{Re_flex}
\end{figure}

\begin{figure}
    \centering
\includegraphics[width=0.65\linewidth]{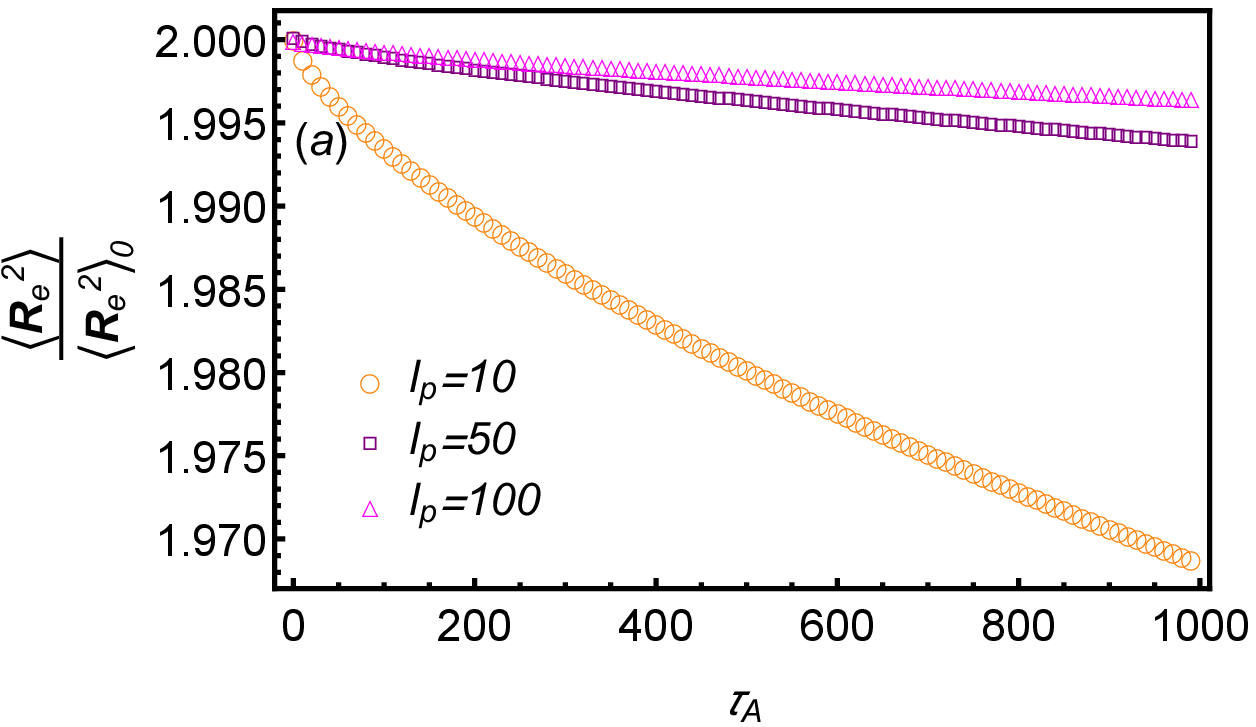}
 \centering
\includegraphics[width=0.65\linewidth]{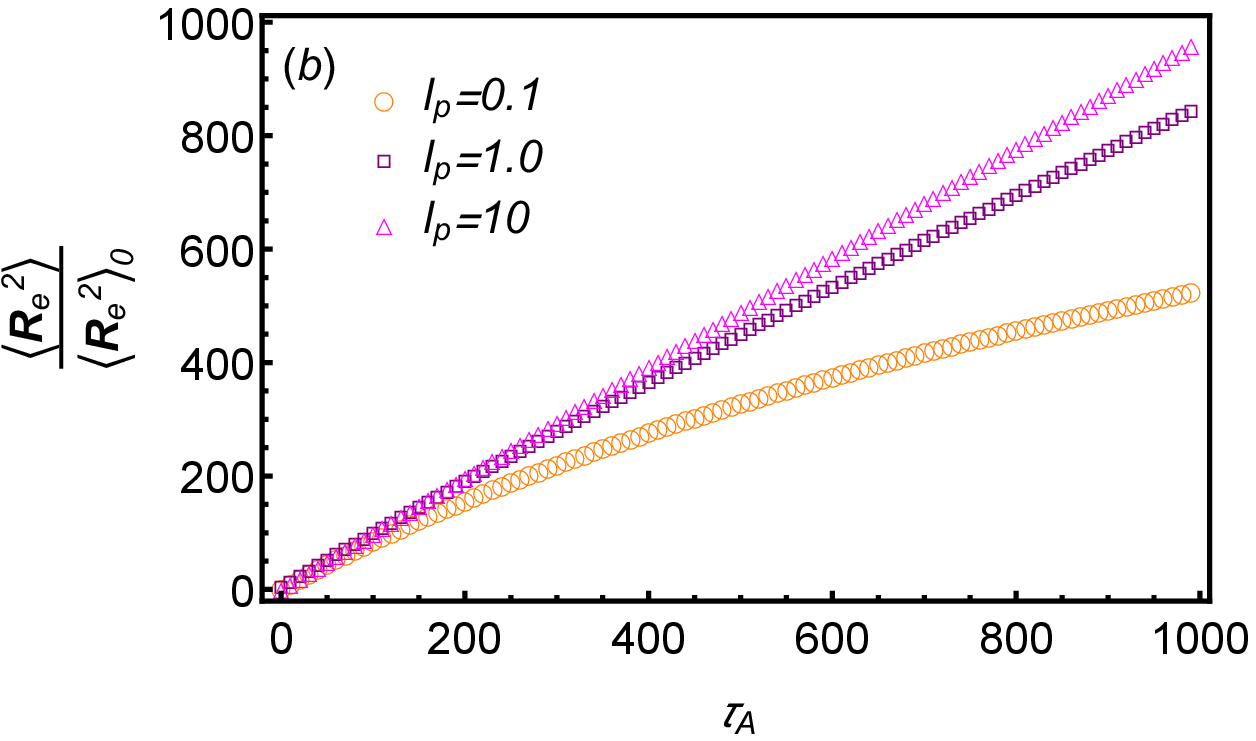}
 \centering
\includegraphics[width=0.65\linewidth]{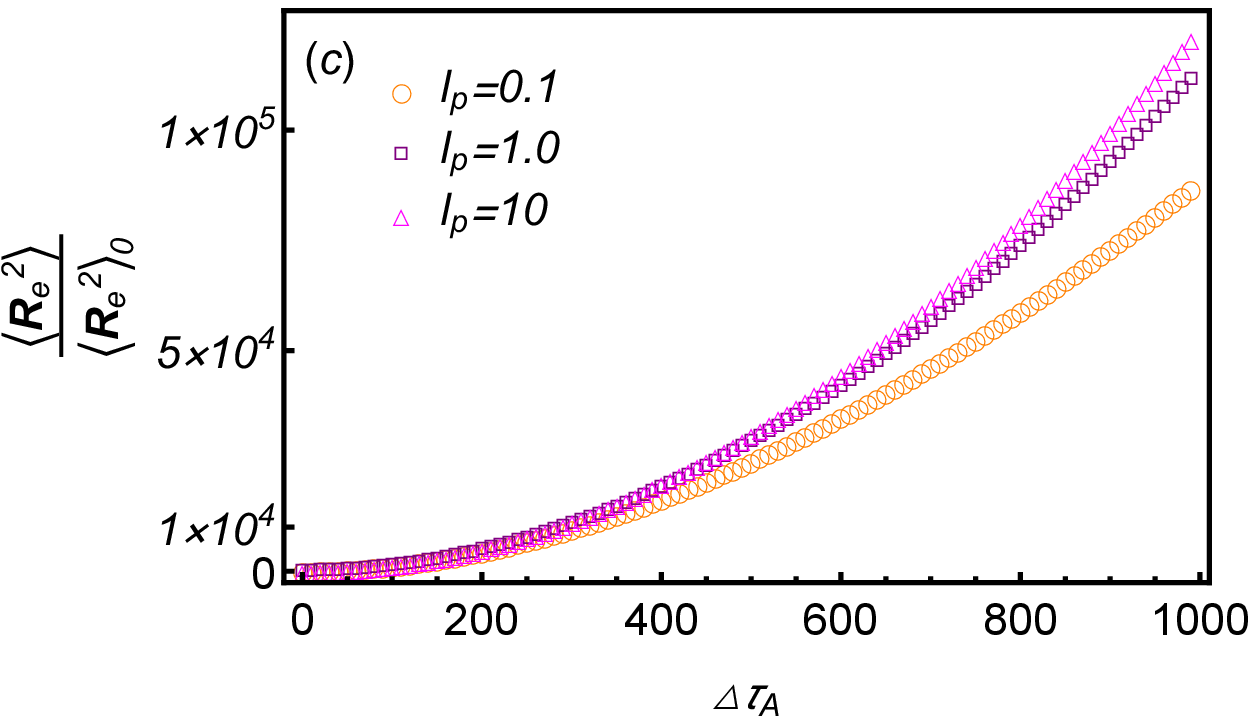}
    \caption{Mean square end-to-end distance versus correlation time plot for different persistence lengths of the semiflexible polymer in the (a) OUP, (b) MOUP and (c) non-Gaussian baths. Here, $L=500.$  }
    \label{Re_semiflex}
\end{figure}

For a flexible polymer in the OUP bath, the radius of gyration can be computed using Eqs. (\ref{Rg}) and (\ref{msd_gauss_p}), and it reads
 \begin{align}
 \langle R_g^2\rangle &=6\sum_{p=1}^{\infty}\,\langle\bm{\psi}_{p}^2(t)\rangle\nonumber\\
 &=6\sum_{p=1}^{\infty}\,\left[\frac{C_{\eta}\tau_1}{2\gamma_p^2}\frac{1}{p^2}+\frac{C_{\sigma_1} \tau _1^2}{\gamma_p^2(p^2\tau_A+\tau_1)}\frac{1}{p^2}\right].   \label{Rg_flex}
\end{align} 
Unlike the single-particle case, the noise amplitudes depend on the number of beads (or length), and these can be given as 
 $C_{\eta}=2\gamma_p k_B T=4N\gamma k_B T,\,C_{\sigma_1}=2 N C_A.$ Using the values, Eq. (\ref{Rg_flex}) can be evaluated to obtain
 \begin{align}
 \langle R_g^2\rangle =\frac16 Nb^2+3 \frac{C_{\sigma_1}\tau_A}{\gamma_p^2}\left(1-\pi\sqrt{\frac{\tau_1}{\tau_A}}\text{coth}\left(\pi\sqrt{\frac{\tau_1}{\tau_A}}\right)\right)+\pi^2\frac{C_{\sigma_1}\tau_1}{\gamma_p^2}\label{rg_ana}.
 \end{align}
 Notice that, without the active noise ($C_{\sigma_1}=0$), the ROG becomes $\langle R_g^2\rangle =\frac16 Nb^2,$ which is the standard result for a flexible polymer in an equilibrium bath. In the presence of active noise, $\langle R_g^2\rangle > \frac16 Nb^2,$ and it is affected by the persistence time $(\tau_A)$ as pictorially depicted in Fig. \ref{ROG_flex}. In the OUP and MOUP baths, it scales asymptotically with $\tau_A$ as $\langle R_g^2\rangle \propto \tau_A^{-1}$ and $\langle R_g^2\rangle \propto \tau_A^0,$ respectively.  The nature of the active noise is reflected to the asymptotic behavior. For the OUP bath, the noise correlation has $1/\tau_A$ dependence in the long-time limit, whereas it has no effect of $\tau_A$ for the MOUP model. Notice that, in the limit $\tau_A\rightarrow 0,$ Eq. (\ref{rg_ana}) modifies to $ \langle R_g^2\rangle \approx \frac16 Nb^2+\pi^2\frac{C_{\sigma_1}\tau_1}{\gamma_p^2} \approx \frac{1}{6}Nb^2\left(1+\frac{C_A}{\gamma k_BT}\right),$ which has a similar form to the passive case, but with an additive contribution arising from the active noise, as can be seen clearly from Fig. \ref{ROG_flex} (a). As mentioned in Sec. \ref{sec-GCN}, the active noise modelled by the OU process behaves like the thermal one in this limit, and this explains the above result. However, for the MOUP model as well the shot-noise case, the active noise vanishes in the limit $\tau_A\,(\text{or}\,\Delta \tau_A)\rightarrow 0,$ and therefore, the ROG approaches to  $ \langle R_g^2\rangle \approx \frac16 Nb^2,$ which  exactly recovers the passive case.
 \\
 \\
 Similarly, for a semiflexible polymer in the OUP bath, one can calculate the ROG using Eqs. (\ref{Rg}) and (\ref{msd_gauss_p}), and it is given by  \begin{align}
 \langle R_g^2\rangle &=6\sum_{p=1}^{\infty}\,\langle\bm{\psi}_{p}^2(t)\rangle\nonumber\\
 &=6\sum_{p=1}^{\infty}\,\left[\frac{C_{\eta}\tau_p}{2\gamma_p^2}+\frac{C_{\sigma_1} \tau _p^2}{\gamma_p^2(\tau_A+\tau_p)}\right]\nonumber\\
 &=\frac13 l_p L+l_p^2\left[\frac{l_p}{L}-\text{coth}\left(\frac{L}{l_p}\right)\right]+6\sum_{p=1}^{\infty}\,\frac{C_{\sigma_1} \tau _p^2}{\gamma_p^2(\tau_A+\tau_p)}.   \label{Rg_semiflex}
\end{align} 
 In the above equation, the second term on the RHS corresponds to the active part which goes to zero if $C_{\sigma_1}=0.$ For $C_{\sigma_1}=0,$ the ROG is $\langle R_g^2\rangle_0=\frac13 l_p L+l_p^2\left[\frac{l_p}{L}-\text{coth}\left(\frac{L}{l_p}\right)\right].$ Eq. (\ref{Rg_semiflex}) is calculated numerically by taking eigen modes up to $p=500,$ and the results are plotted in Figs. \ref{ROG_semiflex} (a)-(b) for the OUP and MOUP baths, respectively.
\\
\\
Another important quantity is the mean square end-to-end distance, which can be calculated using Eqs. (\ref{msd_flex_xp0}) and  (\ref{msd_gauss_p}). For the flexible polymer in the OUP bath, it can be written as 
\begin{align}
 &\langle \bm{R}_e^2(0)  \rangle=\phi(0)\nonumber\\
 &=48\sum_{p=1}^{\infty}\langle\bm{\psi}_{2p-1}^2(0)\rangle=48\sum_{p=1}^{\infty}\left[\frac{k_B T}{2 N \gamma}\frac{ \tau_1}{(2p-1)^2}+\frac{C_{A} }{2 N \gamma^2(\tau_A+\frac{ \tau_1}{(2p-1)^2})}\frac{ \tau_1^2}{(2p-1)^4}\right]\nonumber\\
&=Nb^2+48\sum_{p=1}^{\infty}\frac{C_{A}}{2 N \gamma^2(\tau_A+\frac{ \tau_1}{(2p-1)^2})}\frac{ \tau_1^2}{(2p-1)^4}\nonumber\\
&=Nb^2+\frac{C_A}{\gamma k_BT}Nb^2\left(1-\frac{2}{\pi}\sqrt{\frac{\tau_A}{\tau_1}}\text{tanh}\left(\frac{\pi}{2}\sqrt{\frac{\tau_1}{\tau_A}}\right)\right)\label{msd_flex_end_OUP}.
\end{align}
where $\tau_1=\frac{\gamma N^2 b^2}{3 \pi^2 k_B T }.$ The first term on the RHS of Eq. (\ref{msd_flex_end_OUP}) represents the mean square end-to-end distance for a free Gaussian chain in a thermal bath. The sum in the second term is performed numerically and the results are plotted in Fig.  \ref{Re_flex} (a) and (b) for the OUP and MOUP models of active bath. As one can see from the figures, in the active bath, $\langle \bm{R}_e^2(0)  \rangle \gg Nb^2,$ which means that the polymer swells in the presence of an active noise as each monomer of the polymer  executes longer excursions due to higher activity. 
\\
\\
For the semiflexible polymer, the MSD of  the end-to-end vector can be written as  
\begin{align}
 &\langle \bm{R}_e^2(0)  \rangle=\phi(0)=48\sum_{p=1}^{\infty}\langle\bm{\psi}_{2p-1}^2(0)\rangle\nonumber\\
 &=48\sum_{p=1}^{\infty}\Bigl[\frac{k_B T}{2 L \gamma}\frac{\gamma}{\frac32 l_p k_B T\left(\frac{(2p-1)^4\pi^4}{L^4}+\frac{(2p-1)^2\pi^2}{l_p^2L^2}\right)}\nonumber\\
 &\qquad+\frac{C_A }{2 L \gamma^2\left[\tau_A+\frac{\gamma}{\frac32 l_p k_B T\left(\frac{(2p-1)^4\pi^4}{L^4}+\frac{(2p-1)^2\pi^2}{l_p^2L^2}\right)}\right]}\frac{\gamma^2}{\left(\frac32 l_p k_B T\right)^2\left(\frac{(2p-1)^4\pi^4}{L^4}+\frac{(2p-1)^2\pi^2}{l_p^2L^2}\right)^2}\Bigr]\nonumber\\
&=2Ll_p\left(1-\frac{2l_p}{L}\text{tanh}\left(\frac{L}{2l_p}\right)\right)\nonumber\\
 &\qquad+48\sum_{p=1}^{\infty}\frac{C_A }{2 L \gamma^2\left[\tau_A \frac32 l_p k_B T\left(\frac{(2p-1)^4\pi^4}{L^4}+\frac{(2p-1)^2\pi^2}{l_p^2L^2}\right)+\gamma\right]}\frac{\gamma^2}{\left(\frac32 l_p k_B T\right)\left(\frac{(2p-1)^4\pi^4}{L^4}+\frac{(2p-1)^2\pi^2}{l_p^2L^2}\right)}\label{msd_semiflex_end_OUP}.
\end{align} 
The contribution of the active noise on $\langle \bm{R}_e^2(0)  \rangle$ is reflected in the second term of the RHS, so the distance  between the end vectors is increased due to the presence of active fluctuations \cite{doi:10.1063/1.4891095}. Like the flexible polymer, it also swells, and the degree of swelling greatly depends on the rigidity of the chain as pictorially depicted in Fig. \ref{Re_semiflex};  a stiffer polymer has a larger $\langle R_e^2\rangle$ value, thereby suggesting swelling to a greater extent. This is consistent with the results of  Winkler and co-workers [see Ref.  \cite{eisenstecken2017conformational}] in the low and high activity (or P\'eclet number) regimes. However, in the intermediate limit, shrinkage of polymers which has been seen as reported in Ref.  \cite{eisenstecken2017conformational}  is missing in the present model. As alluded to in the introduction, this model does not allow local segmental relaxation in  its dynamical modes, and  such dissimilar behavior may arise due to this reason. In the limit $L\gg l_p,$  the polymer behaves like a  Rouse chain, and  one can recover the known relation for the thermal bath, $\langle \bm{R}_e^2(0)  \rangle =N b^2$ from the first term of Eq. (\ref{msd_semiflex_end_OUP}) using $b=2 l_p.$

 \subsection{Non-Gaussian active bath}
In the non-Gaussian bath, the parameters we have used here are given by $\gamma_p=2 N \gamma,\,C_{\eta}=4N\gamma k_B T,\,\sigma_A^2=\frac{N}{2} C_A^2.$ Here, all the above formulations follow, and the thermal part of the variables remain the same. but the  active contributions are different. The results are shown below.
\\
\\
For the flexible polymer, one may obtain 
\begin{align}
 \langle R_g^2\rangle &=6\sum_{p=1}^{\infty}\,\langle\bm{\psi}_{p}^2(t)\rangle\nonumber\\
 &=6\sum_{p=1}^{\infty}\,\left[\frac{C_{\eta}\tau_1}{2\gamma_p^2}\frac{1}{p^2}+\frac{\sigma_A^2}{\gamma_p^2}\tau_p^3\nu_A\,\left(\text{exp}\left(-\frac{\Delta \tau_A}{\tau_p}\right)+\frac{\Delta \tau_A}{\tau_p}-1\right)\right]\nonumber\\
 &=\frac16 Nb^2+\frac{3C_A^2}{4N\gamma^2}\sum_{p=1}^{\infty}\,\tau_p^3\nu_A\,\left(\text{exp}\left(-\frac{\Delta \tau_A}{\tau_p}\right)+\frac{\Delta \tau_A}{\tau_p}-1\right), \label{Rg_flex_nong}
\end{align} 
where $\tau_p=\frac{\gamma N^2 b^2}{3 k_B T \pi^2 p^2}.$
 The sum given in the second term has been computed numerically and the result for ROG is depicted in Fig. \ref{ROG_flex}(c). From Eq. (\ref{msd_flex_xp0}), the mean square end-to-end distance can be expressed as 
\begin{align}
& \langle \bm{R}_e^2(0)  \rangle =\phi(0)\nonumber\\
 &=48\sum_{p=1}^{\infty}\langle\bm{\psi}_{2p-1}^2(0)\rangle\nonumber\\
&=48\sum_{p=1}^{\infty}\Bigl[\frac{k_B T}{2 N \gamma}\frac{ \tau_1}{(2p-1)^2}+\frac{C_A^2 \nu_A}{8 N\gamma^2}\frac{\tau_1^3}{(2p-1)^6}\left(\text{exp}\left(-\frac{\Delta \tau_A}{\tau_1}(2p-1)^2\right)-1+\frac{\Delta \tau_A}{\tau_1}(2p-1)^2\right)\Bigr]\nonumber\\
 &=Nb^2+\frac{6\,C_A^2}{N\gamma^2}\sum_{p=1}^{\infty}\,\tau_{2p-1}^3\nu_A\,\left(\text{exp}\left(-\frac{\Delta \tau_A}{\tau_{2p-1}}\right)+\frac{\Delta \tau_A}{\tau_{2p-1}}-1\right). 
\end{align} 
For the semiflexible polymer, the relaxation time for the $p^{th}$ mode is different from the flexible case, and it given by $\tau_p = \frac{\gamma}{\frac32 l_p k_B T\left(\frac{p^4\pi^4}{L^4}+\frac{p^2\pi^2}{l_p^2L^2}\right)}.$ So the conformational variables can be computed as follows:
The radius of gyration  is
 \begin{align}
 \langle R_g^2\rangle &=6\sum_{p=1}^{\infty}\,\langle\bm{\psi}_{p}^2(t)\rangle\nonumber\\
 &=\frac13 l_p L+l_p^2\left[\frac{l_p}{L}-\text{coth}\left(\frac{L}{l_p}\right)\right]+\frac{3C_A^2}{4L\gamma^2}\sum_{p=1}^{\infty}\,\tau_p^3\nu_A\,\left(\text{exp}\left(-\frac{\Delta \tau_A}{\tau_p}\right)+\frac{\Delta \tau_A}{\tau_p}-1\right),
\end{align} 
and the mean square end-to-end distance is
\begin{align}
 &\langle \bm{R}_e^2(0)  \rangle=\phi(0)=48\sum_{p=1}^{\infty}\langle\bm{\psi}_{2p-1}^2(0)\rangle\nonumber\\
&=2Ll_p\left(1-\frac{2l_p}{L}\text{tanh}\left(\frac{L}{2l_p}\right)\right)+\frac{6\,C_A^2}{L\gamma^2}\sum_{p=1}^{\infty}\,\tau_{2p-1}^3\nu_A\,\left(\text{exp}\left(-\frac{\Delta \tau_A}{\tau_{2p-1}}\right)+\frac{\Delta \tau_A}{\tau_{2p-1}}-1\right)\label{msd_semiflex_end_nong}.
\end{align} 
The results are shown in Figs. \ref{ROG_semiflex}(c) and \ref{Re_semiflex}(c). Like in the Gaussian baths, the extent of swelling depends on the amplitude of the active noise \cite{doi:10.1063/1.5086152}.

\section{Dynamical properties \label{sec-6}}
The  mean square COM may be defined as 
\begin{align}
 \langle \bm{R}_0^2(t)  \rangle &=\langle\left(\bm{r}_0(t)-\bm{r}_0(0)\right)^2 \rangle=3\langle\left(\bm{\psi}_0(t)-\bm{\psi}_0(0)\right)^2\rangle\label{msd_flex_x0},
 \end{align}
 where we have used Eq. (\ref{RCOM}). This corresponds to the pure translational motion. 
 
 \begin{figure}
    \centering
\includegraphics[width=0.65\linewidth]{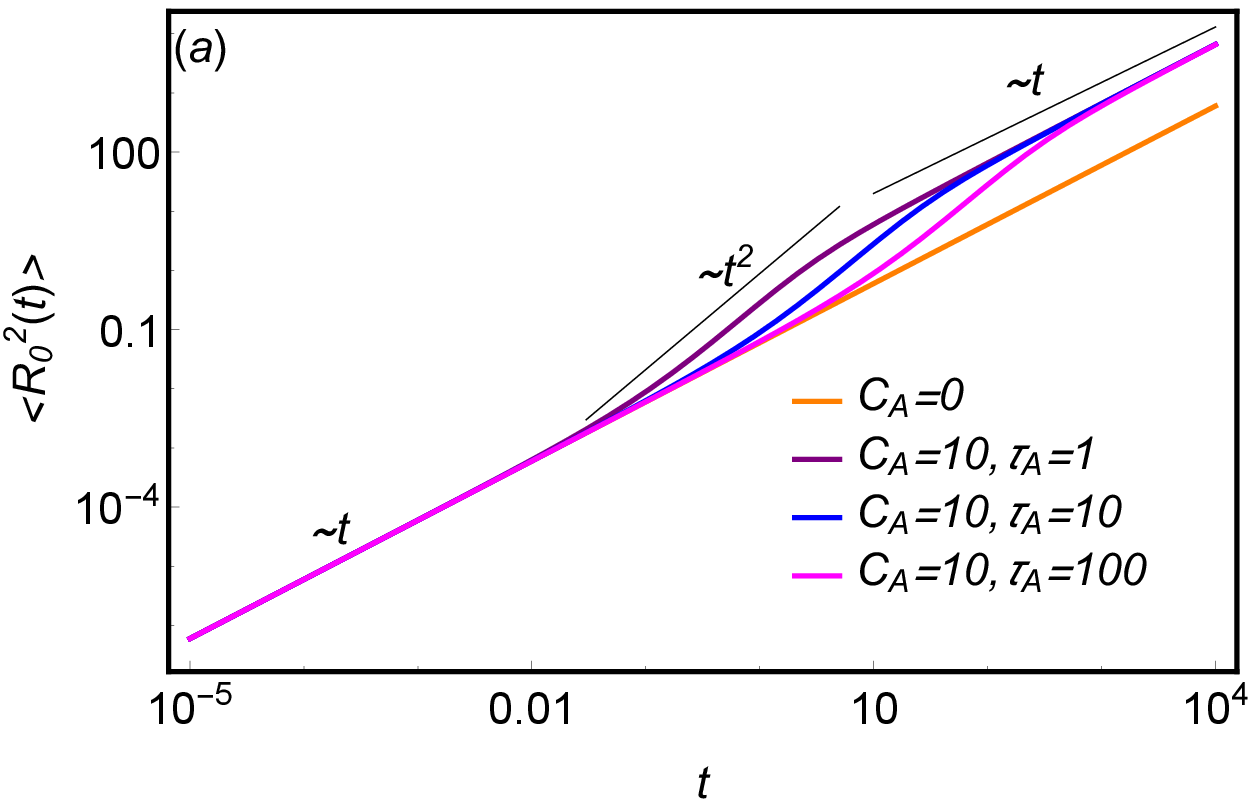}
 \centering
\includegraphics[width=0.65\linewidth]{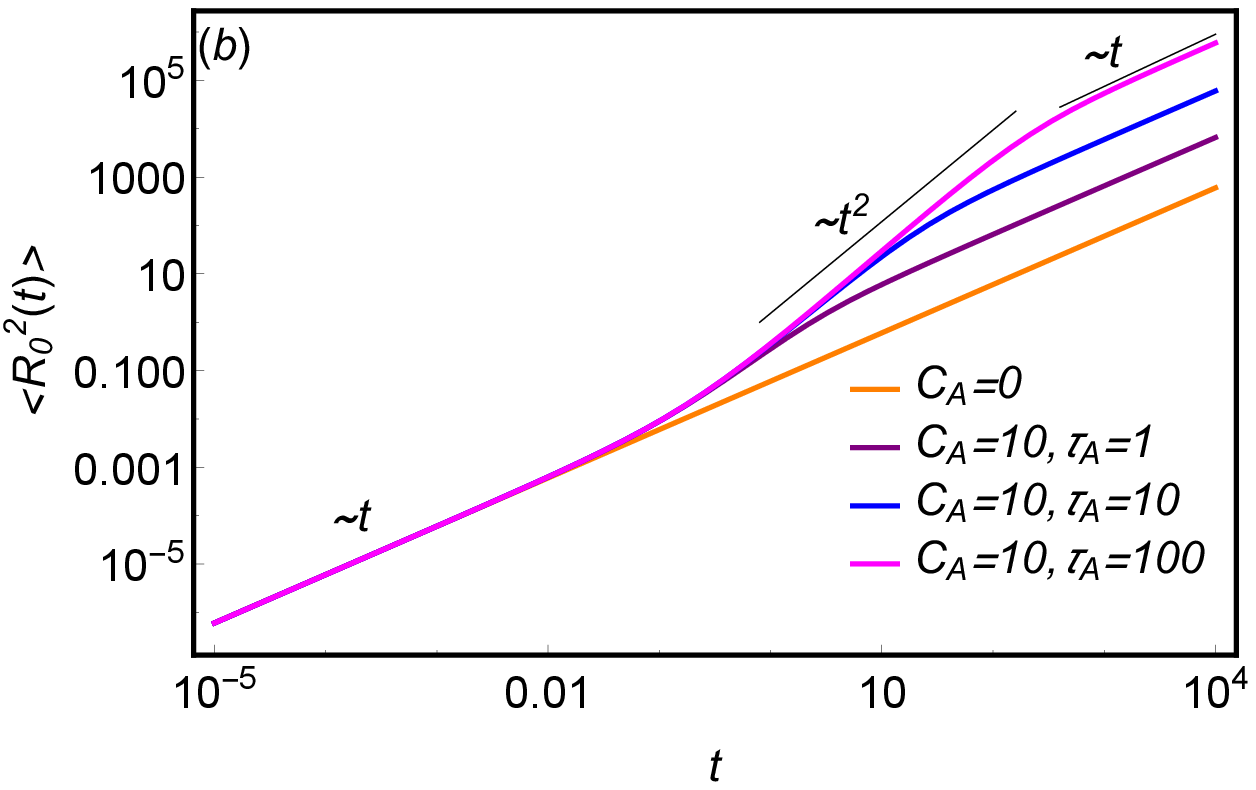}
 \centering
\includegraphics[width=0.65\linewidth]{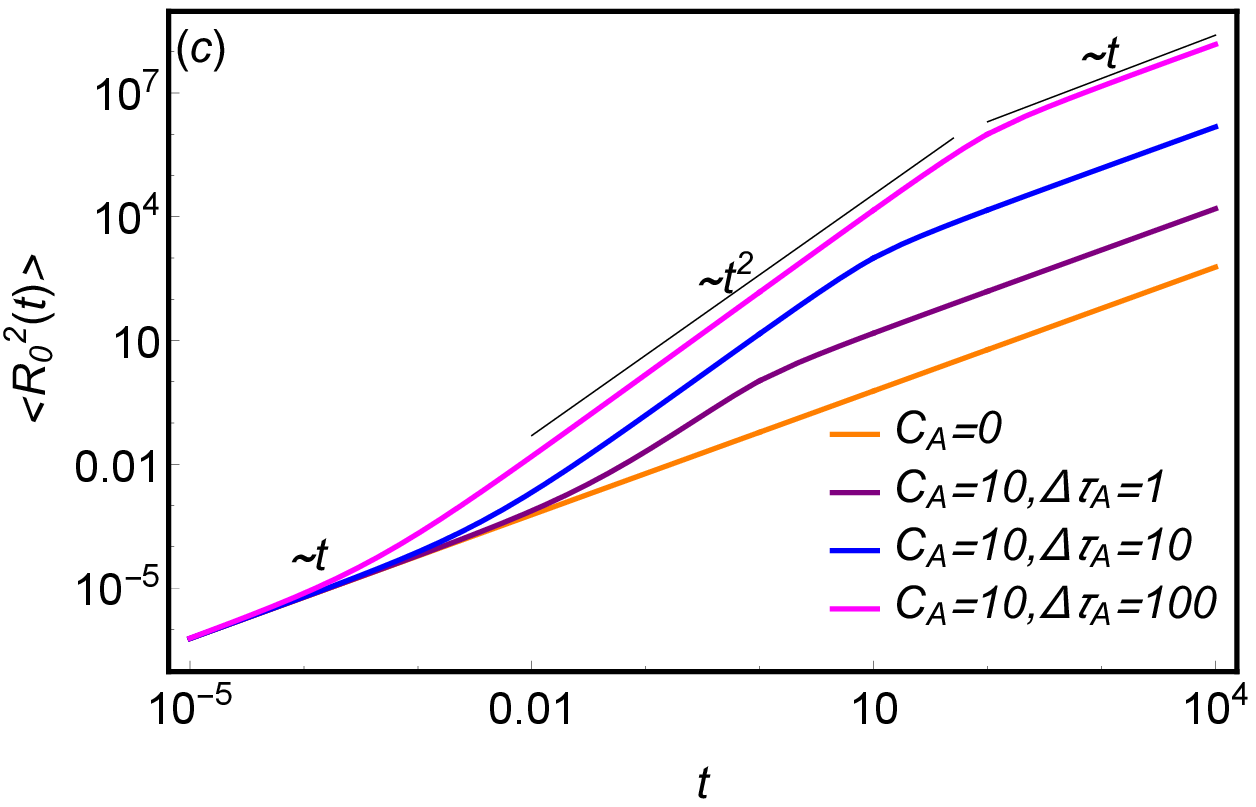}
\caption{Log-Log plot of the MSD of COM for different noise amplitudes and persistence times of active noise  in the (a) OUP, (b) MOUP and (c) non-Gaussian baths. The results are the same for both the polymer models. Here we take $N$(or $L$)$=100.$ Throughout the paper, $\gamma=1,\,k_B T=1.$  }
    \label{msd_com}
\end{figure}

\noindent The auto-correlation function of end-to-end vector is denoted here as $\phi(t,t'),$ and it is given by [see Eq. (\ref{correlation_flex_end})]
 \begin{align}
\phi(t-t')=\langle \bm{R}_e(t)  \bm{R}_e(t') \rangle=     48\sum_{p=1}^{\infty}\,\langle\bm{\psi}_{2p-1}(t)\bm{\psi}_{2p-1}(t')\rangle.
 \end{align}
\\
\\
 The MSD  of a bead (or contour length) at point $r(n,t)$ can be expressed as 
 \begin{align}
   &\Delta^2(n,t)=\langle \bm{R}^2(n,t)  \rangle\nonumber\\ &=\langle\left(\bm{r}(n,t)-\bm{r}(n,0)\right)^2 \rangle\nonumber\\
   &= \langle \bm{R}_0^2(t)  \rangle +4 \sum_{p=1}^{\infty} \left[\text{cos}^2\left(\frac{p \pi n}{N}\right)\left( \langle\bm{\psi}_{p}^2(t)\rangle+ \langle\bm{\psi}_{p}^2(0)\rangle-2  \langle \bm{\psi}_{p}(t)\bm{\psi}_{p}(0)\rangle\right)\right]\label{msd_flex_rs}.  
 \end{align}

\subsection{Gaussian active bath}
The mean square COM can be computed by virtue  of Eqs. (\ref{msdg_x0}) and (\ref{msd_flex_x0}).

\noindent For any given model of polymer in the OUP bath,  it is given by 
\begin{align}
 \langle \bm{R}_0^2(t)  \rangle &=\langle\left(\bm{r}_0(t)-\bm{r}_0(0)\right)^2 \rangle=3\langle\left(\bm{\psi}_0(t)-\bm{\psi}_0(0)\right)^2\rangle\nonumber\\
 &=\frac{6k_B T}{N \gamma}t+\frac{6 C_{A} \tau_A}{N \gamma^2}\left(\frac{t}{\tau_A}-1+e^{-\frac{t}{\tau_A}}\right)\label{msd_flexg_x0}.
\end{align}
In the limit $t\ll \tau_A$, $\langle \bm{R}_0^2(t)  \rangle \approx \frac{6 k_B T}{N \gamma}t+\frac{1}{N}\frac{3 C_{A} }{ \gamma^2\tau_A}t^2.$ Here one can define a timescale $t_c=\frac{2\gamma k_B T}{C_A}\tau_A;$ for $t< t_c,$ the dynamics is diffusive due to thermal noise as shown in Fig. \ref{msd_com} (a); for $t> t_c,$ it approaches the ballistic regime ($\langle \bm{R}_0^2(t)  \rangle \propto t^2$) like the free-particle case. In the long-time limit $i.e.,$ for  $t\gg \tau_A$,  $\langle \bm{R}_0^2(t)  \rangle \approx \left[ \frac{6 k_B T}{N \gamma}+\frac{6 C_{A}\tau_A }{N \gamma^2}\right] t,$ which means that it  exhibits normal diffusion with an enhanced diffusivity \cite{vandebroek2015dynamics,doi:10.1063/1.4891095,Shin_2015}. In the MOUP bath, the result is similar, but only with a difference in the amplitude of the active noise (as a result, the value of $t_c$ also changes). The plots of  $\langle \bm{R}_0^2(t)  \rangle$ are shown in Figs. \ref{msd_com} (a)-(b).
\\
\\
From Eq. (\ref{correlation_flex_end}), the pair correlation function can be computed  using Eq. (\ref{correlationx_Gaussian}) as 
\begin{align}
  & \phi(t-t')=\langle \bm{R}_e(t)  \bm{R}_e(t') \rangle=48\sum_{p=1}^{\infty}\,\langle\bm{\psi}_{2p-1}(t)\bm{\psi}_{2p-1}(t')\rangle\nonumber\\
 &=48\sum_{p=1}^{\infty}\,\left[\frac{C_{\eta}\tau_{2p-1}}{2\gamma_{2p-1}^2}e^{-\frac{|t-t'|}{\tau_{2p-1}}}+\frac{C_{\sigma_1}}{\gamma_{2p-1}^2}\frac{ \tau _{2p-1}^2 \left(\tau _A e^{-\frac{|t-t'|}{\tau _A}}-\tau_{2p-1} e^{-\frac{|t-t'|}{\tau _{2p-1}}}\right)}{\tau _A^2-\tau _{2p-1}^2}\right]\label{phi_general},
\end{align}
For the flexible polymer in the OUP bath,
\begin{align}
 \phi(t-t') =48\sum_{p=1}^{\infty}\,\left[\frac{\tau_1 k_B T}{2 N \gamma (2p-1)^2}e^{-\frac{|t-t'|}{\tau_1}(2p-1)^2}+\frac{C_{A}}{2N\gamma^2}\frac{ \left(\tau _A e^{-\frac{|t-t'|}{\tau _A}}-\frac{\tau _1}{(2p-1)^2} e^{-\frac{|t-t'|}{\tau _1}(2p-1)^2}\right)}{\frac{\tau _A^2}{\tau _1^2}(2p-1)^4-1}\right]  ,
\end{align}
where  $\tau_1=\frac{\gamma N^2 b^2}{3 \pi^2 k_B T }.$ The value of $\tau_p$ should be replaced with $ \frac{\gamma}{\frac32 l_p k_B T\left(\frac{p^4\pi^4}{L^4}+\frac{p^2\pi^2}{l_p^2L^2}\right)}$ in Eq. (\ref{phi_general}) for the semiflexible polymer. 
\\
\\
The MSD of the  $n^{th}$ tagged monomer of a chain in the OUP bath can be calculated with the help of Eqs. (\ref{msd_flex_rs}) and (\ref{msd_gauss_p}) as shown below.
\begin{align}
  &\Delta^2(n,t)=\langle \bm{R}^2(n,t)  \rangle\nonumber\\ 
 &= \langle \bm{R}_0^2(t)  \rangle +4 \sum_{p=1}^{\infty} \left[\text{cos}^2\left(\frac{p \pi n}{N}\right)\left( \langle\bm{\psi}_{p}^2(t)\rangle+ \langle\bm{\psi}_{p}^2(0)\rangle-2  \langle \bm{\psi}_{p}(t)\bm{\psi}_{p}(0)\rangle\right)\right]\nonumber\\
  &=\frac{6k_B T}{N \gamma}t+\frac{6 C_{A} \tau_A}{N \gamma^2}\left(\frac{t}{\tau_A}-1+e^{-\frac{t}{\tau_A}}\right)+ 8 \sum_{p=1}^{\infty} \text{cos}^2\left(\frac{p \pi n}{N}\right)\left[\frac{C_{\eta}\tau_p}{2\gamma_p^2}+\frac{C_{\sigma_1} \tau _p^2}{\gamma_p^2(\tau_A+\tau_p)}\right]\nonumber\\
  &\quad - 8 \sum_{p=1}^{\infty} \text{cos}^2\left(\frac{p \pi n}{N}\right)\left[\frac{C_{\eta}\tau_p}{2\gamma_p^2}e^{-\frac{t}{\tau_p}}+\frac{C_{\sigma_1}}{\gamma_p^2}\frac{ \tau _p^2 \left(\tau _A e^{-\frac{t}{\tau _A}}-\tau _p e^{-\frac{t}{\tau _p}}\right)}{\tau _A^2-\tau _p^2}\right]\label{msd_rs1}.  
 \end{align}
For a  flexible polymer in the OUP bath,  Eq. (\ref{msd_rs1}) can be recast as
\begin{align}
  &\Delta^2(n,t)=\langle \bm{R}^2(n,t)  \rangle\nonumber\\ 
  &=\frac{6k_B T}{N \gamma}t+\frac{6 C_{A} \tau_A}{N \gamma^2}\left(\frac{t}{\tau_A}-1+e^{-\frac{t}{\tau_A}}\right)+ 8 \sum_{p=1}^{\infty} \text{cos}^2\left(\frac{p \pi n}{N}\right)\left[\frac{C_{\eta}\tau_1}{2\gamma_p^2}\frac{1}{p^2}+\frac{C_{\sigma_1} \tau _1^2}{\gamma_p^2(p^2\tau_A+\tau_1)}\frac{1}{p^2}\right]\nonumber\\
  &\quad - 8 \sum_{p=1}^{\infty} \text{cos}^2\left(\frac{p \pi n}{N}\right)\left[\frac{C_{\eta}\tau_1}{2\gamma_p^2 p^2}e^{-\frac{t p^2}{\tau_1}}+\frac{C_{\sigma_1}}{p^2\gamma_p^2}\frac{ \tau _1^2 \left(\tau _A p^2 e^{-\frac{t}{\tau _A}}-\tau_1 e^{-\frac{t p^2}{\tau _1}}\right)}{\tau _A^2 p^4-\tau _1^2}\right]\label{msd_flex_rs1}.  
 \end{align}
For the end bead of the flexible polymer, $i.e.,$ for $n=N$, the above equation further simplifies to  
  \begin{align}
   & \Delta^2(N,t)=\langle \bm{R}^2(N,t)  \rangle\nonumber\\ 
  & =\frac{6k_B T}{N \gamma}t+\frac{6 C_{A} \tau_A}{N \gamma^2}\left(\frac{t}{\tau_A}-1+e^{-\frac{t}{\tau_A}}\right)+\frac{2}{9}Nb^2\left(1+\frac{C_A}{ k_B T}\right)+\frac{2C_A}{N\gamma^2}\tau_A\left(1-\pi\sqrt{\frac{\tau_1}{\tau_A}}\text{coth}\left(\pi\sqrt{\frac{\tau_1}{\tau_A}}\right)\right)\nonumber\\
   &\quad - 8 \sum_{p=1}^{\infty} \left[\frac{Nb^2}{6\pi^2 p^2}e^{-\frac{t p^2}{\tau_1}}+\frac{C_A}{2N \gamma^2p^2}\frac{ \tau _1^2 \left(\tau _A p^2 e^{-\frac{t}{\tau _A}}-\tau_1 e^{-\frac{t p^2}{\tau _1}}\right)}{\tau _A^2 p^4-\tau _1^2}\right].
  \end{align}
 For other beads which are close to the COM, the term $\text{cos}^2\left(\frac{p \pi n}{N}\right)$ in the summation of Eq. (\ref{msd_flex_rs1}) may be substituted with $1/2$ for its rapidly oscillating nature. In the limit $\tau_A\ll t\ll \tau_1,$ the summation of Eq. (\ref{msd_flex_rs1}) can be transformed to integration \cite{khokhlov1994statistical}, which yields
  \begin{align}
 \Delta^2(n,t) & \approx \frac{2C_{\eta}\tau_1}{\gamma_p^2} \int_{0}^{\infty}\frac{dp}{p^2}\left[1-e^{-\frac{t}{\tau_1}p^2}\right]+ \frac{4C_{\sigma_1}\tau_1^2}{\gamma_p^2} \int_{0}^{\infty}\frac{dp}{p^2}\frac{1}{(p^2\tau_A+\tau_1)}\left[1-\frac{\tau _A p^2 e^{-\frac{t}{\tau _A}}-\tau_1 e^{-\frac{t p^2}{\tau _1}}}{p^2\tau_A-\tau_1}\right]\nonumber\\
& \approx \frac{2C_{\eta}\sqrt{\pi\tau_1}}{\gamma_p^2}t^{1/2}+\frac{4C_{\sigma_1}\tau_1}{\gamma_p^2}\int_{0}^{\infty}dp\left[\frac{1}{p^2}-\frac{\tau_A}{(p^2\tau_A+\tau_1)}\right]\left[1-\frac{\tau _A p^2 e^{-\frac{t}{\tau _A}}-\tau_1 e^{-\frac{t p^2}{\tau _1}}}{p^2\tau_A-\tau_1}\right]\nonumber\\
& \approx \frac{2C_{\eta}\sqrt{\pi\tau_1}}{\gamma_p^2}t^{1/2}+\frac{4C_{\sigma_1}\tau_1}{\gamma_p^2}\int_{0}^{\infty}\frac{dp}{p^2}\left[1-e^{-\frac{t}{\tau_1}p^2}\right]-\frac{4C_{\sigma_1}\tau_1}{\gamma_p^2}\int_{0}^{\infty}\frac{dp}{p^2+\frac{\tau_1}{\tau_A}}\left[1-e^{-\frac{t}{\tau_1}p^2}\right]\nonumber\\
& \approx \frac{2C_{\eta}\sqrt{\pi\tau_1}}{\gamma_p^2}t^{1/2}+\frac{4C_{\sigma_1}\sqrt{\pi\tau_1}}{\gamma_p^2}t^{1/2}+\frac{\pi C_{\sigma_1}\sqrt{\tau_1\tau_A}}{\gamma_p^2}  \left(e^{\frac{t}{\tau_A}} \text{erfc}\left(\sqrt{\frac{t}{\tau_A}}\right)-2\right). \label{approx_delta}
  \end{align}
 In this limit, the third term on the RHS can be ignored as $e^{\frac{t}{\tau_A}} \text{erfc}\left(\sqrt{\frac{t}{\tau_A}}\right) \approx \sqrt{\frac{\tau_A}{\pi t}}$ for $\frac{t}{\tau_A}\gg 1.$ Therefore, one has $\Delta^2(n,t)\propto \frac{2\sqrt{\pi\tau_1}}{\gamma_p^2} \left(C_{\eta}+2C_{\sigma_1}\right)t^{1/2},$ which is independent of $N$ and scales as $t^{1/2},$ implying a subdiffusive motion. For the passive polymer $(C_{\sigma_1}=0)$,  $\Delta^2(n,t)=\sqrt{\frac{4b^2 k_BT}{3\pi\gamma }t},$ dictating a a subdiffusive regime over the entire time scales shorter than the Rouse relaxation time $\tau_1,$ as one can clearly see this in Fig.  \ref{figure9}. 
 \\
 \\
 For the semiflexible polymer, we have  calculated Eq. (\ref{msd_rs1})  numerically, and the results are shown in panels (a) and (b) of Fig. \ref{msd-mono_semiflex} for the OUP and MOUP baths, respectively. The discussions are given towards the end of this section.

 \subsection{Non-Gaussian active bath}
Using the values of the parameters as given here,  $\gamma_0=N\gamma,\gamma_p=2 N\gamma,\,\,C_{\eta}=4N\gamma k_B T,\,\sigma_A^2=\frac{N}{2} C_A^2,$ and  substituting Eq. (\ref{msd_x0_nong}) in Eq. (\ref{msd_flex_x0}), one can obtain the MSD of COM as shown below. 
\begin{align}
 \langle \bm{R}_0^2(t)  \rangle &=\langle\left(\bm{r}_0(t)-\bm{r}_0(0)\right)^2 \rangle=3\langle\left(\bm{\psi}_0(t)-\bm{\psi}_0(0)\right)^2\rangle\nonumber\\
 &=\frac{6k_B T}{N \gamma}t+\frac{3\sigma_A^2\nu_A}{3\gamma_0^2} \left(\left(\Delta \tau _A-t\right)^3 \Theta \left(\Delta \tau _A-t\right)+  \Delta \tau _A^2(3t-\Delta \tau _A)\right)\nonumber\\
 &=\frac{6k_B T}{N \gamma}t+\frac{C_A^2\nu_A}{2N\gamma^2} \left(\left(\Delta \tau _A-t\right)^3 \Theta \left(\Delta \tau _A-t\right)+ \Delta \tau _A^2(3t-\Delta \tau _A)\right).\label{msd_flexng_x0}
\end{align}
In the limit $\Delta \tau _A\gg t$, 
$\langle \bm{R}_0^2(t) \rangle \approx \frac{6k_B T}{N \gamma}t+\frac{3 C_A^2\nu_A \Delta \tau _A}{2N\gamma^2}t^2 ,$ and  in the limit $t\rightarrow \infty,$ 
$\langle \bm{R}_0^2(t) \rangle  \approx \frac{6k_B T}{N \gamma}t+\frac{3 C_A^2\nu_A \Delta \tau _A^2}{2N\gamma^2}t.$  Like in the Gaussian baths, it exhibits a short-time ballistic regime (for $t_c<t\ll \tau_A$) arising due to the presence of active noise and a long-time diffusive behavior (for $t\gg \tau_A$).  \\
\\
The correlation between the end-to-end distance at two times $t$ and $t'(<t)$ can be expressed from Eq. (\ref{correlation_flex_end}) as  
\begin{align}
  &\langle \bm{R}_e(t)  \bm{R}_e(t') \rangle=\phi(t-t')=48\sum_{p=1}^{\infty}\,\langle\bm{\psi}_{2p-1}(t)\bm{\psi}_{2p-1}(t')\rangle \nonumber\\
 &=48\sum_{p=1}^{\infty}\,\Bigl[\frac{C_{\eta}\tau_p}{2\gamma_{2p-1}^2}e^{-\frac{t-t'}{\tau_{2p-1}}}+\frac{\sigma_A^2}{\gamma_{2p-1}^2}\tau_{2p-1}^3\nu_A\,e^{-\frac{|t-t'|}{\tau_{2p-1}}}\left(\text{cosh}\left(\frac{\Delta \tau_A}{\tau_{2p-1}}\right)-1\right)\nonumber\\
&+\frac{\sigma_A^2}{\gamma_{2p-1}^2}\tau_{2p-1}^2\nu_A\,\Theta\left(\Delta \tau_A-|t-t'|\right)\left(\Delta \tau_A-|t-t'|\right)\left(1-\frac{\text{sinh}\left(\frac{1}{\tau_{2p-1}}\left(\Delta \tau_A-|t-t'|\right)\right)}{\frac{1}{\tau_{2p-1}}\left(\Delta \tau_A-|t-t'|\right)}\right)\Bigr].\label{correlation_semiflex_end1}
 \end{align}
 For the flexible polymer, the auto-correlation is given by 
 \begin{align}
 & \phi(t-t')= \langle \bm{R}_e(t)  \bm{R}_e(t') \rangle\nonumber\\
&=48\sum_{p=1}^{\infty}\,\Bigl[\frac{\tau_1 k_B T}{2 N \gamma (2p-1)^2}e^{-\frac{t-t'}{\tau_1}(2p-1)^2}+\frac{C_A^2 \nu_A}{8 N\gamma^2}\frac{\tau_1^3}{(2p-1)^6}\,e^{-\frac{t-t'}{\tau_1}(2p-1)^2}\left(\text{cosh}\left(\frac{\Delta \tau_A}{\tau_1}(2p-1)^2\right)-1\right)\nonumber\\
& +\frac{C_A^2 \nu_A}{8 N\gamma^2}\frac{\tau_1^2}{(2p-1)^4}\,\Theta\left(\Delta \tau_A-(t-t')\right)\left(\Delta \tau_A-(t-t')\right)\left(1-\frac{\text{sinh}\left(\frac{(2p-1)^2}{\tau_1}\left(\Delta \tau_A-|t-t'|\right)\right)}{\frac{(2p-1)^2}{\tau_1}\left(\Delta \tau_A-|t-t'|\right)}\right)\Bigr].
\end{align}
The auto-correlation functions for the flexible and semiflexible polymers are shown in Figs. \ref{phi_flex} and \ref{phi_semiflex}, respectively. For both the polymer models, one can see that the correlation decays exponentially in  the passive case. But with the increase of persistence time of the active noise, it exhibits multi-exponential decay, which means that it decays at slower rates.  The initial rate is mainly determined by $\tau_A,$ which can be clearly understood from the plots for $\tau_A=50,\,100,\,400.$ At longer times ($t\gg \tau_A$), the correlation (or persistence) gradually vanishes due to thermal fluctuations, and consequently, the decay is then dictated by thermal relaxation times of the normal modes. Compared to the flexible case, the semiflexible polymer has more correlated motion possibly provided by the extra spatial persistence due to backbone rigidity, and this can be comprehended from Fig. \ref{phi_semiflex}. 
\\
\\
The MSD for bead $n$ (or contour length $s$) is given by Eq. (\ref{msd_flex_rs}), and it can be calculated using Eqs. (\ref{msd_flexng_x0}), (\ref{msd_nong}) and  (\ref{correlation_nong}) as follows: 
  \begin{align}
   &\Delta^2(n,t)=\langle \bm{R}^2(n,t)  \rangle\nonumber\\ 
   &= \langle \bm{R}_0^2(t)  \rangle +4 \sum_{p=1}^{\infty} \left[\text{cos}^2\left(\frac{p \pi n}{N}\right)\left( \langle\bm{\psi}_{p}^2(t)\rangle+ \langle\bm{\psi}_{p}^2(0)\rangle-2  \langle \bm{\psi}_{p}(t)\bm{\psi}_{p}(0)\rangle\right)\right]\nonumber\\
   &=\frac{6k_B T}{N \gamma}t+\frac{C_A^2\nu_A}{2N\gamma^2} \left(\left(\Delta \tau _A-t\right)^3 \Theta \left(\Delta \tau _A-t\right)+ \Delta \tau _A^2(3t-\Delta \tau _A)\right) +8 \sum_{p=1}^{\infty} \text{cos}^2\left(\frac{p \pi n}{N}\right)\nonumber\\
   &\Bigl[ \frac{C_{\eta}\tau_p}{2\gamma_p^2}+\frac{\sigma_A^2}{\gamma_p^2}\tau_p^3\nu_A\,\left(\text{exp}\left(-\frac{\Delta \tau_A}{\tau_p}\right)+\frac{\Delta \tau_A}{\tau_p}-1\right)-\frac{C_{\eta}\tau_p}{2\gamma_p^2}e^{-\frac{t}{\tau_p}}-\frac{\sigma_A^2}{\gamma_p^2}\tau_p^3\nu_A\,e^{-\frac{t}{\tau_p}}\left(\text{cosh}\left(\frac{\Delta \tau_A}{\tau_p}\right)-1\right)\nonumber\\
&-\frac{\sigma_A^2}{\gamma_p^2}\tau_p^2\nu_A\,\Theta\left(\Delta \tau_A-t\right)\left(\Delta \tau_A-t\right)\left(1-\frac{\text{sinh}\left(\frac{1}{\tau_p}\left(\Delta \tau_A-t\right)\right)}{\frac{1}{\tau_p}\left(\Delta \tau_A-t\right)}\right) \Bigr].  
 \end{align}

\begin{figure}
    \centering
\includegraphics[width=0.65\linewidth]{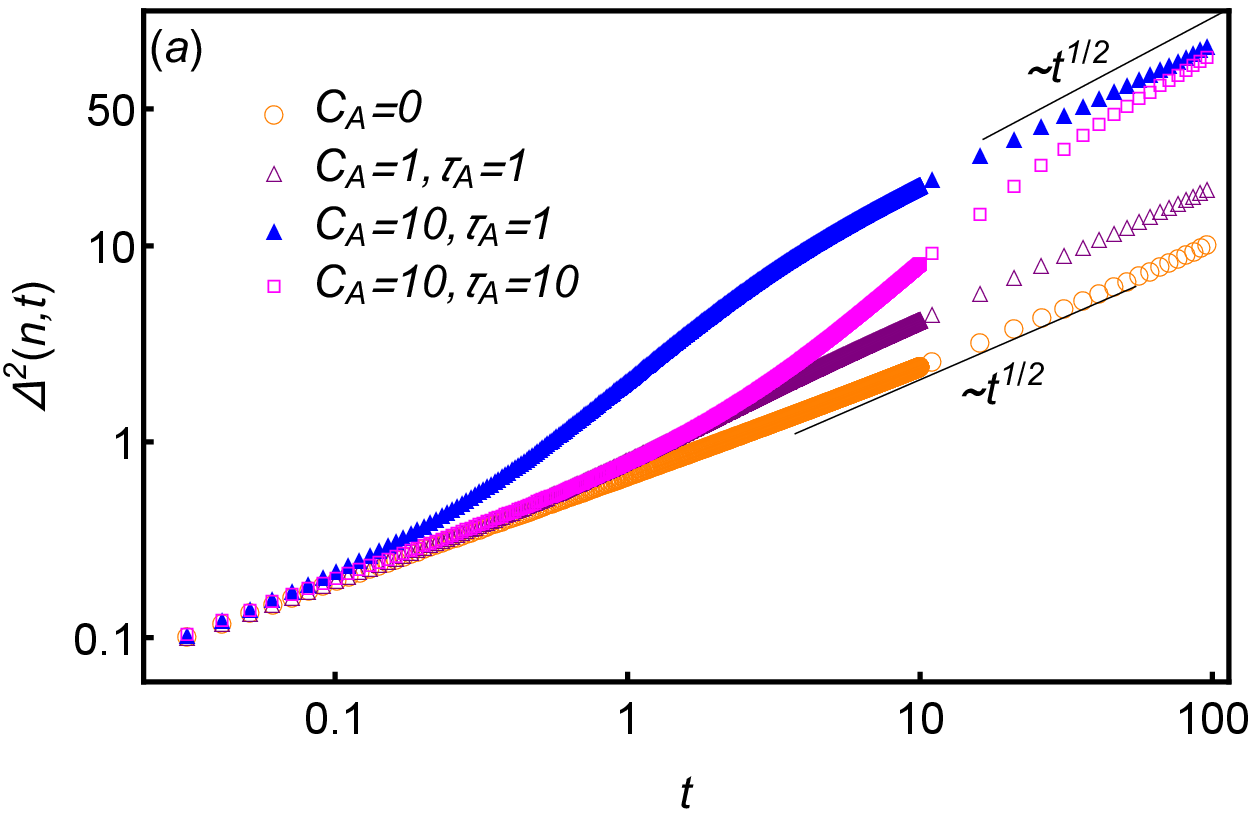}
 \centering
\includegraphics[width=0.65\linewidth]{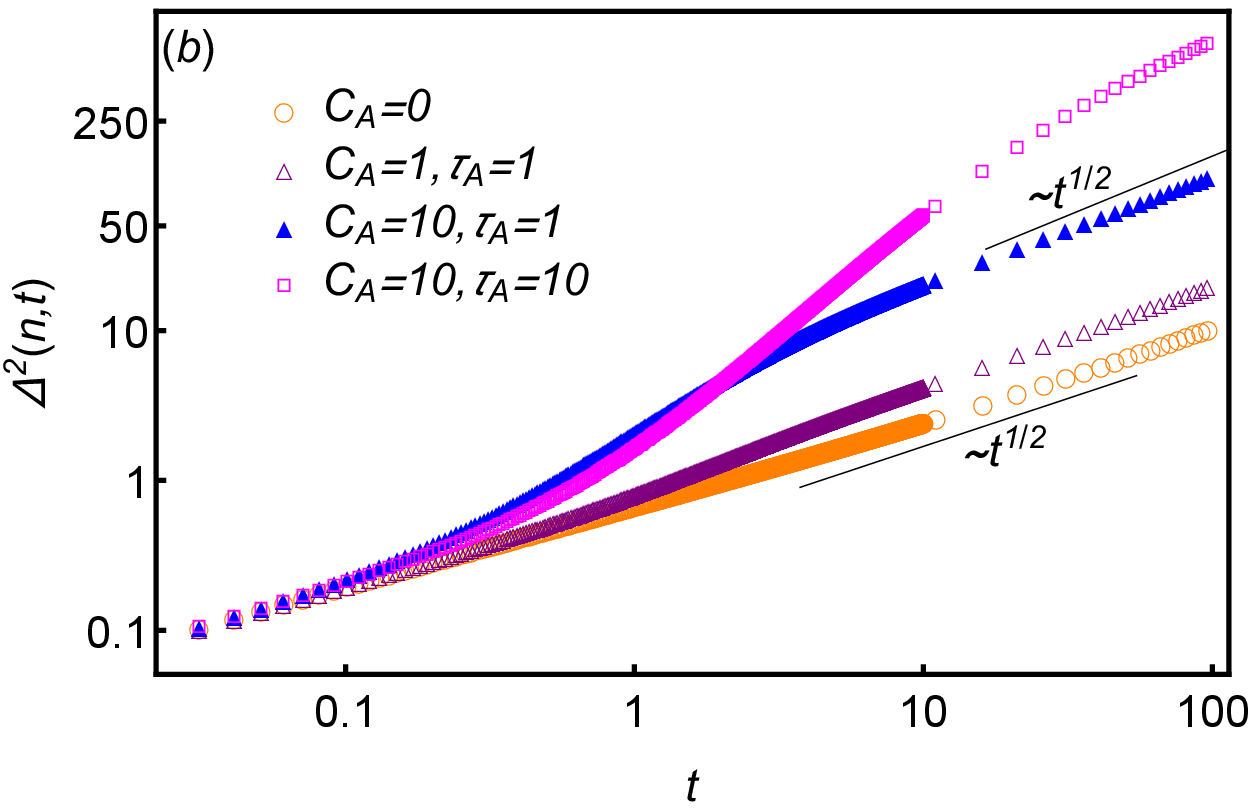}
 \centering
\includegraphics[width=0.65\linewidth]{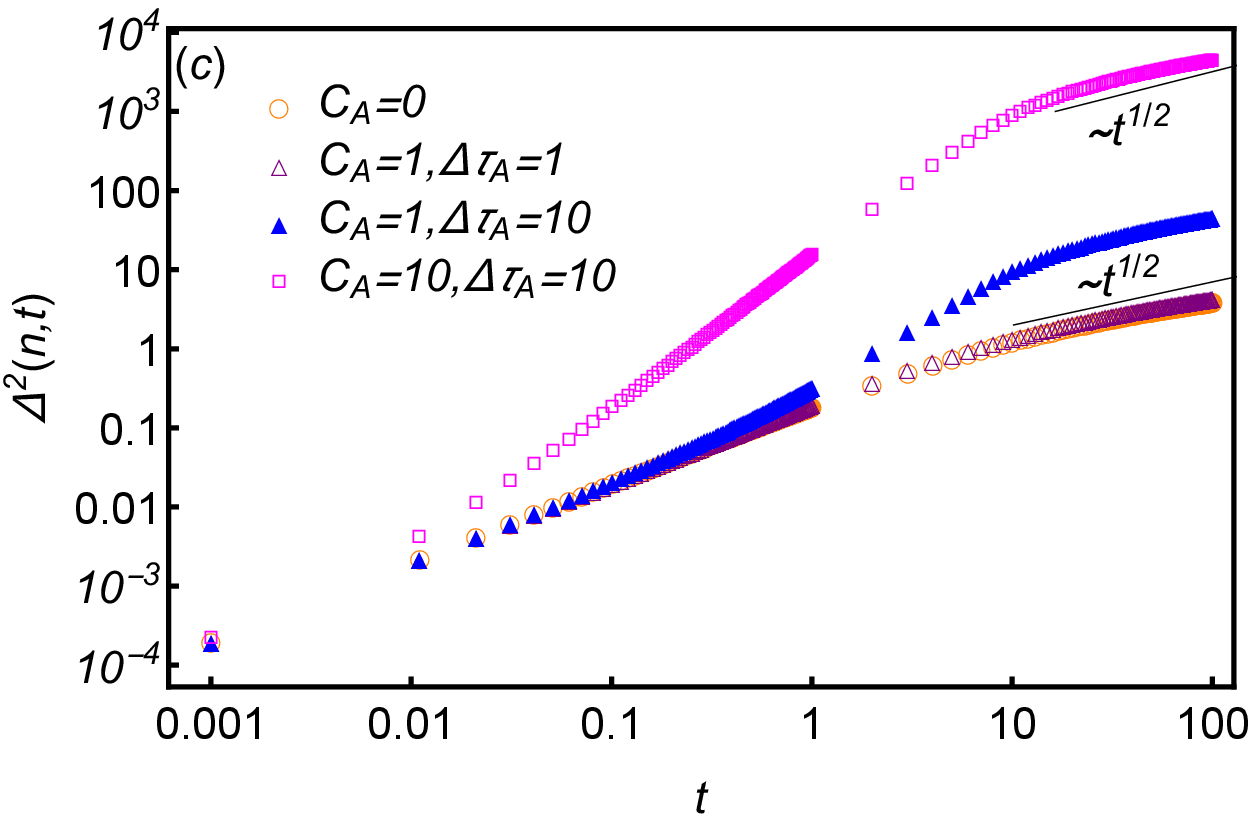}
 \caption{Plot of the MSD of a monomer in the flexible chain as a function of time $t$ in logarithmic scales on both the axes for different values of noise amplitude of the active noise  modelled as  (a) OU process, (b) MOU process and (c) shot noise. Here, $N=100,\,n=30.$ }
    \label{figure9}
\end{figure}

\begin{figure}
    \centering
\includegraphics[width=0.65\linewidth]{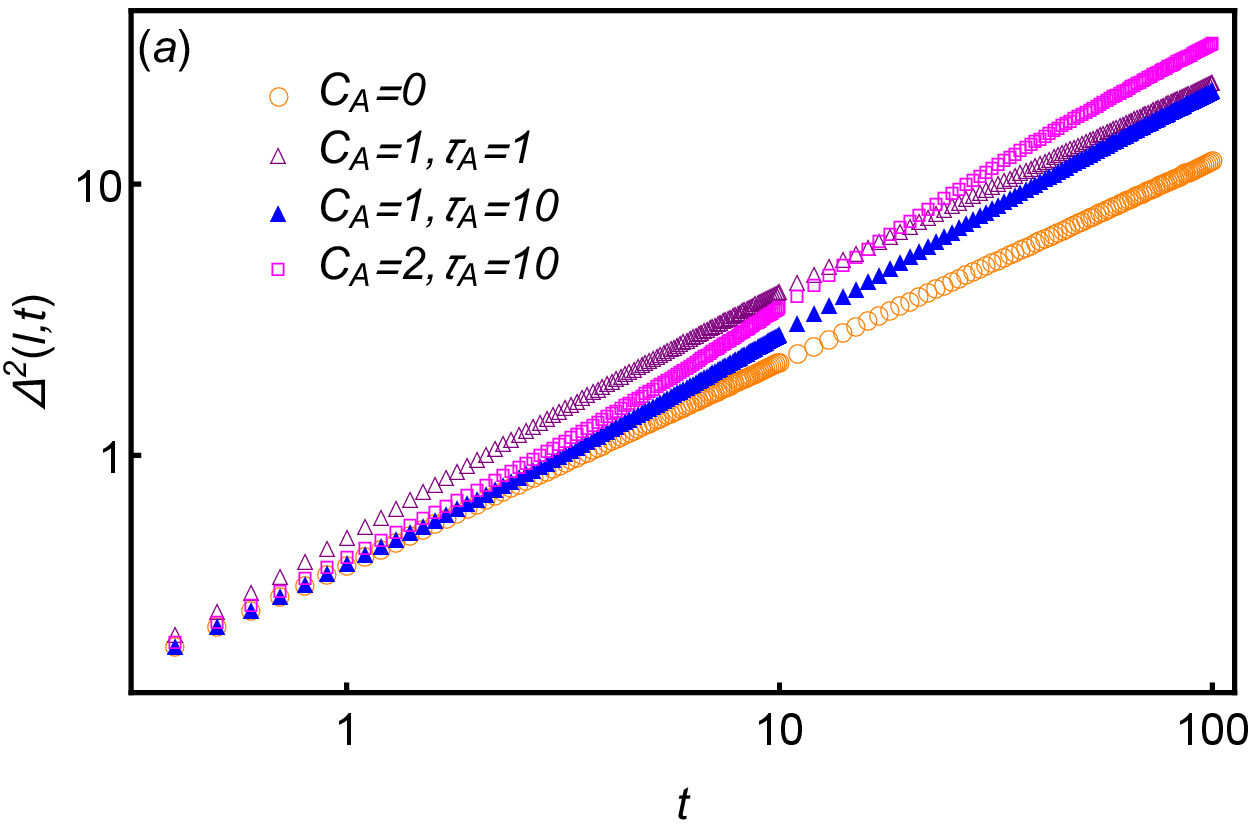}
 \centering
\includegraphics[width=0.65\linewidth]{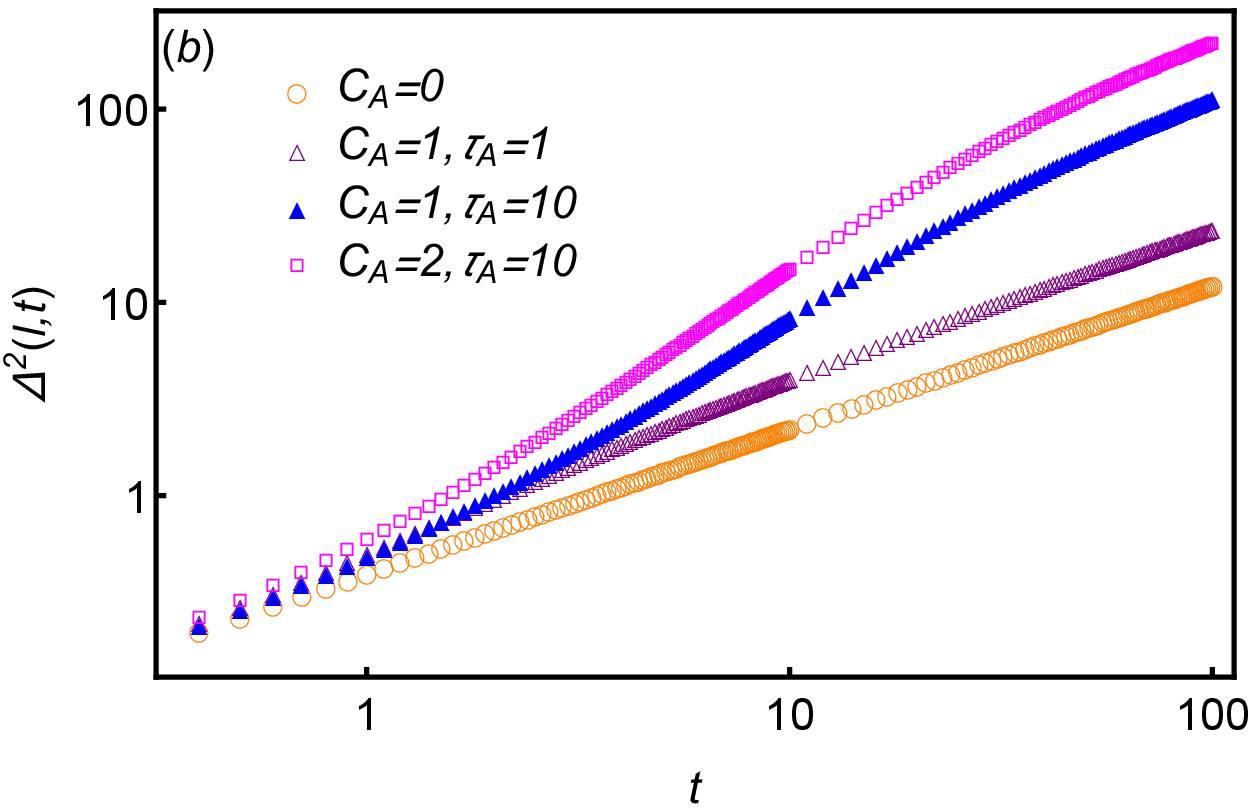}
 \centering
\includegraphics[width=0.65\linewidth]{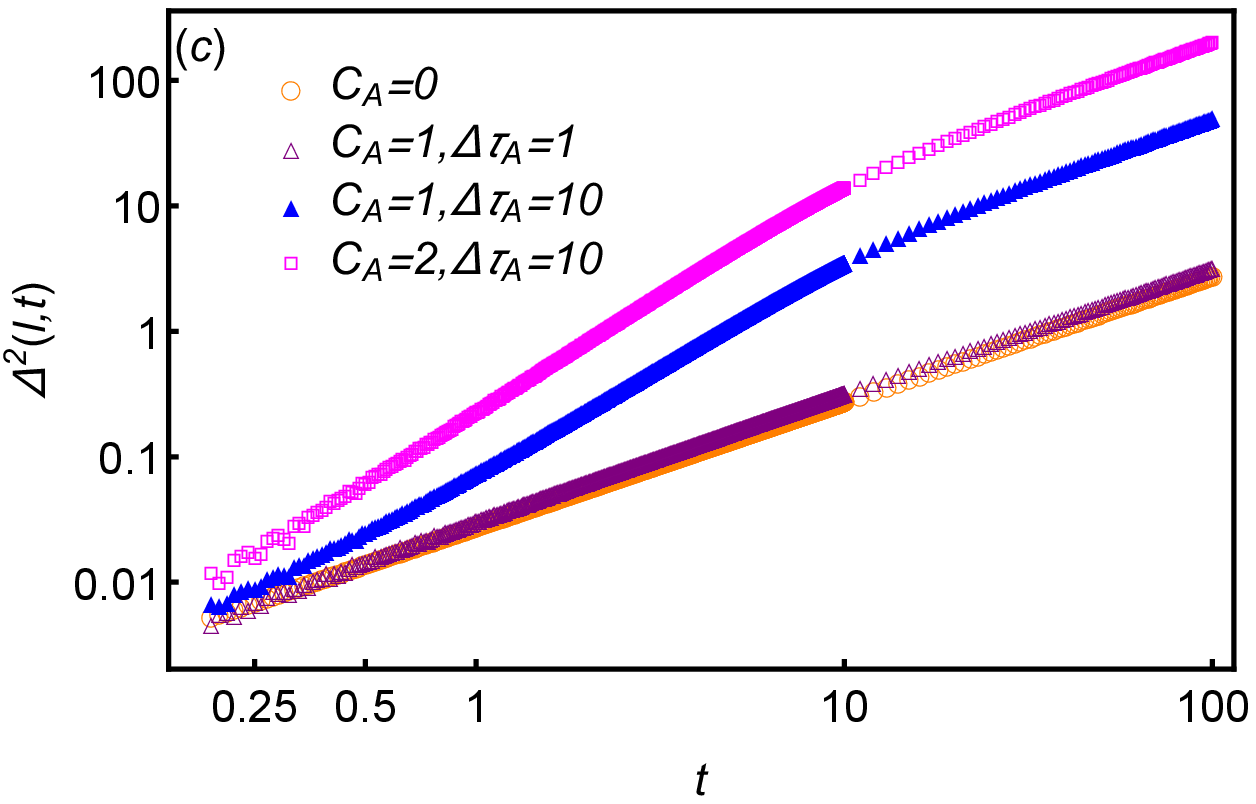}
\caption{Log-Log plot of the MSD of a position $\bm{r}(l=150,t)$ on a semiflexible chain  as a function of time $t$  for different values of  amplitude of active noise $C_A$ and persistence time $\tau_A$  in the (a) OUP, (b) MOUP and (c) non-Gaussian baths. Here, the total length and  persistence length of the polymer are taken as $L=500$ and $l_p=10$, respectively. }
    \label{msd-mono_semiflex}
\end{figure}

\noindent In the case of a passive flexible polymer in a thermal bath, the MSD of a tagged monomer scales with time as $t^{1/2}$  \cite{khokhlov1994statistical}. However with introducing the activity, it shows an intermediate super-diffusive behavior ($i.e.$, $\Delta^2(n,t)\propto t^{\nu}$, where $\nu>1$)  between two passive-like regimes characterized by $t^{1/2}$ scaling [cf. Eq. (\ref{approx_delta})], as illustrated in Fig. \ref{figure9} \cite{Samanta_2016,doi:10.1063/1.5086152}. In the very long-time limit $i.e.,$ for $t\gg\tau_1,$ the dynamics can be described by the center-of-mass motion, which is diffusive (scales as $t^1$) with an elevated diffusivity. The length of the super-diffusive regime strongly depends on the parameters $C_A$ and $\tau_A.$ In both   MOUP and non-Gaussian baths, with increasing $C_A$ and $\tau_A,$ the superdiffusive motion can be seen for longer period of times as displayed in Figs. \ref{figure9} and \ref{msd-mono_semiflex}. However, in the OUP bath, at initial times, it has inverse dependence on $\tau_A$ as the amplitude of the active noise is reduced with increasing $\tau_A.$ For the semiflexible chain, $\tau_1\propto L^4,$ which suggests that the thermal contributions play a role at a  much  later time compared to a flexible polymer of the same length, and so the long-diffusive limit governed by the center-of-mass motion also occurs at a comparatively longer time. However, other features of the MSD remain the same, $viz.$ an intermediate superdiffusive region flanked by two passive-like regions, but  $\Delta^2(l,t)$ for  the  passive regimes  grows with time as $t^{3/4},$ unlike a passive flexible case \cite{eisenstecken2017internal}. It is to note here that, in Ref. \cite{eisenstecken2017internal} the authors have seen a Rouse-like regime ($ \Delta^2(l,t) \propto \sqrt{t}$) in the limit $\tau_A<t<\tau_1$ for a semiflexible polymer at a relatively high P\'eclet number. However, such feature disappears if one considers the stretching coefficient ($\mu$) to be independent of activity, which is the case for the present model.  

\begin{figure}
    \centering
\includegraphics[width=0.65\linewidth]{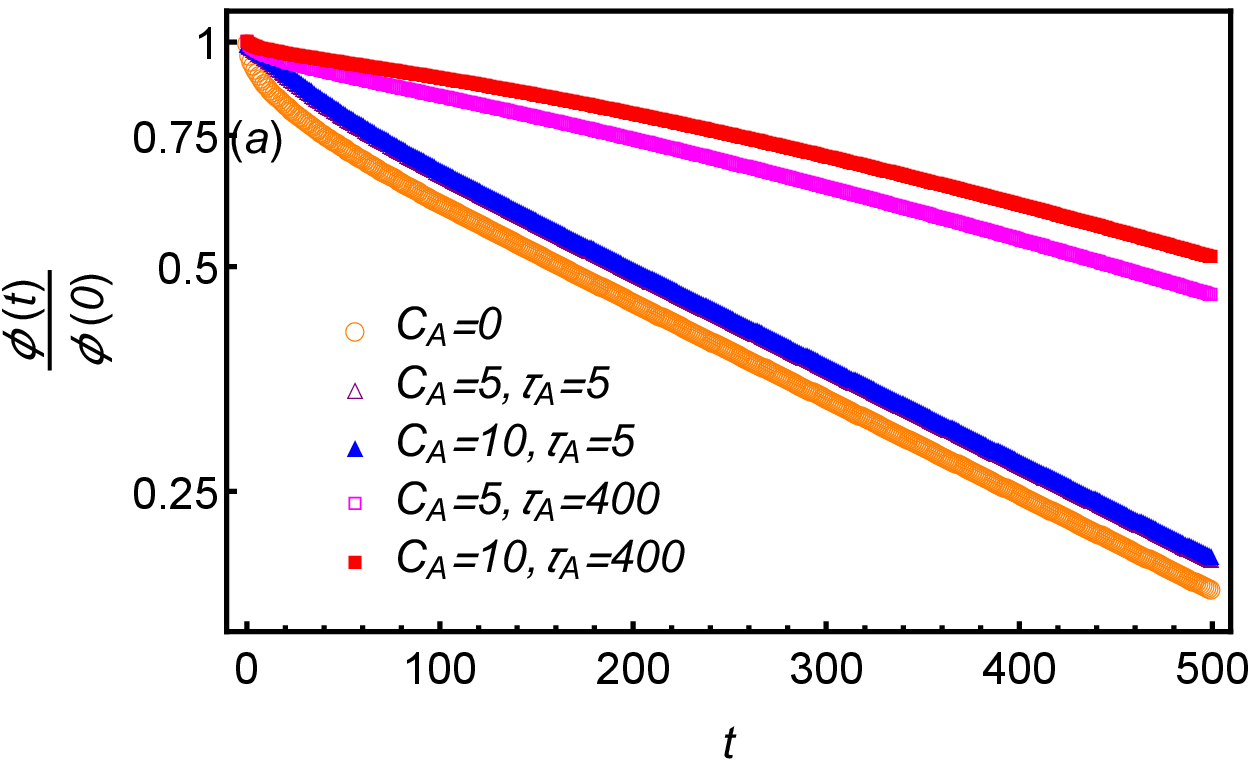}
 \centering
\includegraphics[width=0.65\linewidth]{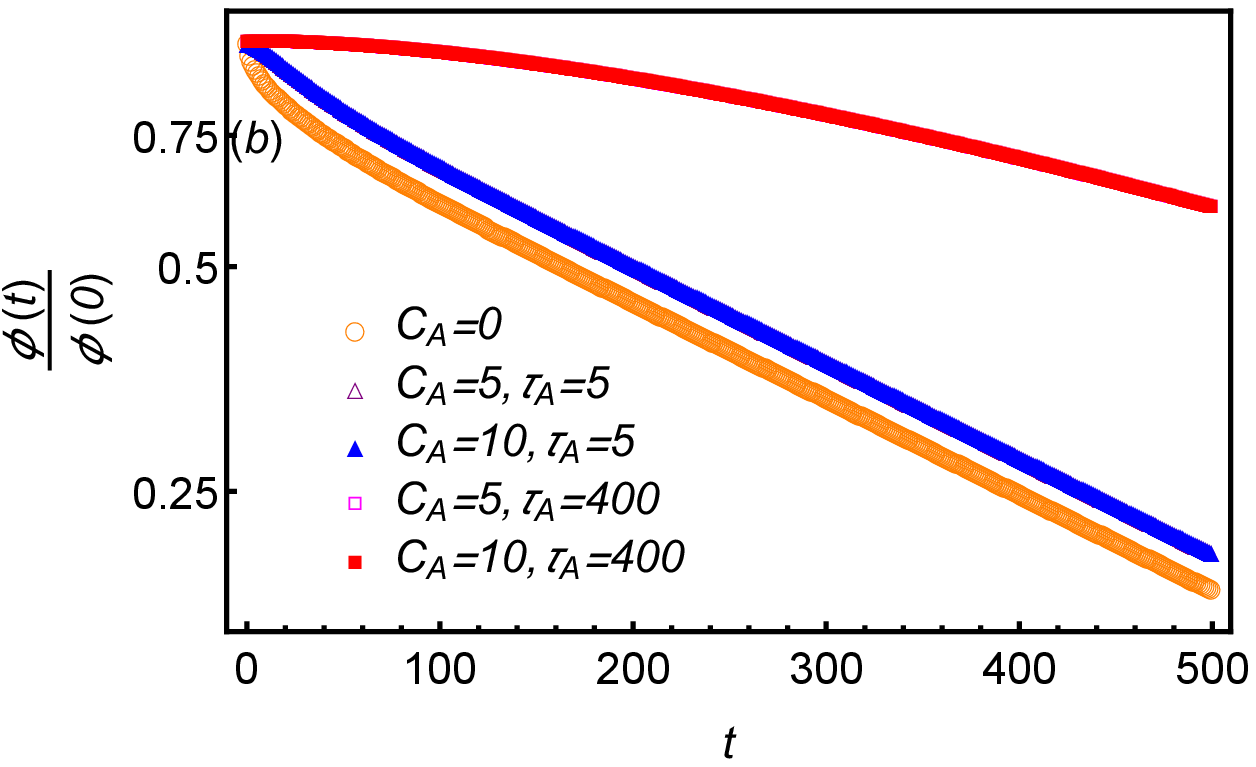}
 \centering
\includegraphics[width=0.65\linewidth]{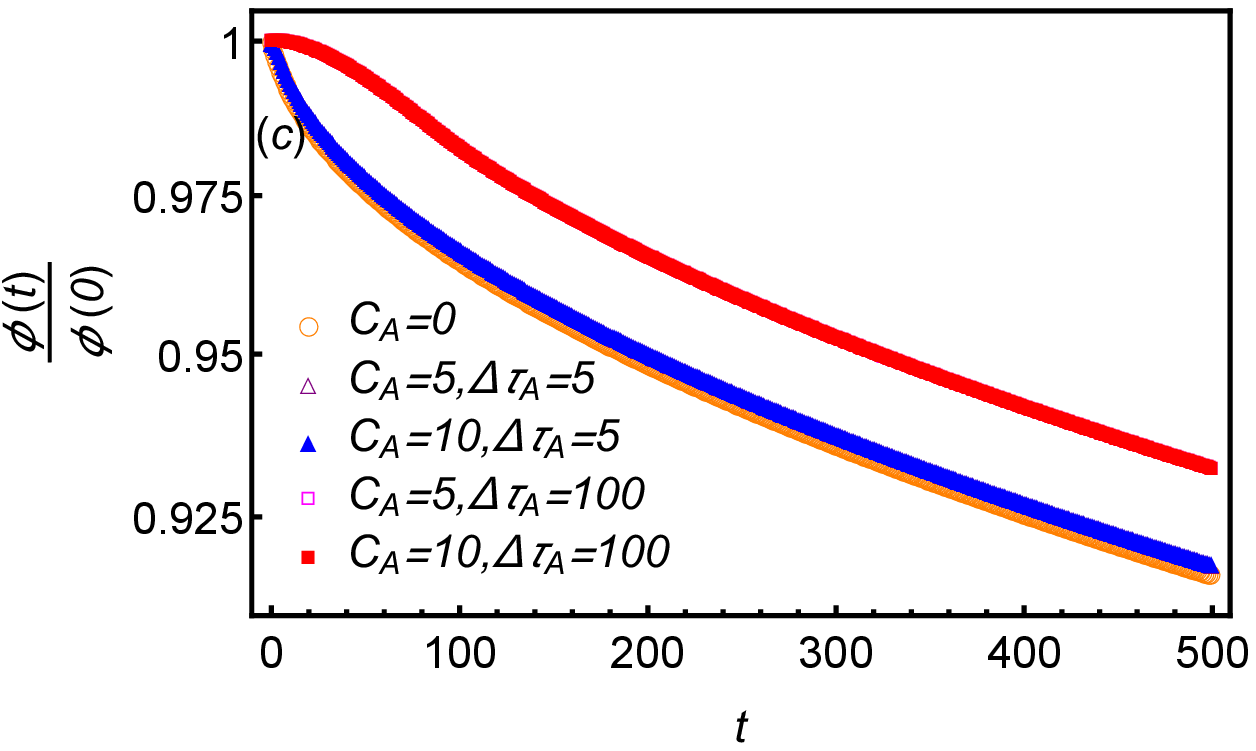}
    \caption{The auto-correlation function of the end-to-end vector scaled by the same time correlation as a function of time $t$ in the case of the flexible polymer model for different values of amplitude of active noise  in the (a) OUP, (b) MOUP and (c) non-Gaussian baths. Here, N=100. }
    \label{phi_flex}
\end{figure}

\begin{figure}
    \centering
\includegraphics[width=0.65\linewidth]{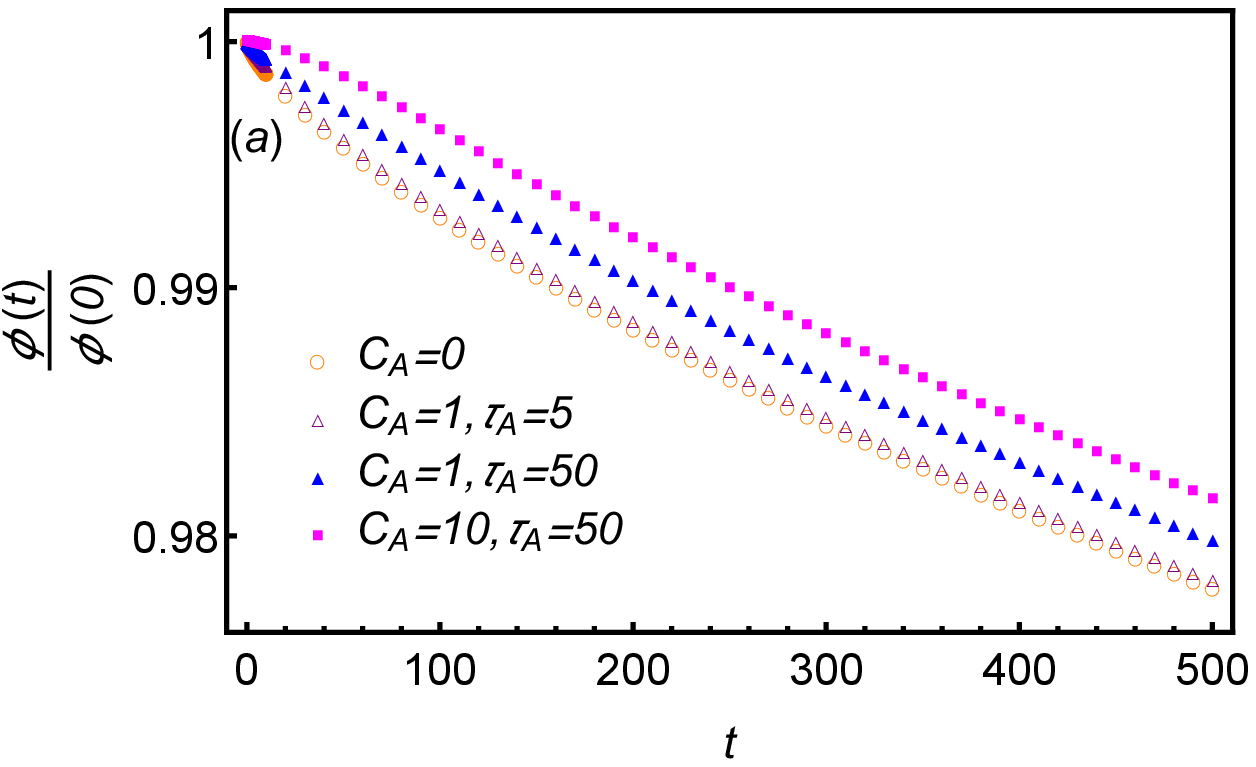}
 \centering
\includegraphics[width=0.65\linewidth]{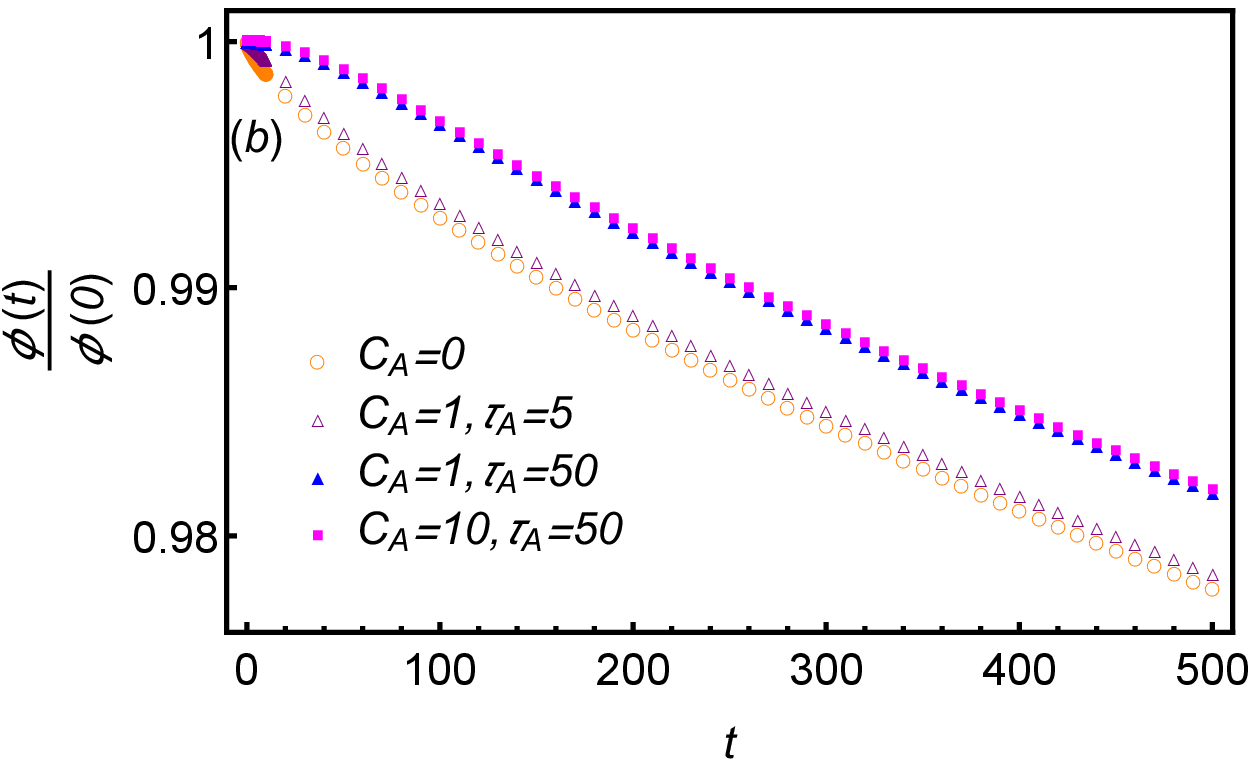}
 \centering
\includegraphics[width=0.65\linewidth]{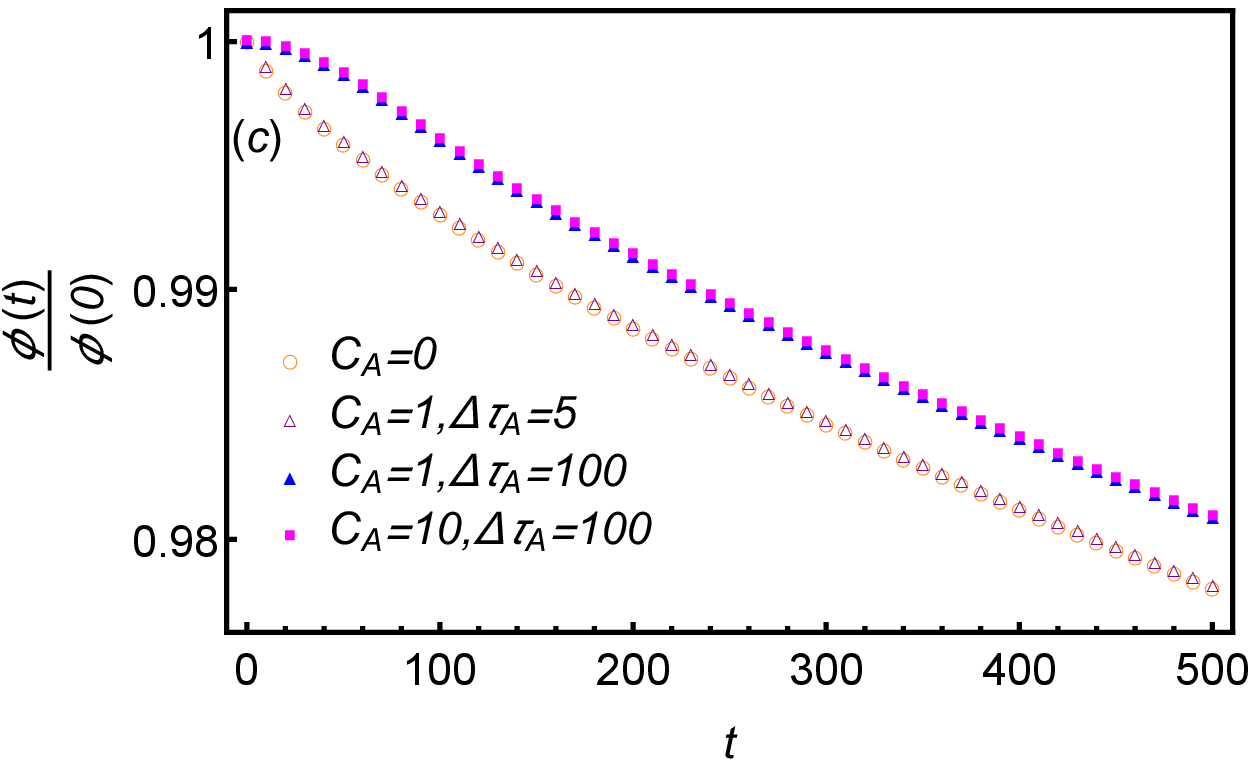}
\caption{The auto-correlation function of the end-to-end vector scaled by the same time correlation as a function of time $t$ in the case of the semiflexible polymer model for different values of  amplitude of active noise  in the (a) OUP, (b) MOUP and (c) non-Gaussian baths. Here, $L=500,\,l_p=10.$  }
    \label{phi_semiflex}
\end{figure}


\section{Chain reconfiguration \label{sec-7}}
One can get an insight about the charactereistic timescale for the fluctuations of the end-to-end vector from the reconfiguration time $\tau_e$ which is defined as \cite{samanta2015looping,Samanta_2016}
\begin{align}
    \tau_{e}=\int_{0}^{\infty}dt\,\frac{\phi(t)}{\phi(0)}.\label{recon_time}
\end{align}
It is useful to understand the rate of conformational change of an unfolded protein molecule. Also, it is of experimental relevance  as  the long-time decay  of the intensity correlation function for two tagged regions of a polymer is related to $\tau_{e}.$ 
\\
\\
The effective mode relaxation time can be defined as 
\begin{align}
\tau_m(p)=\int_{0}^{\infty}dt\,\frac{\langle\bm{\psi}_{p}(t)\bm{\psi}_{p}(0)\rangle}{\langle\bm{\psi}_{p}(0)\bm{\psi}_{p}(0)\rangle},\label{mode_time}
\end{align}
which gives idea about how the normal modes relaxes with the mode numbers. For  a passive flexible polymer, $\tau_m(p)$ scales as $p^{-2}$ for all modes.  For the passive semiflexible polymer, one can observe two scalings for  two extreme regimes, $viz.$, $(i)\,\tau_m(p)\propto p^{-4}$ for $L\leq \pi l_p$ and $(ii)\,\tau_m(p)\propto p^{-2}$ for $L\gg \pi l_p.$ With activity, the relaxation time is also modified the results of which are shown in the next section.

\subsection{Gaussian active bath}
\begin{figure}
    \centering
\includegraphics[width=0.65\linewidth]{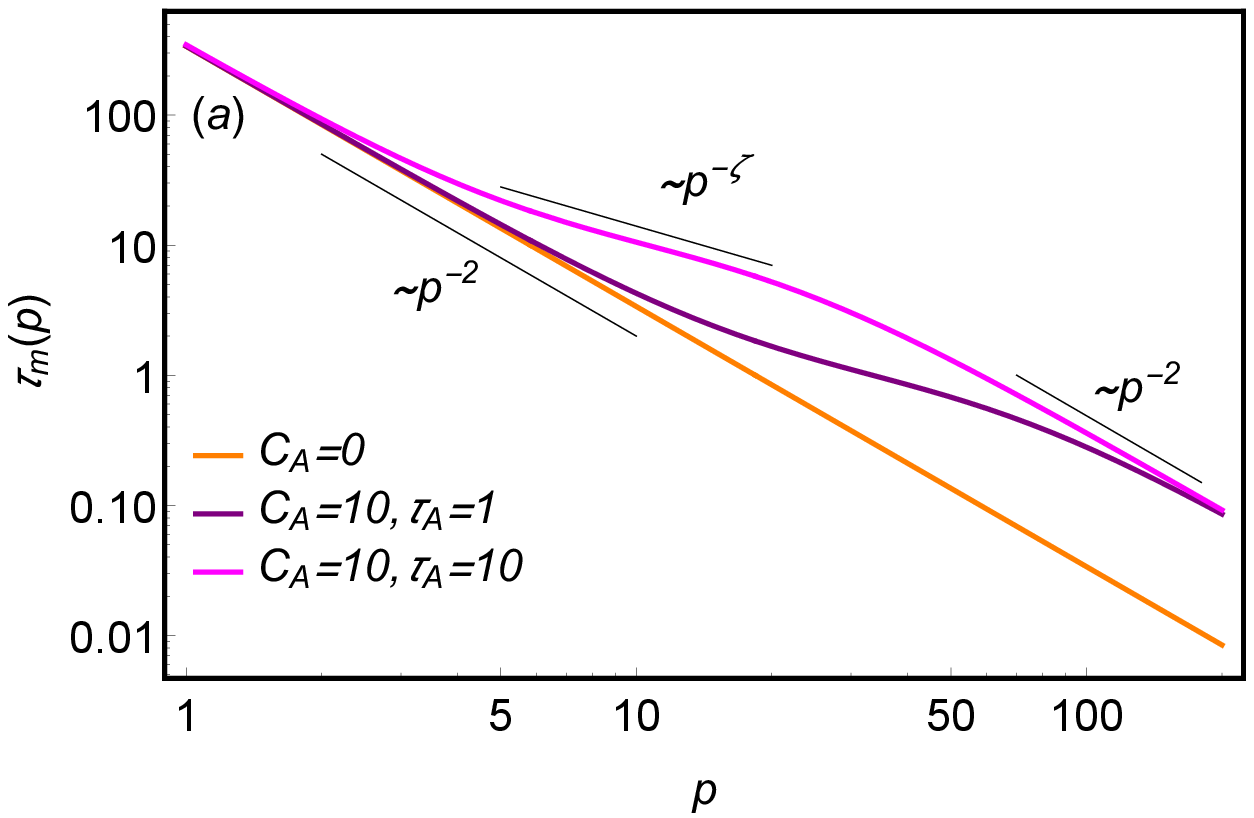}
 \centering
\includegraphics[width=0.65\linewidth]{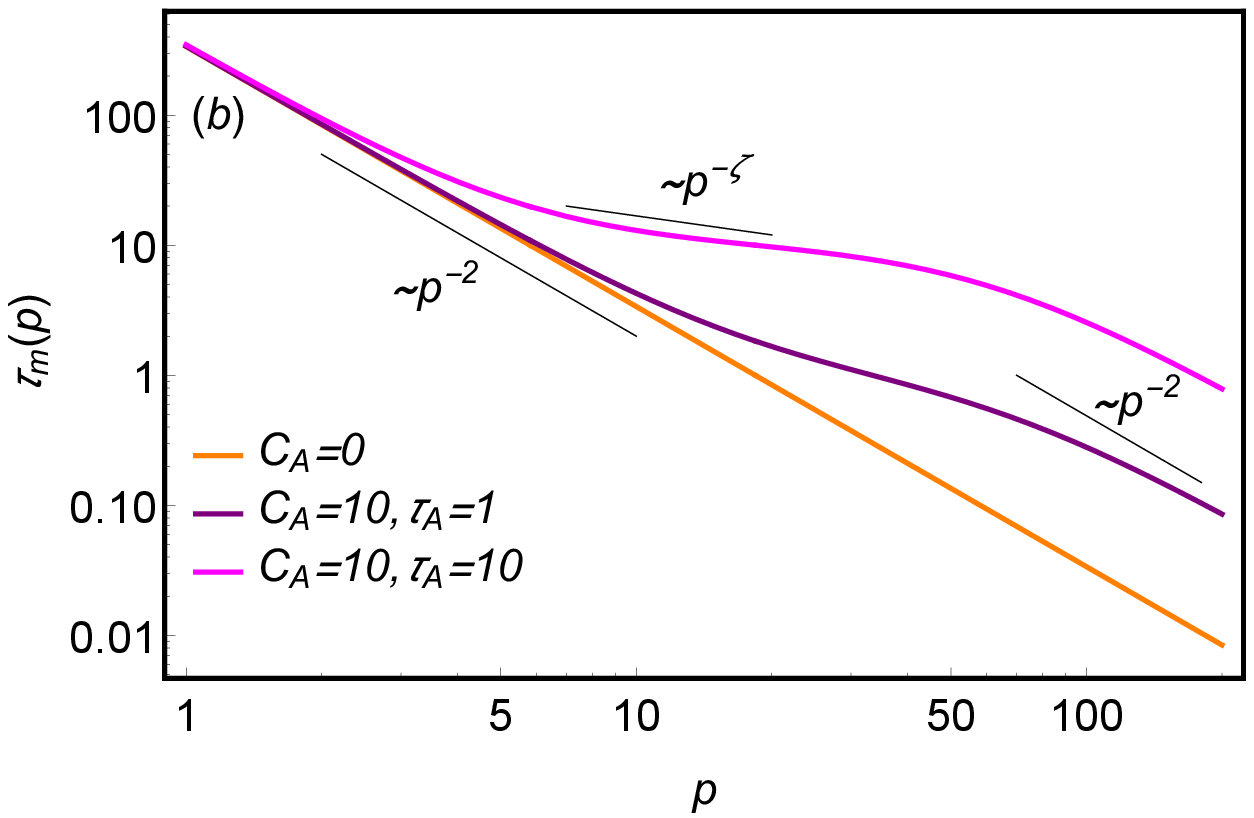}
 \centering
\includegraphics[width=0.65\linewidth]{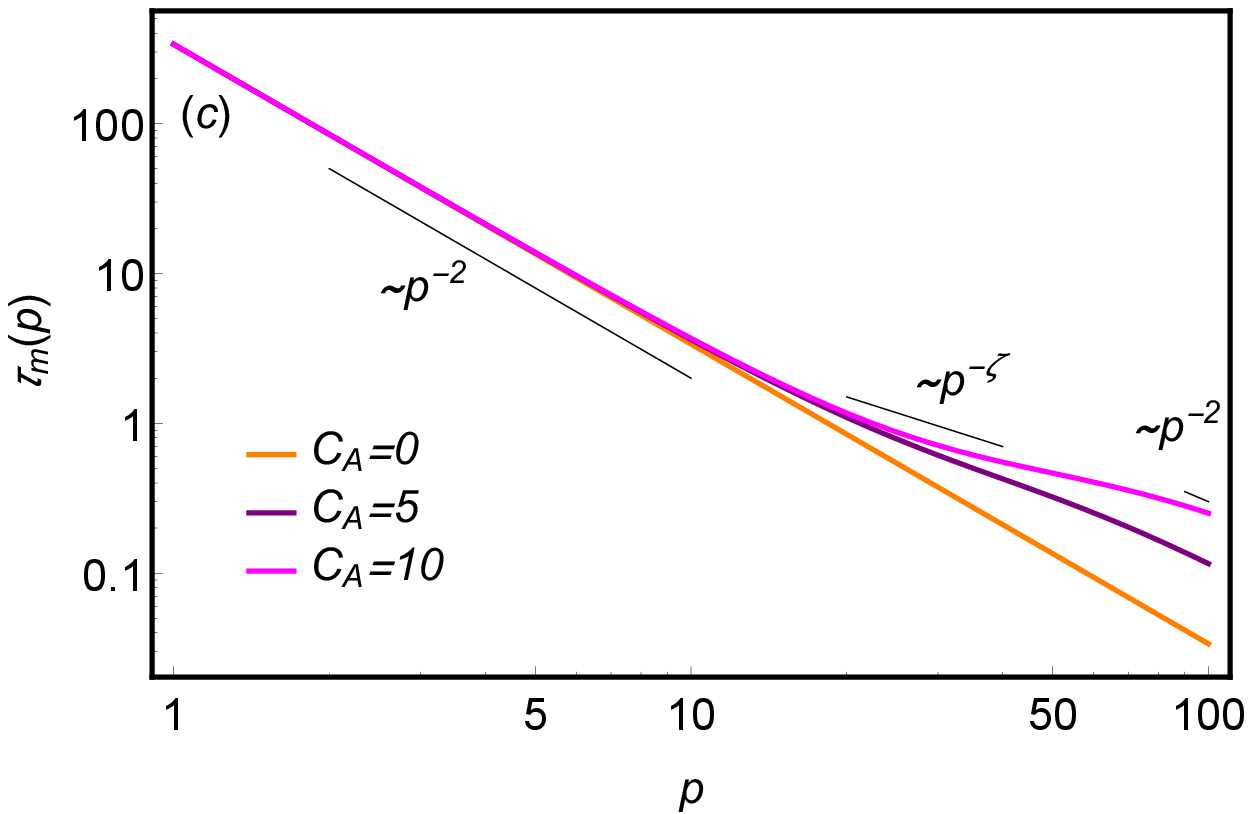}
    \caption{Log-Log plot of  effective relaxation time of normal modes  versus mode number $p$ for different values of noise amplitude and persistence time of active noise in the case of the flexible polymer model   in the (a) OUP, (b) MOUP and (c) non-Gaussian baths. Here, $N=100.$ }
    \label{tm_flex}
\end{figure}

\begin{figure}
    \centering
\includegraphics[width=0.65\linewidth]{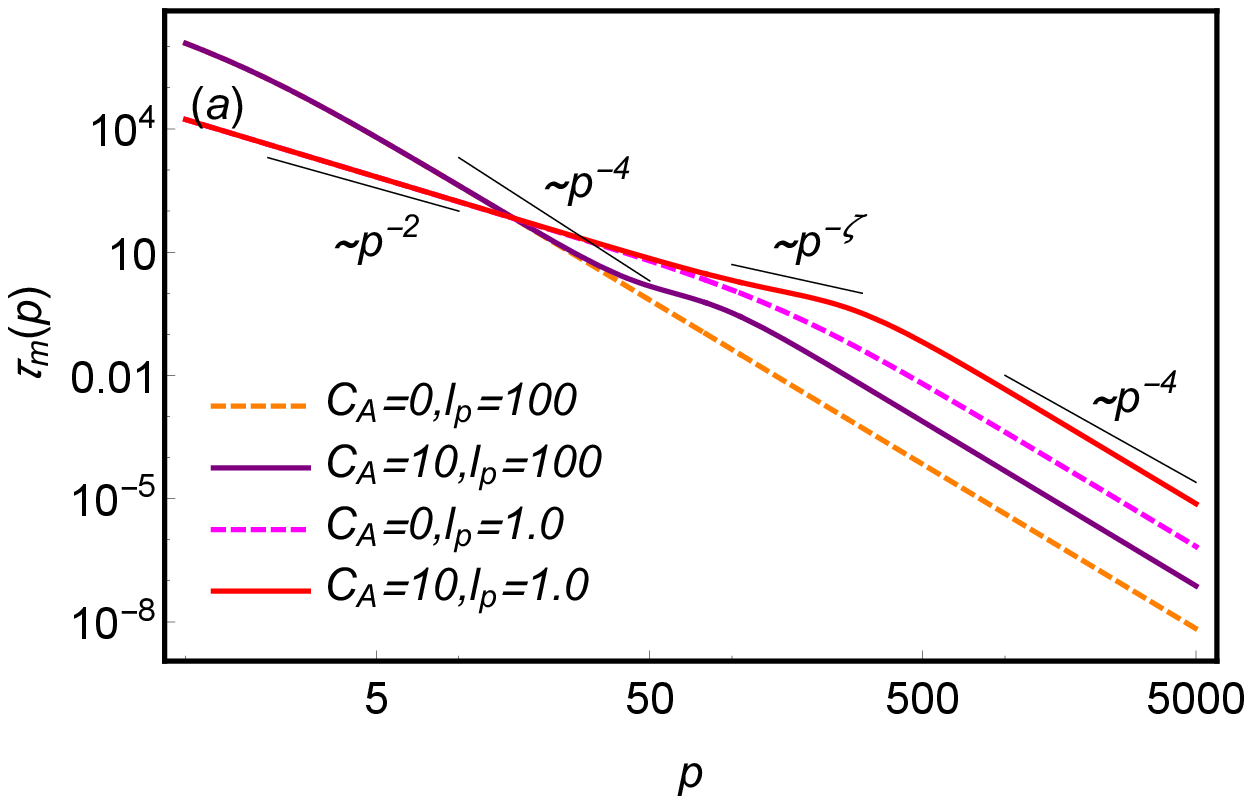}
 \centering
\includegraphics[width=0.65\linewidth]{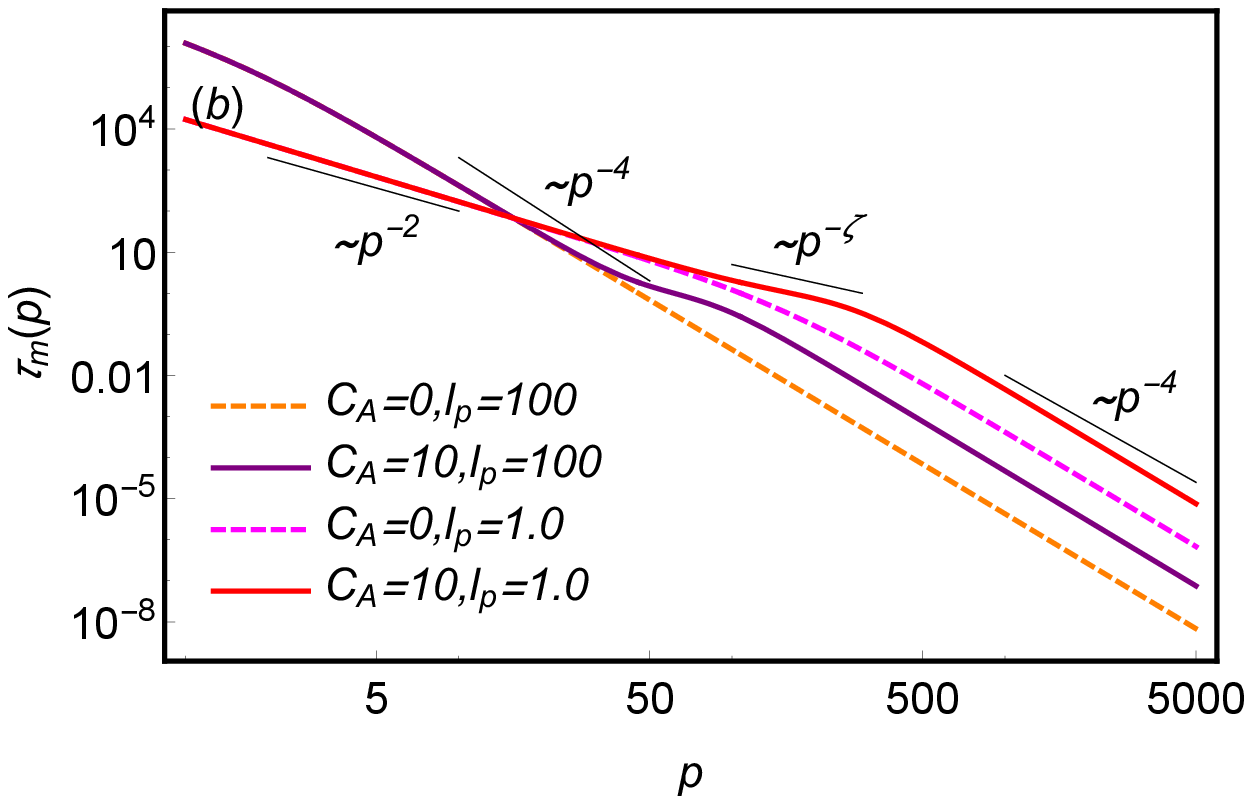}
 \centering
\includegraphics[width=0.65\linewidth]{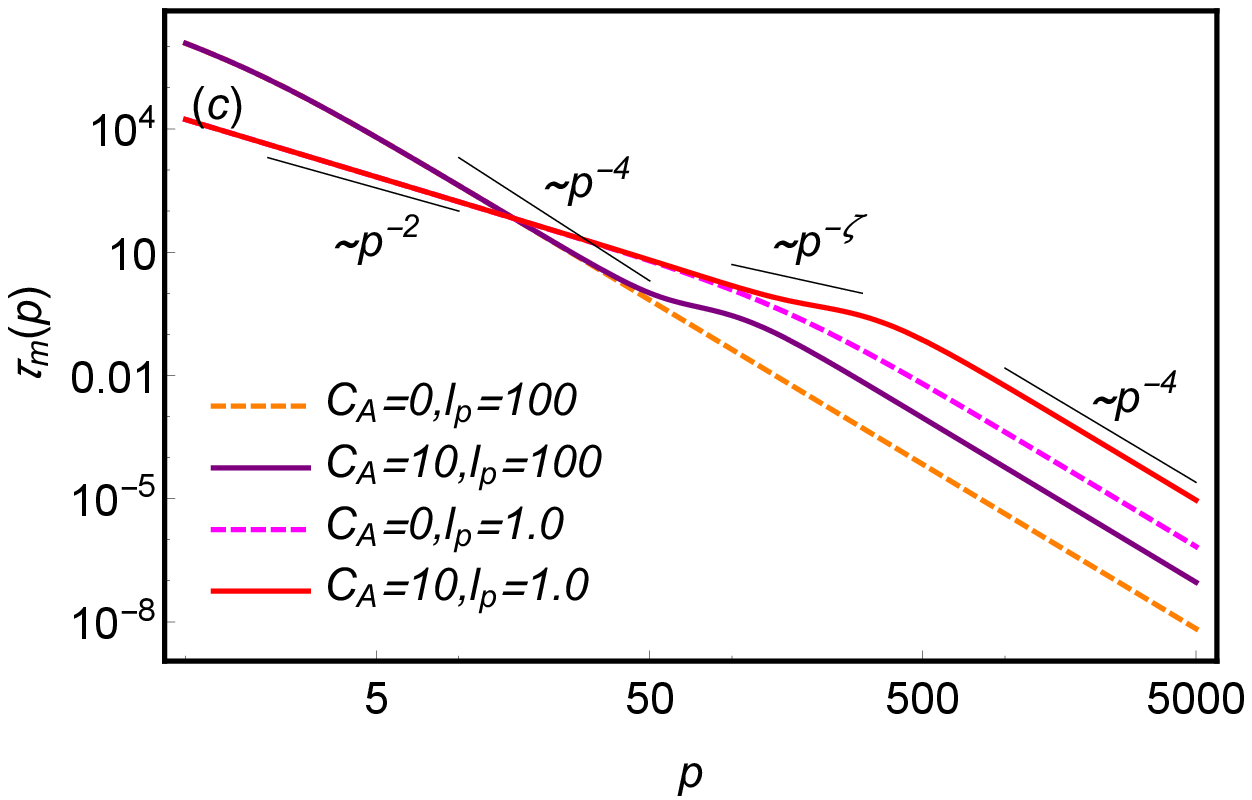}
\caption{Log-Log plot of  effective relaxation time of normal modes  versus mode number $p$ for different values of amplitude of active noise in the case of the semiflexible polymer model   in the (a) OUP, (b) MOUP and (c) non-Gaussian baths. Here, $L=500.$ }
    \label{tm_semiflex}
\end{figure}

Using Eq. (\ref{correlationx_Gaussian}), the effective mode relaxation time [Eq. (\ref{mode_time})] for a polymer in the OUP bath can be calculated exactly, which is given by
\begin{align}
    \tau_m(p)=\tau_p+\frac{2C_{\sigma_1}\tau_A}{2C_{\sigma_1}+C_{\eta}\left(1+\frac{\tau_A}{\tau_p}\right)}\label{taum_gauss}.
\end{align}
It is very much evident from the above equation that $\tau_m(p)=\tau_p$ without any active noise. But as the activity increases, the second term on the RHS contributes which leads to a different scaling for intermediate eigenmodes than $\tau_p$ does. In the high $p$ limits, the contribution of the second term again diminishes, and as a result, $\tau_m(p)$ and $\tau_p$ scales similarly with $p.$ This has been illustrated in Fig. \ref{tm_flex} for the flexible polymer where $\tau_m(p)\propto p^{-\zeta},$ with $\zeta<2,$ is observed after a certain mode number $p_c,$ and $p_c$ takes a smaller value as one increases $\tau_A.$  For the semiflexible polymer,  $\tau_m(p)\propto p^{-4}$ at very high $p$ value. However, the behavior $\tau_m(p)$ at smaller $p$  depends on the persistence length $l_p;$ If $L\gg \pi l_p$ then  $\tau_m(p)\propto p^{-2},$ displaying  the flexible characteristics, and in the other limit $L\leq \pi l_p$,  $\tau_m(p)\propto p^{-4},$  dominating the bending modes. The effect of activity is similar to the former case, though the scaling may differ, $viz.$, $\tau_m(p)\propto p^{-\zeta},$ where  $\zeta<4,$ as shown in Fig. \ref{tm_semiflex}.  

\noindent Using Eq. (\ref{phi_general}), one can compute
\begin{align}
  &\int_{0}^{\infty} dt\,\phi(t)=48\int_{0}^{\infty} dt\,\sum_{p: \text{ odds}}^{}\, \langle \psi_p(t) \psi_p(0) \rangle\nonumber\\
 &=48\sum_{p: \text{ odds}}^{}\int_{0}^{\infty} dt\,\left[ \frac{C_{\eta}\tau_p}{2\gamma_p^2}e^{-\frac{t}{\tau_p}}+\frac{C_{\sigma_1}}{\gamma_p^2}\frac{ \tau _p^2 \left(\tau _A e^{-\frac{t}{\tau _A}}-\tau _p e^{-\frac{t}{\tau _p}}\right)}{\tau _A^2-\tau _p^2}\right]\nonumber\\
 &=48\sum_{p: \text{ odds}}^{}\frac{C_{\eta}\tau_p^2}{2\gamma_p^2}+\frac{C_{\sigma_1}\,\tau _p^2}{\gamma_p^2}.
\end{align}
For the flexible polymer in the OUP bath, the above can be further simplified to
\begin{align}
 \int_{0}^{\infty} dt\,\phi(t)   =48\sum_{p=1}^{\infty}\frac{\tau_1^2}{(2p-1)^4}\left[\frac{k_B T }{2N\gamma}+\frac{C_A }{2N\gamma^2}\right]=\frac{\tau_1^2\pi^4}{4 N \gamma}\left[k_B T +\frac{ C_A}{\gamma}\right].
\end{align}
From Eq. (\ref{msd_flex_end_OUP}), one has
$\phi(0)=Nb^2+\frac{C_A}{\gamma k_BT}Nb^2\left(1-\frac{2}{\pi}\sqrt{\frac{\tau_A}{\tau_1}}\text{tanh}\left(\frac{\pi}{2}\sqrt{\frac{\tau_1}{\tau_A}}\right)\right).$
Putting all the above results together  and using Eq. (\ref{recon_time}), $ \tau_{e}$  for the flexible polymer in the OUP bath can be found to be 
\begin{align}
\tau_e=\frac{\frac{\tau_1^2\pi^4}{4 N \gamma}\left[k_B T +\frac{ C_A}{\gamma}\right]}{Nb^2+\frac{C_A}{\gamma k_BT}Nb^2\left(1-\frac{2}{\pi}\sqrt{\frac{\tau_A}{\tau_1}}\text{tanh}\left(\frac{\pi}{2}\sqrt{\frac{\tau_1}{\tau_A}}\right)\right)}.\label{reconfiguration_flex}
\end{align}
Without the active noise, the configuration time can be calculated easily, which is given by
\begin{align}
    \tau_{e}&=\int_{0}^{\infty}dt\,\frac{\phi(t)}{\phi(0)}= \frac{\tau_1^2\pi^4\,k_B T}{4 N \gamma Nb^2}=\frac{N^2b^2\gamma}{36k_BT}=\frac{\langle R_e^2 \rangle}{36 D_0},
\end{align}
with $D_0$ being the diffusivity of the Rouse polymer, defined as $D_0=\frac{k_BT}{N\gamma},$ and the mean square end-to-end distance $\langle R_e^2 \rangle=Nb^2 .$
\\
\\
 For the semiflexible polymer, one can write 
 \begin{align}
     \tau_e=\frac{48}{\phi(0)} \sum_{p: \text{ odds}}^{}\left[\frac{C_{\eta}\tau_p^2}{2\gamma_p^2}+\frac{C_{\sigma_1}}{\gamma_p^2}\frac{\tau _p^2 \left(\tau _A^2-\tau _p^2\right)}{\tau _A^2-\tau _p^2}\right],\label{reconfiguration_semiflex}
 \end{align}
where $\phi(0)$ is given by Eq. (\ref{msd_semiflex_end_OUP}), and $\tau_p = \frac{\gamma}{\frac32 l_p k_B T\left(\frac{p^4\pi^4}{L^4}+\frac{p^2\pi^2}{l_p^2L^2}\right)}.$ We have performed the summation numerically and plotted the results in Fig.  \ref{te_semiflex}.

\begin{figure}
    \centering
\includegraphics[width=0.65\linewidth]{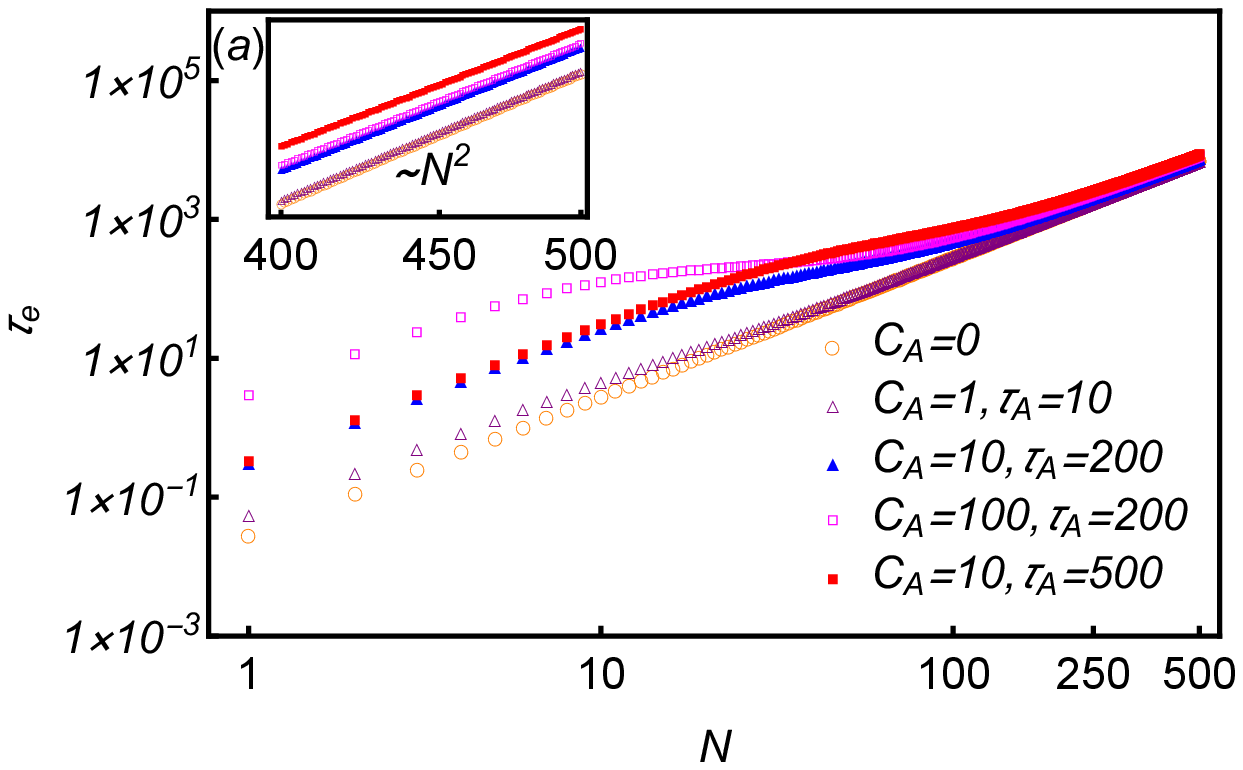}
 \centering
\includegraphics[width=0.65\linewidth]{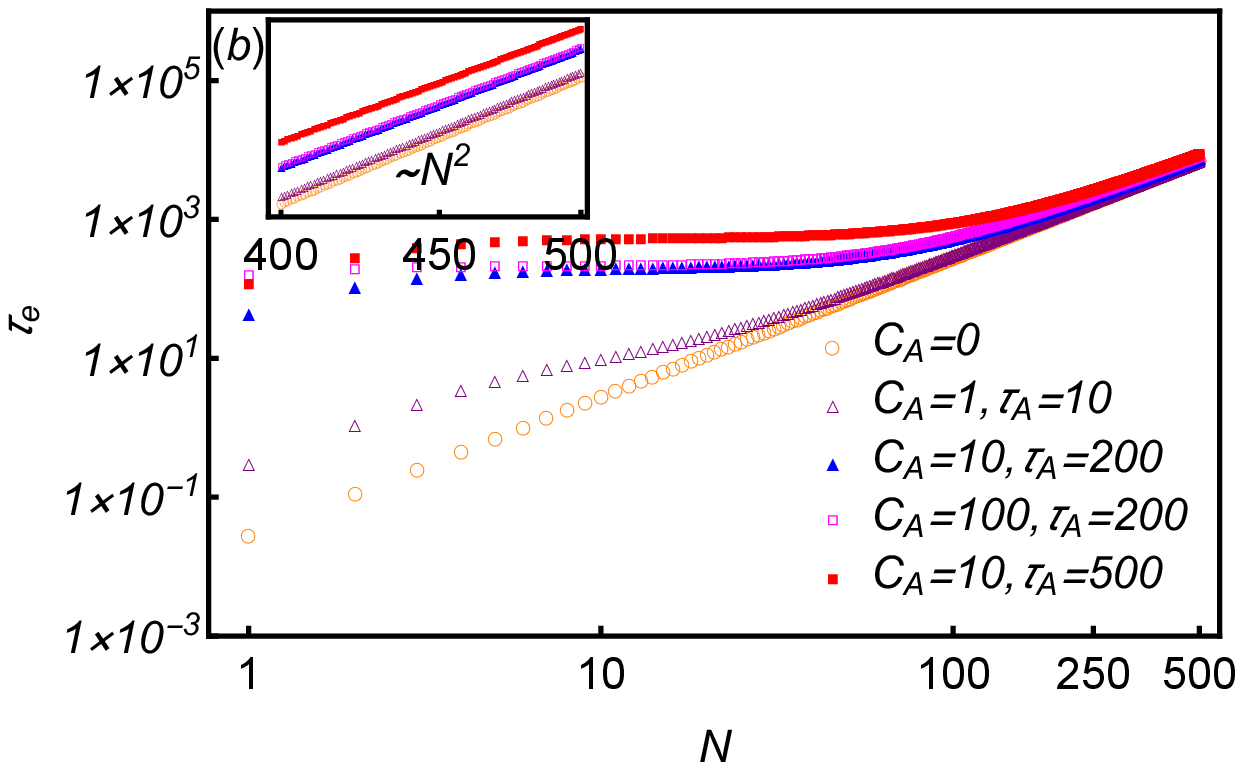}
 \centering
\includegraphics[width=0.65\linewidth]{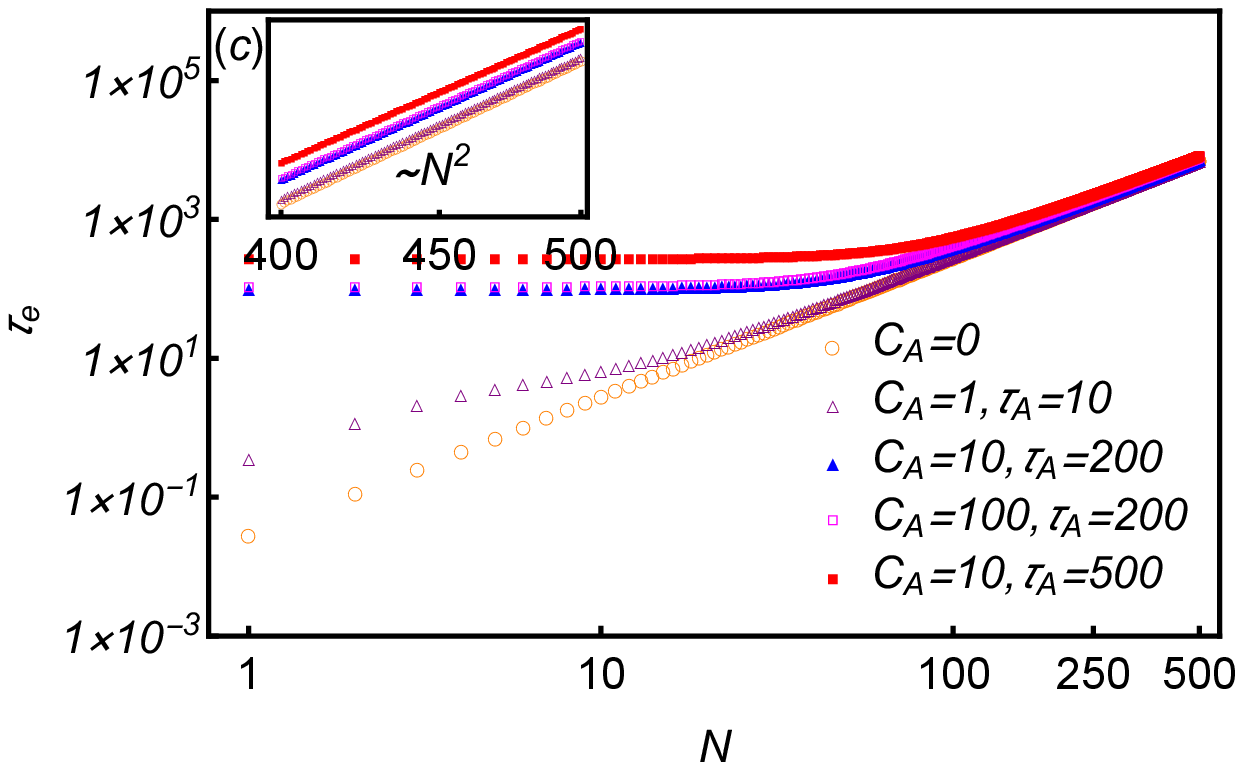}
    \caption{Log-Log plot of  reconfiguration time of the end-to-end vector  versus total  number of monomers $N$ for different values  of noise amplitude and persistence time of active noise in the case of the flexible polymer model  in the (a) OUP, (b) MOUP and (c) non-Gaussian baths.  }
    \label{te_flex}
\end{figure}

\begin{figure}
    \centering
\includegraphics[width=0.65\linewidth]{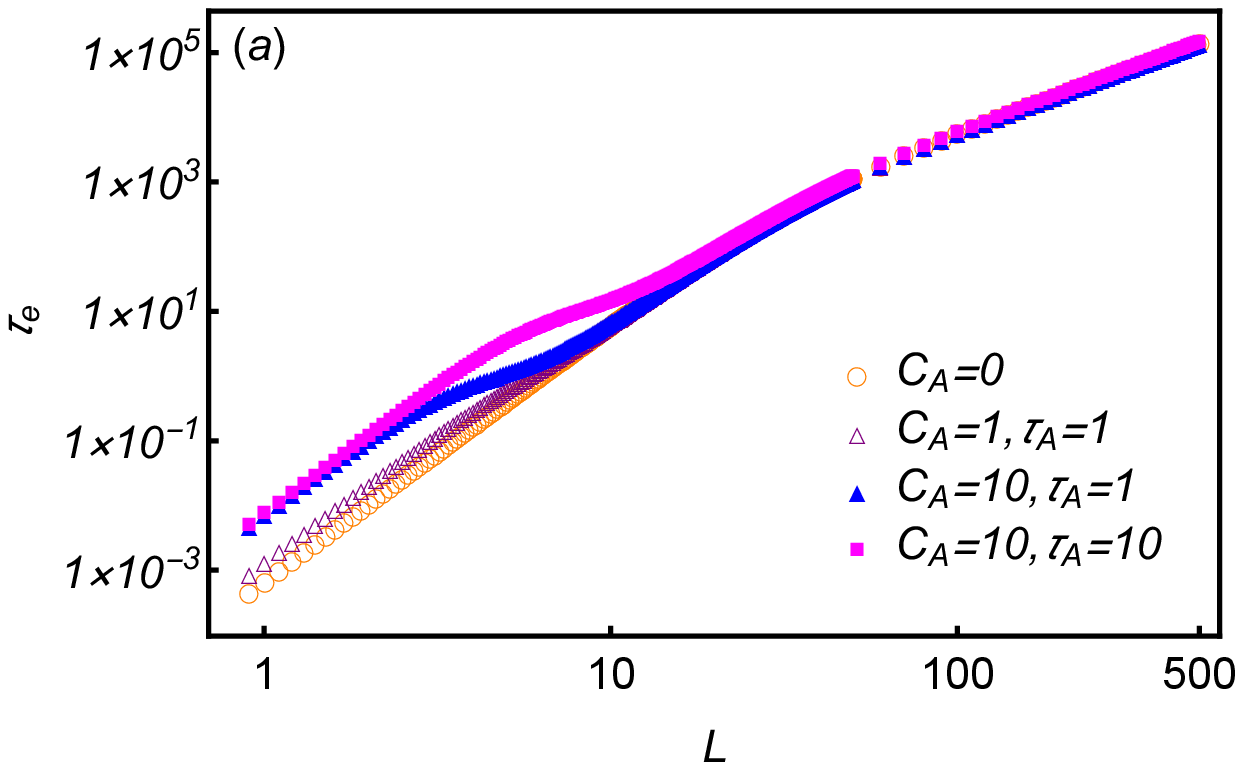}
 \centering
\includegraphics[width=0.65\linewidth]{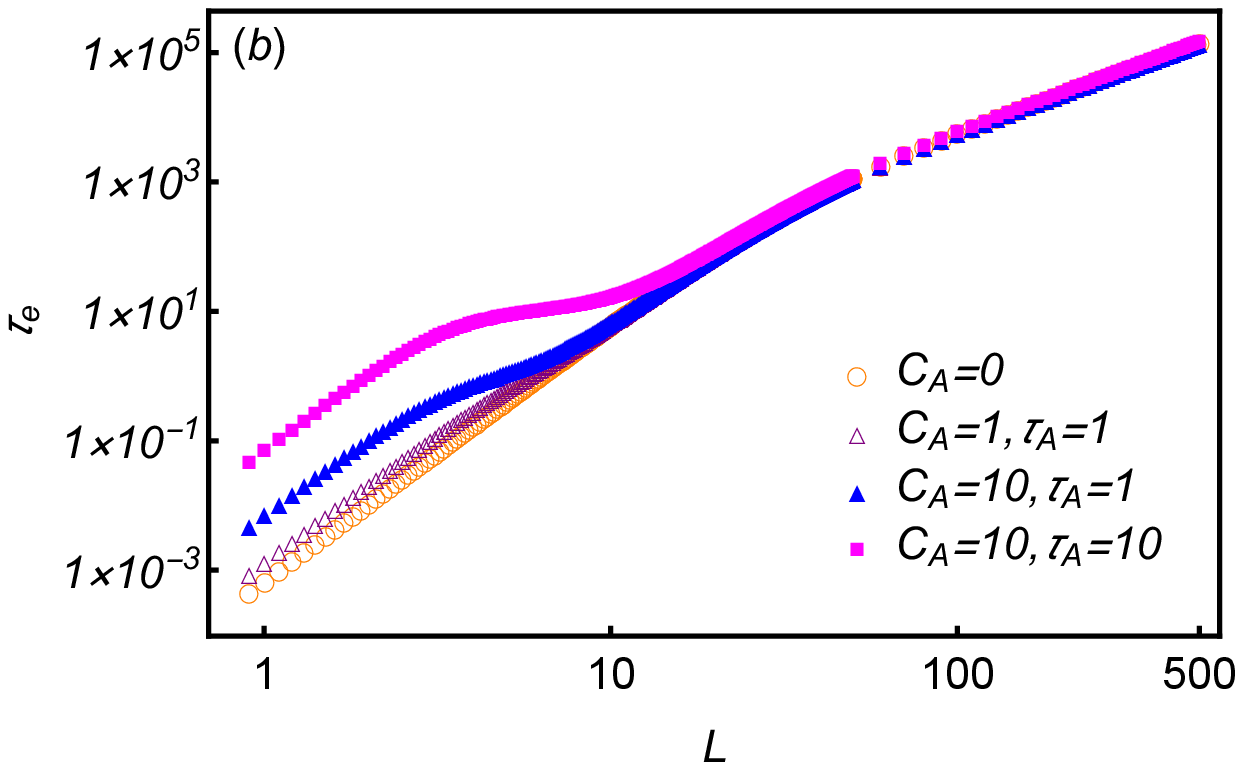}
 \centering
\includegraphics[width=0.65\linewidth]{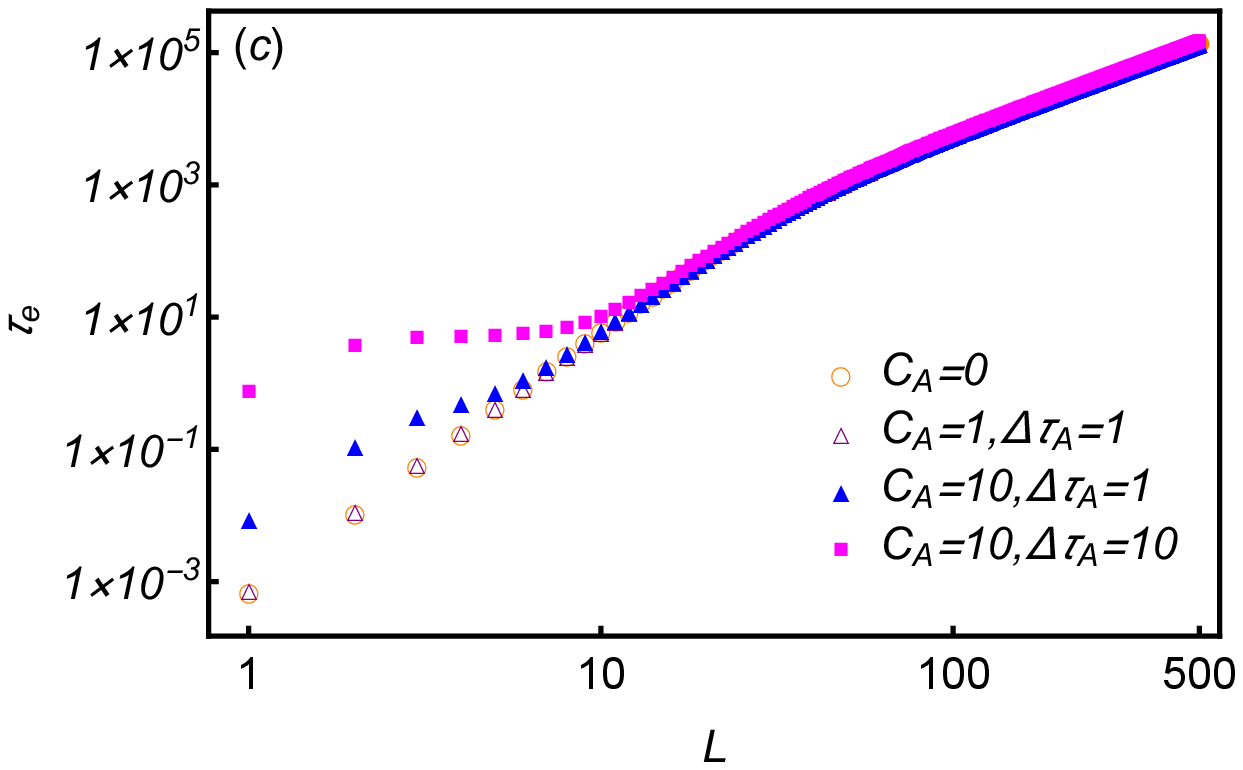}
    \caption{Log-Log plot of  reconfiguration time of the end-to-end vector  versus total  contour length $L$ for different values of noise amplitude and persistence time of active noise in the case of the semiflexible polymer model  in the (a) OUP, (b) MOUP and (c) non-Gaussian baths. Here, we take $l_p=10.$  }
    \label{te_semiflex}
\end{figure}

\subsection{Non-Gaussian active bath}
Using Eq. (\ref{correlation_nong}), the mode relaxation time given in Eq. (\ref{mode_time}) can be calculated as 
\begin{align}
    \tau_m(p)=\tau_p+\frac{\sigma_A^2\tau_p^3\nu_A\,\left(1+(1-\frac{\Delta \tau_A}{\tau_p})^2-2\text{exp}\left(-\frac{\Delta \tau_A}{\tau_p}\right)\right)}{ C_{\eta}+2\sigma_A^2\tau_p^2\nu_A\,\left(\text{exp}\left(-\frac{\Delta \tau_A}{\tau_p}\right)+\frac{\Delta \tau_A}{\tau_p}-1\right)}.
\end{align}
In the limit $\Delta \tau_A\rightarrow 0$, $\tau_m(p)\sim\tau_p.$ With increasing activity, $\tau_m(p)$ does not scale like $\tau_p$ with $p$ as shown in Fig. \ref{tm_flex}. The results are equivalent to the one in the Gaussian baths, as can easily be comprehended from Fig. \ref{tm_semiflex}.

For the flexible polymer in the non-Gaussian bath, one has
\begin{align}
  &\int_{0}^{\infty} dt\,\phi(t)=48\int_{0}^{\infty} dt\,\sum_{p: \text{ odds}}^{}\, \langle \psi_p(t) \psi_p(0) \rangle\nonumber\\
 &=48\sum_{p=1}^{\infty}\,\Bigl[\frac{\tau_1^2 k_B T}{2 N \gamma (2p-1)^4}+\frac{C_A^2 \nu_A}{8 N\gamma^2}\frac{\tau_1^4}{(2p-1)^8}\left(\text{cosh}\left(\frac{\Delta \tau_A}{\tau_1}(2p-1)^2\right)-1\right)\nonumber\\
& \quad+\frac{C_A^2 \nu_A}{8 N\gamma^2}\frac{\tau_1^4}{(2p-1)^8}\,\left(\frac12 \frac{(2p-1)^4}{\tau_1^2}\Delta \tau_A^2-\text{cosh}\left(\frac{(2p-1)^2}{\tau_1}\Delta \tau_A\right)+1\right)\Bigr]\nonumber\\
&=48\sum_{p=1}^{\infty}\,\Bigl[\frac{\tau_1^2 k_B T}{2 N \gamma (2p-1)^4}+\frac{C_A^2 \nu_A}{16 N\gamma^2}\frac{\tau_1^2 \Delta \tau_A^2}{(2p-1)^4}\Bigr]\nonumber\\
&=\frac{\tau_1^2\pi^4}{4 N \gamma}\left[ k_B T+\frac{C_A^2 \Delta \tau_A^2 \nu_A}{8 \gamma}\right].
\end{align} 
So the configuration time [Eq. (\ref{recon_time})] can be expressed as 
\begin{align}
    \tau_e=\frac{\frac{\tau_1^2\pi^4}{4 N \gamma}\left[ k_B T+\frac{C_A^2 \Delta \tau_A^2 \nu_A}{8 \gamma}\right]}{Nb^2+\frac{6\,C_A^2}{N\gamma^2}\sum_{p=1}^{\infty}\,\tau_{2p-1}^3\nu_A\,\left(\text{exp}\left(-\frac{\Delta \tau_A}{\tau_{2p-1}}\right)+\frac{\Delta \tau_A}{\tau_{2p-1}}-1\right)}.
\end{align}
 For the semiflexible polymer in the non-Gaussian bath, from Eq. (\ref{correlation_semiflex_end1})  one gets
\begin{align}
  &\int_{0}^{\infty}dt\, \phi(t)= 48\sum_{p=1}^{\infty}\,\Bigl[\frac{C_{\eta}\tau_p^2}{2\gamma_p^2}+\frac{\sigma_A^2}{\gamma_p^2}\tau_p^4\nu_A\,\left(\text{cosh}\left(\frac{\Delta \tau_A}{\tau_p}\right)-1\right)\Bigr]\nonumber\\
  &=\frac{16 \gamma}{135}\frac{l_p^2L^3}{k_B T}-\frac{32\gamma}{9}\frac{l_p^4L}{k_B T}-16\gamma\frac{l_p^6}{L k_B T}+\frac{40 \gamma}{3}\frac{l_p^5}{k_B T}\text{coth}\left(\frac{L}{l_p}\right)+\frac{8 \gamma}{3}\frac{l_p^4 L}{k_B T}\text{csch }\left(\frac{L}{l_p}\right)\nonumber\\
  &\quad + 48\sum_{p=1}^{\infty}\,\frac{\sigma_A^2}{\gamma_p^2}\tau_p^4\nu_A\,\left(\text{cosh}\left(\frac{\Delta \tau_A}{\tau_p}\right)-1\right).
\end{align}
So the configuration time is given by \begin{align}
    \tau_e=\frac{48\sum_{p=1}^{\infty}\,\Bigl[\frac{C_{\eta}\tau_p^2}{2\gamma_p^2}+\frac{\sigma_A^2}{\gamma_p^2}\tau_p^4\nu_A\,\left(\text{cosh}\left(\frac{\Delta \tau_A}{\tau_p}\right)-1\right)\Bigr]}{2Ll_p\left(1-\frac{2l_p}{L}\text{tanh}\left(\frac{L}{2l_p}\right)\right)+\frac{6\,C_A^2}{N\gamma^2}\sum_{p=1}^{\infty}\,\tau_{2p-1}^3\nu_A\,\left(\text{exp}\left(-\frac{\Delta \tau_A}{\tau_{2p-1}}\right)+\frac{\Delta \tau_A}{\tau_{2p-1}}-1\right)}.
\end{align}
The results of the reconfiguration time are illustrated in figures \ref{te_flex} and \ref{te_semiflex} for the flexible and semiflexible polymers, respectively. For the flexible polymer, $\tau_e\propto N^2$ for the passive case. But in the presence of active noise, short chains ($N=10-200$) have different scaling, $viz.$, $\tau_e \propto N^{\xi},$ where $\xi<2.$ With increasing activity, the value of $\xi$ decreases, which suggests that the reconfiguration time of short chains of variable lengths becomes of similar magnitude at a very high activity. For longer chains, the modes are dominated by the thermal noise, and thus the scaling $\tau_e \propto N^2$ is observed \cite{Samanta_2016,doi:10.1063/1.5086152}. For the semiflexible polymer, at least two distinct regimes can be recognized, namely for $(i)\,L\leq \pi l_p$ and $(ii)\,L\gg \pi l_p,$ as can be seen in Fig. \ref{te_semiflex}. In the first case ($L\leq \pi l_p$), $\tau_e$ is determined by the bending modes, and so $\tau_e \propto L^4,$ reflecting the behavior of $\tau_1.$ Like the flexible case, with increasing activity, the dependence of $L$ on $\tau_e$ for a short chain decreases, $i.e.$, $\tau_e \propto L^{\chi},$ where $\chi<4$ for $C_A>0.$ In the limit $L\gg \pi l_p$, the polymer behaves like a Rouse chain, and thus reconfiguration is governed by the Rouse modes with relaxation time $\tau_1\propto L^2.$ So one can see $\tau_e \propto L^2$ for $L\gg \pi l_p.$ 

\section{Conclusions and future directions\label{conclusion}}
\noindent We have presented an overview of the analytically solvable models to study the conformational and dynamical properties of active flexible and semiflexible polymers. We have considered a continuum representation of an ideal polymer with a certain number of active monomers or a linear polymer subjected to active force. Each of the monomers is following an active Ornstein-Uhlenbeck process where the active noise is modeled as the Gaussian and exponentially correlated noise. In most of the previous theoretical attempts to investigate the dynamics of active polymer, active noise were modeled as Gaussian. As an important extension of the previous studies \cite{Samanta_2016, ghosh2014dynamics, osmanovic2017dynamics, vandebroek2015dynamics, winkler2020physics}, recently we have taken into account the non-Gaussian
nature of active noise arising from the run-and-tumble motion of bacteria \cite{chaki2019effects}. In our model, the non-Gaussian active noise is represented by a series of random pulses and modeled as shot noise. In the presence of both the Gaussian and non-Gaussian noise, the dynamics of a harmonically trapped particle is superdiffusive on the timescale of the persistence time of the active noise and reaches the steady state at the long time limit. However, the tail behavior of the non-Gaussian probability distribution can be obtained from the fourth moment, through the non-Gaussian parameter (NGP). For a harmonically bound particle, the NGP is negative in the steady state for the large persistence time whereas it is zero for the Gaussian distribution.  For small persistence time, the NGP is positive in the steady state. 
\\
\\
In the presence of both the Gaussian and non-Gaussian active noise, activity leads to the monotonic swelling of a flexible polymer over a wide range of persistence times in a power-law manner. Swelling is caused by enhanced fluctuations to the monomers by activity. The semiflexible polymer also exhibits swelling as flexible polymer if the chain length exceeds the persistence length of the polymer. However, bending rigidity increases the extent of swelling of the semiflexible polymer compared to the flexible  one. The center of mass motion of both the flexible and semiflexible polymer  is similar to the one of an active Brownian particle where the dynamics shows a three step growth: short time thermal diffusion, intermediate superdiffusion and enhanced translational diffusion at longer times. This sort of growth has recently been seen experimentally for colloidal chains subjected to active fluctuations \cite{mousavi2021active, biswas2017}. However, the translational diffusion coefficient of the center of mass is $N$ times smaller compared to the active Brownian particle where $N$ is the number of monomers. The dynamics of a tagged monomer exhibits four step growth as the connectivity between the monomers adds another time scale in the system. Before the Rouse relaxation time, the initial and long time growth of a tagged monomer exhibit Rouse-like (subdiffusive) behaviour, whereas it shows transient superdiffusion at intermediate times. At times larger than the the Rouse relaxation time, the dynamics of the tagged monomer is entirely governed by the center of mass of the chain. 
\\
\\
Polymer exhibits a spectrum of timescales of fluctuation  or relaxation modes and each mode associated with a different length scale.  As a result, the reconfiguration dynamics will depend on the end-to-end distance of the chain. For short polymers, the persistence time of the active noise will be greater than the Rouse relaxation time, and the end-to-end autocorrelation function will be governed by the activity. Hence, the reconfiguration time increases with increasing the persistence time of the active noise. However, for long polymers, Rouse relaxation time will dominate over the the persistence time of the active noise and the reconfiguration time becomes independent of the activity. For the lower modes of a flexible polymer, the mode relaxation time coincides with Rouse relaxation time. With increasing the number of modes, it starts deviating from Rouse relaxation time. However for very large mode number, it again shows quadratic scaling but has higher value than Rouse relaxation time. For semiflexible polymer, the mode relaxation time exhibits a crossover from a quadratic to a quartic dependence on the mode number with increasing mode number. 
\\
\\
 \noindent  The nucleus of a cell is a nonequilibrium system where several active processes dictate the three-dimensional spatiotemporal organization of the chromosome for the proper functioning of the cell \cite{eaton2020structural}. Recent experiments have demonstrated two types of chromatin motion: a fast local motion, observed by tracking single genes \cite{weber2012nonthermal}, and a large-scale coherent motion \cite{zidovska2013micron} which are indications of ATP-powered activity \cite{agrawal2017chromatin}. Such observations invoke a number of theoretical models  which are based on the active polymer models discussed here. In some of the contexts of chromosomal dynamics, the effect of ATP-driven active processes is incorporated via delta-correlated noise but with higher amplitude than the thermal noise to explain the phase separation between heterochromatin and euchromatin \cite{liu2018chain,agrawal2020nonequilibrium}. There are other ways to model the effect of activity on the chromosomal dynamics, such as by introducing tangential dipolar active forces to explain the large-scale coherent motion \cite{saintillan2018extensile,chaki2022polymer}. 
However, full treatment of these topics are beyond the scope of this topical review.


\section*{Acknowledgements\label{acknowledements}}
\noindent R.C. acknowledges SERB (Project No. MTR/2020/000230 under MATRICS scheme) and IRCC-IIT Bombay (Project No. RD/0518-IRCCAW0-001) for funding. K.G. acknowledges IIT Bombay for support through the institute postdoctoral fellowship. S.C. acknowledges DST INSPIRE for a fellowship. The authors would like to acknowledge Dr. Nairhita Samanta for her earlier collaborative works with the group. The authors would like to thank Sanaa Sharma for reading the manuscript.
\section*{Data availability statement}
All data that support the findings of this study are included within the article (and any supplementary files)
\appendix
\section{Computation of  Eq. (\ref{correlationx_Gaussian}) \label{appen1}}
The first term on the right-hand side (RHS) of Eq. (\ref{correlationx_Gaussian}) for $t>t'$ is 
\begin{align}
&\frac{1}{\gamma_p^2}\int_{-\infty}^{t}dt_1 \int_{-\infty}^{t'}dt_2\,e^{-\frac{(t-t_1)}{\tau_p}-\frac{(t'-t_2)}{\tau_p}}\langle \eta_p(t_1)\eta_p(t_2)\rangle\nonumber\\
&=\frac{C_{\eta}}{\gamma_p^2}\int_{-\infty}^{t}dt_1 \int_{-\infty}^{t}dt_2\,e^{-\frac{(t-t_1)}{\tau_p}-\frac{(t'-t_2)}{\tau_p}}\delta(t_1-t_2)\Theta(t'-t_2)\Theta(t-t')\nonumber\\
&=\frac{C_{\eta}}{\gamma_p^2}\int_{-\infty}^{t}dt_1\,e^{-\frac{(t+t'-2t_1)}{\tau_p}}\Theta(t'-t_1)\Theta(t-t')\nonumber\\
&=\frac{C_{\eta}}{\gamma_p^2}\int_{-\infty}^{t'}dt_1\,e^{-\frac{(t+t'-2t_1)}{\tau_p}}=\frac{C_{\eta}\tau_p}{2\gamma_p^2}e^{-\frac{(t-t')}{\tau_p}}.
\end{align}

The second term on the RHS for $t>t'$ is 
\begin{align}
&\frac{1}{\gamma_p^2}\int_{-\infty}^{t}dt_1 \int_{-\infty}^{t'}dt_2\,e^{-\frac{(t-t_1)}{\tau_p}-\frac{(t'-t_2)}{\tau_p}}\langle \sigma_p(t_1)\sigma_p(t_2)\rangle\nonumber\\  
&=\frac{C_{\sigma_1}}{\tau_A\gamma_p^2}\int_{-\infty}^{t}dt_1 \int_{-\infty}^{t}dt_2\,e^{-\frac{(t-t_1)}{\tau_p}-\frac{(t'-t_2)}{\tau_p}}e^{-\frac{|t_1-t_2|}{\tau_A}}\Theta(t'-t_2)\Theta(t-t')\nonumber\\
&=\frac{C_{\sigma_1}}{\gamma_p^2 \tau_A}\int_{-\infty}^{t}dt_1 \int_{-\infty}^{t}dt_2\,e^{-\frac{(t-t_1)}{\tau_p}-\frac{(t'-t_2)}{\tau_p}}e^{-\frac{(t_1-t_2)}{\tau_A}}\Theta(t'-t_2)\Theta(t-t')\Theta(t_1-t_2)\nonumber\\
&+\frac{C_{\sigma_1}}{\gamma_p^2 \tau_A}\int_{-\infty}^{t}dt_1 \int_{-\infty}^{t}dt_2\,e^{-\frac{(t-t_1)}{\tau_p}-\frac{(t'-t_2)}{\tau_p}}e^{-\frac{(t_2-t_1)}{\tau_A}}\Theta(t'-t_2)\Theta(t-t')\Theta(t_2-t_1)\nonumber\\
&=\frac{C_{\sigma_1}}{\gamma_p^2 \tau_A}\int_{-\infty}^{t}dt_2\int_{t_2}^{t}dt_1\,e^{-\frac{(t-t_1)}{\tau_p}-\frac{(t'-t_2)}{\tau_p}}e^{-\frac{(t_1-t_2)}{\tau_A}}\Theta(t'-t_2)\Theta(t-t')\nonumber\\
&+\frac{C_{\sigma_1}}{\gamma_p^2 \tau_A}\int_{-\infty}^{t}dt_2 \int_{-\infty}^{t_2}dt_1\,e^{-\frac{(t-t_1)}{\tau_p}-\frac{(t'-t_2)}{\tau_p}}e^{-\frac{(t_2-t_1)}{\tau_A}}\Theta(t'-t_2)\Theta(t-t')\nonumber\\
&=\frac{C_{\sigma_1}}{\gamma_p^2 \tau_A}\int_{-\infty}^{t'}dt_2\,\left[\frac{\tau _A \tau _p \left(e^{t_2 \left(\frac{1}{\tau _A}-\frac{1}{\tau _p}\right)}-e^{t \left(\frac{1}{\tau _A}-\frac{1}{\tau _p}\right)}\right) e^{-\frac{t}{\tau _A}-\frac{t'-2 t_2}{\tau _p}}}{\tau _A-\tau _p}+\frac{\tau _A \tau _p e^{-\frac{(t'+t-2 t_2)}{\tau _p}}}{\tau _A+\tau _p}\right]\nonumber\\
&=\frac{C_{\sigma_1}}{\gamma_p^2}\frac{ \tau _p^2 \left(\tau _A e^{-\frac{|t-t'|}{\tau _A}}-\tau _p e^{-\frac{|t-t'|}{\tau _p}}\right)}{\tau _A^2-\tau _p^2}.
\end{align}

\section{ Correlations of noise $\sigma_p(t)$ \label{appen2}}
The noise given in Eq. (\ref{sigmap}) is 
\begin{align}
\sigma_p(t)=\sum_i\,h_i\,g(t-t_i),
\end{align}
where 
$g(t)=\sigma_A [\Theta(t)-\Theta(t-\Delta \tau_A)]$ and $h_i=\pm 1.$
Using the Campbell's theorem, we can find the correlations of the noise as follows:
 
\noindent Considering $t_1>t_2,$
the second-order correlation can be given as 
\begin{align}
&\langle\sigma_p(t_1)\sigma_p(t_2)\rangle\nonumber\\
&=\nu_A\int_{0}^{\infty}dt'\,g(t_1-t')g(t_2-t')\nonumber\\
&=\nu_A\sigma_A^2\int_{0}^{\infty}dt'\,[\Theta(t_1-t')-\Theta(t_1-\Delta \tau_A-t')][\Theta(t_2-t')-\Theta(t_2-\Delta \tau_A-t')]\nonumber\\
&=\nu_A\sigma_A^2\int_{0}^{\infty}dt'\,\Theta(t_1-t')\Theta(t_2-t')-\nu_A\sigma_A^2\int_{0}^{\infty}dt'\,\Theta(t_1-t')\Theta(t_2-\Delta \tau_A-t')\nonumber\\
&-\nu_A\sigma_A^2\int_{0}^{\infty}dt'\,\Theta(t_2-t')\Theta(t_1-\Delta \tau_A-t')+\nu_A\sigma_A^2\int_{0}^{\infty}dt'\,\Theta(t_1-\Delta \tau_A-t')\Theta(t_2-\Delta \tau_A-t')\nonumber\\
&=\nu_A\sigma_A^2\left[\int_{0}^{t_2}dt'-\int_{0}^{t_2-\Delta \tau_A}dt'-\int_{0}^{t_1-\Delta \tau_A}dt'+\int_{0}^{t_2-\Delta \tau_A}dt'\right],\,\text{for }  \Delta \tau_A>|t_1-t_2|\nonumber\\
&=\nu_A\sigma_A^2\left(\Delta \tau_A-|t_1-t_2|\right)\Theta(\Delta \tau_A-|t_1-t_2|)\label{correlation_shot}.
\end{align}

\noindent Considering $t_1>t_2>t_3>t_4,$ the fourth-order correlation can be calculated as follows:
\begin{align}
&\langle\sigma_p(t_1)\sigma_p(t_2)\sigma_p(t_3)\sigma_p(t_4)\rangle\nonumber\\
&=\nu_A\int_{0}^{\infty}dt'\,g(t_1-t')g(t_2-t')g(t_3-t')g(t_4-t')\nonumber\\
&=\nu_A\sigma_A^4\int_{0}^{\infty}dt'\,[\Theta(t_1-t')-\Theta(t_1-\Delta \tau_A-t')][\Theta(t_2-t')-\Theta(t_2-\Delta \tau_A-t')]\nonumber\\
&\quad\quad \times [\Theta(t_3-t')-\Theta(t_3-\Delta \tau_A-t')][\Theta(t_4-t')-\Theta(t_4-\Delta \tau_A-t')]\nonumber\\
&=\nu_A\sigma_A^4\int_{0}^{\infty}dt'\,\Bigl(\Theta(t_1-t')\Theta(t_2-t')-\Theta(t_1-t')\Theta(t_2-\Delta \tau_A-t')-\Theta(t_1-\Delta \tau_A-t')\Theta(t_2-t')\nonumber\\&\quad\quad+ \Theta(t_1-\Delta \tau_A-t')\Theta(t_2-\Delta \tau_A-t')\Bigr)\Bigl(\Theta(t_3-t')\Theta(t_4-t')-\Theta(t_3-t')\Theta(t_4-\Delta \tau_A-t')\nonumber\\&\quad\quad-\Theta(t_3-\Delta \tau_A-t')\Theta(t_4-t')+ \Theta(t_3-\Delta \tau_A-t')\Theta(t_4-\Delta \tau_A-t')
\Bigr)\nonumber\\
&=\nu_A\sigma_A^4\Bigl(\int_{0}^{t_4}dt'-\int_{0}^{t_4-\Delta \tau_A}dt'-\int_{0}^{\text{min}(t_4,\,t_3-\Delta \tau_A)}dt'+\int_{0}^{t_4-\Delta \tau_A}dt'-\int_{0}^{\text{min}(t_4,\,t_2-\Delta \tau_A)}dt'\nonumber\\
&\quad\quad + \int_{0}^{t_4-\Delta \tau_A}dt' +\int_{0}^{\text{min}(t_4,\,t_3-\Delta \tau_A)}dt'-\int_{0}^{t_4-\Delta \tau_A}dt' -\int_{0}^{\text{min}(t_4,\,t_1-\Delta \tau_A)}dt'+\int_{0}^{t_4-\Delta \tau_A}dt'\nonumber\\
&\quad\quad +\int_{0}^{\text{min}(t_4,\,t_3-\Delta \tau_A)}dt'-\int_{0}^{t_4-\Delta \tau_A}dt' +\int_{0}^{\text{min}(t_4,\,t_2-\Delta \tau_A)}dt'-\int_{0}^{t_4-\Delta \tau_A}dt'-\int_{0}^{\text{min}(t_4,\,t_3-\Delta \tau_A)}dt'\nonumber\\
&\quad\quad +\int_{0}^{t_4-\Delta \tau_A}dt'\Bigr)\nonumber\\
&=\nu_A\sigma_A^4\Bigl(\int_{0}^{t_4}dt'-\int_{0}^{\text{min}(t_4,\,t_1-\Delta \tau_A)}dt'\Bigr)=\nu_A\sigma_A^4\Bigl(t_4-\text{min}(t_4,\,t_1-\Delta \tau_A)\Bigr)\nonumber\\
&=\nu_A\sigma_A^4\left(\Delta \tau_A-|t_1-t_4|\right)\Theta(\Delta \tau_A-|t_1-t_4|).
\end{align}

\section{Solving Eq. (\ref{msd_x0_nong}) \label{appen3}}
The second term on the RHS  can calculated using the  Laplace transformation defined as $\mathcal{L}[f(t)]=\Tilde{f}(s)=\int_{0}^{\infty}dt\,e^{-st}f(t),$ and it reads
\begin{align}
&\mathcal{L}\left[\int_{0}^{t}dt_1\int_{0}^{t}dt_2\,\langle\sigma_0(t_1)\sigma_0(t_2)\rangle\right]\nonumber\\
&=\sigma_A^2\nu_A\,\mathcal{L}\left[\int_{0}^{t}dt_1\int_{0}^{t}dt_2\, \left(\Delta \tau_A-|t_1-t_2|\right)\,\Theta\left(\Delta \tau_A-|t_1-t_2|\right)\right]\nonumber\\
&=\sigma_A^2\nu_A\,\mathcal{L}\Bigl[\int_{0}^{t}dt_1\int_{0}^{t_1}dt_2\, \left(\Delta \tau_A-(t_1-t_2)\right)\,\Theta\left(\Delta \tau_A-(t_1-t_2)\right)\nonumber\\
&\quad\quad\quad+\int_{0}^{t}dt_2\int_{0}^{t_2}dt_1\, \left(\Delta \tau_A-(t_2-t_1)\right)\,\Theta\left(\Delta \tau_A-(t_2-t_1)\right)\Bigr]\nonumber\\
&=2\sigma_A^2\nu_A\,\Bigl[\frac{\Delta \tau_A  \left(1-e^{-\Delta\tau_A   s}\right)}{s^3}-\frac{e^{-\Delta \tau_A s} \left(-\Delta \tau_A s+e^{\Delta \tau_A  s}-1\right)}{s^4}\Bigr].\nonumber\\
\end{align}
 On doing the inverse Laplace transform, one obtains
 \begin{align}
&\int_{0}^{t}dt_1\int_{0}^{t}dt_2\,\langle\sigma_0(t_1)\sigma_0(t_2)\rangle\nonumber\\
&=2\sigma_A^2\nu_A\,\mathcal{L}^{-1}\Bigl[\frac{\Delta \tau_A  \left(1-e^{-\Delta\tau_A   s}\right)}{s^3}-\frac{e^{-\Delta \tau_A s} \left(-\Delta \tau_A s+e^{\Delta \tau_A  s}-1\right)}{s^4}\Bigr]\nonumber\\
&=\frac{2\sigma_A^2\nu_A}{6} \left(\left(t-\Delta \tau _A\right)^3 \Theta \left(t-\Delta \tau _A\right)+3 t^2 \Delta \tau _A-t^3\right)\nonumber\\
&=\frac{\sigma_A^2\nu_A}{3} \left(\left(\Delta \tau _A-t\right)^3 \Theta \left(\Delta \tau _A-t\right)+ \Delta \tau _A^2(3t-\Delta \tau _A)\right).
 \end{align}

\section{Solving Eq. (\ref{correlation_nong})  \label{appen4}}
   
 The integration involving $\sigma_A(t)$ on the RHS of Eq. (\ref{correlation_nong})  can be performed as given below. 
\begin{align}
&\int_{-\infty}^{t}dt_1 \int_{-\infty}^{t'}dt_2\,e^{-\frac{(t-t_1)}{\tau_p}-\frac{(t'-t_2)}{\tau_p}}\langle \sigma_p(t_1)\sigma_p(t_2)\rangle\nonumber\\
&=\sigma_A^2\nu_A\,\int_{-\infty}^{t}dt_1\int_{-\infty}^{t'}dt_2\,e^{-\frac{(t-t_1)}{\tau_p}-\frac{(t'-t_2)}{\tau_p}}\, \left(\Delta \tau_A-|t_1-t_2|\right)\,\Theta\left(\Delta \tau_A-|t_1-t_2|\right)\nonumber\\
&=\sigma_A^2\nu_A\,\int_{-\infty}^{+\infty}dt_1\int_{-\infty}^{+\infty}dt_2\,e^{-\frac{(t-t_1)}{\tau_p}-\frac{(t'-t_2)}{\tau_p}}\, \left(\Delta \tau_A-|t_1-t_2|\right)\,\Theta\left(\Delta \tau_A-|t_1-t_2|\right)\,\Theta\left(t-t_1\right)\,\Theta\left(t'-t_2\right)\nonumber\\
&=\sigma_A^2\nu_A\,\int_{-\infty}^{+\infty}dt_1\int_{-\infty}^{+\infty}dt_2\,e^{-\frac{(t-t_1)}{\tau_p}-\frac{(t'-t_2)}{\tau_p}}\, \left(\Delta \tau_A-(t_1-t_2)\right)\,\Theta\left(\Delta \tau_A-(t_1-t_2)\right)\,\Theta\left(t-t_1\right)\nonumber\\
&\,\qquad\qquad\qquad\qquad\qquad\qquad\qquad\qquad\qquad\qquad\qquad\qquad\qquad\qquad\Theta\left(t'-t_2\right)\,\Theta\left(t_1-t_2\right)\nonumber\\
&+\sigma_A^2\nu_A\,\int_{-\infty}^{+\infty}dt_1\int_{-\infty}^{+\infty}dt_2\,e^{-\frac{(t-t_1)}{\tau_p}-\frac{(t'-t_2)}{\tau_p}}\, \left(\Delta \tau_A-(t_2-t_1)\right)\Theta\left(\Delta \tau_A-(t_2-t_1)\right)\,\nonumber\\
& \qquad\qquad\qquad\qquad\qquad\qquad\qquad\qquad\qquad\qquad\qquad\qquad\qquad \Theta\left(t-t_1\right)\,\Theta\left(t'-t_2\right)\,\Theta\left(t_2-t_1\right)\nonumber\\
&=\sigma_A^2\tau_p^3\nu_A\,\Theta\left(\Delta \tau_A-|t-t'|\right)\left(\frac{1}{\tau_p}\left(\Delta \tau_A-|t-t'|\right)-\text{sinh}\left(\frac{1}{\tau_p}\left(\Delta \tau_A-|t-t'|\right)\right)\right)\nonumber\\
&\quad+\sigma_A^2\tau_p^3\nu_A\,e^{-\frac{|t-t'|}{\tau_p}}\left(\text{cosh}\left(\frac{\Delta \tau_A}{\tau_p}\right)-1\right).
\end{align}

\section{Calculation of fourth moment}

The Fourth moment of the position  described by Eq. (\ref{langevin}) subjected to the non-Gaussian noise can be evaluated in the the steady-state limit using the correlation functions of the noises. The calculation is as follows:

\begin{align}
&\langle x_p^4(t) \rangle\nonumber\\
&=\frac{1}{\gamma_p^4}\int_{-\infty}^{t}dt_1 \int_{-\infty}^{t}dt_2\int_{-\infty}^{t}dt_3\int_{-\infty}^{t}dt_4\,e^{-\frac{(t-t_1)}{\tau_p}-\frac{(t-t_2)}{\tau_p}-\frac{(t-t_3)}{\tau_p}-\frac{(t-t_4)}{\tau_p}}\langle (\eta_p(t_1)+\sigma_p(t_1))(\eta_p(t_2)+\sigma_p(t_2))\nonumber\\
&\quad\quad\quad\quad\quad\quad\quad\quad\quad\quad\quad\quad\quad\quad\quad\quad(\eta_p(t_3)+\sigma_p(t_3))(\eta_p(t_4)+\sigma_p(t_4))\rangle\nonumber\\
&=\frac{1}{\gamma_p^4}\int_{-\infty}^{t}dt_1 \int_{-\infty}^{t}dt_2\int_{-\infty}^{t}dt_3\int_{-\infty}^{t}dt_4\,e^{-\frac{(t-t_1)}{\tau_p}-\frac{(t-t_2)}{\tau_p}-\frac{(t-t_3)}{\tau_p}-\frac{(t-t_4)}{\tau_p}}\langle \eta_p(t_1)\eta_p(t_2)\eta_p(t_3)\eta_p(t_4)\rangle\nonumber\\
&+\frac{2}{\gamma_p^4}\int_{-\infty}^{t}dt_1 \int_{-\infty}^{t}dt_2\int_{-\infty}^{t}dt_3\int_{-\infty}^{t}dt_4\,e^{-\frac{(t-t_1)}{\tau_p}-\frac{(t-t_2)}{\tau_p}-\frac{(t-t_3)}{\tau_p}-\frac{(t-t_4)}{\tau_p}}\langle \eta_p(t_1)\eta_p(t_2)\sigma_p(t_3)\sigma_p(t_4)\rangle\nonumber\\
&+\frac{1}{\gamma_p^4}\int_{-\infty}^{t}dt_1 \int_{-\infty}^{t}dt_2\int_{-\infty}^{t}dt_3\int_{-\infty}^{t}dt_4\,e^{-\frac{(t-t_1)}{\tau_p}-\frac{(t-t_2)}{\tau_p}-\frac{(t-t_3)}{\tau_p}-\frac{(t-t_4)}{\tau_p}}\langle \sigma_p(t_1)\sigma_p(t_2)\sigma_p(t_3)\sigma_p(t_4)\rangle\nonumber
\end{align}

\begin{align}
& \frac{1}{\gamma_p^4}\int_{-\infty}^{t}dt_1 \int_{-\infty}^{t}dt_2\int_{-\infty}^{t}dt_3\int_{-\infty}^{t}dt_4\,e^{-\frac{(t-t_1)}{\tau_p}-\frac{(t-t_2)}{\tau_p}-\frac{(t-t_3)}{\tau_p}-\frac{(t-t_4)}{\tau_p}}\langle \eta_p(t_1)\eta_p(t_2)\eta_p(t_3)\eta_p(t_4)\rangle\nonumber\\
&=3\frac{C_{\eta}^2}{\gamma_p^4}\int_{-\infty}^{t}dt_1 \int_{-\infty}^{t}dt_2\int_{-\infty}^{t}dt_3\int_{-\infty}^{t}dt_4\,e^{-\frac{(t-t_1)}{\tau_p}-\frac{(t-t_2)}{\tau_p}-\frac{(t-t_3)}{\tau_p}-\frac{(t-t_4)}{\tau_p}}\delta(t_1-t_2)\delta(t_3-t_4)\nonumber\\
 &=3\frac{C_{\eta}^2\tau_p^2}{4\gamma_p^4}.
\end{align}
Now one can compute the integration involving the non-Gaussian noise as follows:
\begin{align}
 &\frac{1}{\gamma_p^4}\int_{-\infty}^{t}dt_1 \int_{-\infty}^{t}dt_2\int_{-\infty}^{t}dt_3\int_{-\infty}^{t}dt_4\,e^{-\frac{(t-t_1)}{\tau_p}-\frac{(t-t_2)}{\tau_p}-\frac{(t-t_3)}{\tau_p}-\frac{(t-t_4)}{\tau_p}}\langle \sigma_p(t_1)\sigma_p(t_2)\sigma_p(t_3)\sigma_p(t_4)\rangle\nonumber\\
 &=\frac{\nu_A\sigma_A^4}{\gamma_p^4}\int_{-\infty}^{t}dt_1 \int_{-\infty}^{t}dt_2\int_{-\infty}^{t}dt_3\int_{-\infty}^{t}dt_4\,e^{-\frac{(t-t_1)}{\tau_p}-\frac{(t-t_2)}{\tau_p}-\frac{(t-t_3)}{\tau_p}-\frac{(t-t_4)}{\tau_p}}\left(\Delta \tau_A-|t_1-t_4|\right)\Theta(\Delta \tau_A-|t_1-t_4|)\nonumber\\
 &=3\nu_A\tau_p^5\,\frac{\sigma_A^4}{\gamma_p^4}\left(\text{exp}\left(-\frac{\Delta \tau_A}{\tau_p}\right)+\frac{\Delta \tau_A}{\tau_p}-1\right)
\end{align}

\noindent The integration involving white noise can be done easily, and it reads
\begin{align}
& \frac{2}{\gamma_p^4}\int_{-\infty}^{t}dt_1 \int_{-\infty}^{t}dt_2\int_{-\infty}^{t}dt_3\int_{-\infty}^{t}dt_4\,e^{-\frac{(t-t_1)}{\tau_p}-\frac{(t-t_2)}{\tau_p}-\frac{(t-t_3)}{\tau_p}-\frac{(t-t_4)}{\tau_p}}\langle \eta_p(t_1)\eta_p(t_2)\sigma_p(t_3)\sigma_p(t_4)\rangle\nonumber\\
&=\frac{6}{\gamma_p^4}\int_{-\infty}^{t}dt_1 \int_{-\infty}^{t}dt_2\int_{-\infty}^{t}dt_3\int_{-\infty}^{t}dt_4\,e^{-\frac{(t-t_1)}{\tau_p}-\frac{(t-t_2)}{\tau_p}-\frac{(t-t_3)}{\tau_p}-\frac{(t-t_4)}{\tau_p}}\langle \eta_p(t_1)\eta_p(t_2)\rangle\langle\sigma_p(t_3)\sigma_p(t_4)\rangle\nonumber\\
&=6\frac{C_{\eta}\tau_p}{2\gamma_p^2}\frac{\sigma_A^2}{\gamma_p^2}\tau_p^3\nu_A\,\left(\text{exp}\left(-\frac{\Delta \tau_A}{\tau_p}\right)+\frac{\Delta \tau_A}{\tau_p}-1\right).
\end{align}

\noindent Therefore, the fourth moment is given by
\begin{align}
& \langle x_p^4(t) \rangle= 3\left(\frac{C_{\eta}\tau_p}{2\gamma_p^2}+\frac{\sigma_A^2}{\gamma_p^2}\tau_p^3\nu_A\,\left(\text{exp}\left(-\frac{\Delta \tau_A}{\tau_p}\right)+\frac{\Delta \tau_A}{\tau_p}-1\right)\right)^2\nonumber\\
&\quad\quad +3\nu_A\tau_p^5\,\frac{\sigma_A^4}{\gamma_p^4}\left(\text{exp}\left(-\frac{\Delta \tau_A}{\tau_p}\right)+\frac{\Delta \tau_A}{\tau_p}-1\right)\left[1-\nu_A\tau_p\left(\text{exp}\left(-\frac{\Delta \tau_A}{\tau_p}\right)+\frac{\Delta \tau_A}{\tau_p}-1\right)\right]\nonumber\\
&=3\langle x_p^2(t) \rangle^2+3\nu_A\tau_p^5\,\frac{\sigma_A^4}{\gamma_p^4}\left(\text{exp}\left(-\frac{\Delta \tau_A}{\tau_p}\right)+\frac{\Delta \tau_A}{\tau_p}-1\right)\left[1+\nu_A(\tau_p-\Delta \tau_A)-\nu_A\tau_p\text{exp}\left(-\frac{\Delta \tau_A}{\tau_p}\right)\right]\label{x^4_nongauss}.
\end{align}

\providecommand{\newblock}{}

\end{document}